\documentclass[useAMS,usenatbib,letterpaper,usegraphicx]{mn2e}

\usepackage{times,amssymb,hyperref,subfigure,epsfig,aas_macros,longtable,threeparttablex}

\usepackage[totalwidth=480pt,totalheight=680pt,layoutvoffset=0.5cm]{geometry}

\newcommand{\rt}{$\mathcal{R}_T$}
\newcommand{\rf}{$\mathcal{R}_f$}
\newcommand{\rn}{$\mathcal{R}_n$}

\begin{document}


\title[Two Temperature Debris Disks: Multiple Belts?]{Do Two Temperature Debris Disks
  Have Multiple Belts?}

\author[G. M. Kennedy \& M. C. Wyatt]{G. M. Kennedy\thanks{Email:
    \href{mailto:gkennedy@ast.cam.ac.uk}{gkennedy@ast.cam.ac.uk}}, M. C. Wyatt \\
  Institute of Astronomy, University of Cambridge, Madingley Road, Cambridge CB3 0HA, UK}
\maketitle

\begin{abstract}
  We present a study of debris disks whose spectra are well modelled by dust emission at
  two different temperatures. These disks are typically assumed to be a sign of multiple
  belts, which in only a few cases have been confirmed via high resolution
  observations. We first compile a sample of two-temperature disks to derive their
  properties, summarised by the ratios of the warm and cool component temperatures and
  fractional luminosities. The ratio of warm to cool temperatures is constant in the
  range 2-4, and the temperatures of both warm and cool components increases with stellar
  mass. We then explore whether this emission can arise from dust in a single narrow
  belt, with the range of temperatures arising from the size variation of grain
  temperatures. This model can produce two-temperature spectra for Sun-like stars, but is
  not supported where it can be tested by observed disk sizes and far-IR/mm spectral
  slopes. Therefore, while some two-temperature disks arise from single belts, it is
  probable that most have multiple spatial components. These disks are plausibly similar
  to the outer Solar System's configuration of Asteroid and Edgeworth-Kuiper belts
  separated by giant planets. Alternatively, the inner component could arise from inward
  scattering of material from the outer belt, again due to intervening planets. In either
  case, we suggest that the ratio of warm/cool component temperatures is indicative of
  the scale of outer planetary systems, which typically span a factor of about ten in
  radius.
\end{abstract}


\section{Introduction}\label{s:intro}

Debris disks are a sign of successful planetesimal formation. The radial structure of
most planetesimal belts is unknown; they may lie in multiple rings analogous to the
Asteroid and Edgeworth-Kuiper belts, but may also be significantly extended in a way
similar to gaseous protoplanetary disks
\citep[e.g.][]{2005Natur.435.1067K,2009ApJ...705..314S,2012MNRAS.424.1206W}. Because they
are generally detected by excess emission above the photospheric level at infra-red (IR)
wavelengths \citep[e.g.][]{1984ApJ...278L..23A}, and with high resolution imaging
detections being relatively rare, discerning radial structure is in general
difficult. The major difficulty is that the equilibrium temperature of a dust grain
depends on both distance from the star and the size and optical properties of that dust
grain. Thus, the radius of an unresolved debris disk cannot be unambiguously determined
from the temperature of the observed emission, as the temperature is degenerate with the
sizes of grains in the disk.

Infra-red excess detections are generally well approximated by a single blackbody. This
property is in part due to the emission properties of circumstellar dust, but also due to
a lack of a high disk signal to noise ratio over a wide range of wavelengths. However, an
increasing number of more complex systems are being discovered with the help of mid-IR
spectra, which when combined with far-IR photometry show emission at more than one
temperature, and therefore may be indicative of dust that resides at a range of
stellocentric distances
\citep[e.g.][]{2009ApJ...690.1522B,2009ApJ...701.1367C,2009ApJ...699.1067M,2013ApJ...775...55B}. Such
disks can be modelled in different ways, but a promising approach is simply to add a
second blackbody component
\citep[e.g.][]{2009ApJ...701.1367C,2011ApJ...730L..29M,2014ApJS..211...25C,2013ApJ...775...55B}. These
``two-temperature'' disks may be analogous to the Solar System, because a possible
interpretation of two temperatures is an origin in dust emission from two distinct radial
locations. Again by analogy with the Solar System, a further question is then whether the
intervening region between the two belts contains planets, and if so, whether dynamical
clearing by these planets is the reason for two-belt structure. Circumstantial evidence
for such a picture is given by systems with planets that reside between two dust
components, such as HR~8799 and HD~95086
\citep{2008Sci...322.1348M,2009A&A...503..247R,2013ApJ...779L..26R,2013ApJ...775L..51M}.

An alternative interpretation is that the two belts may be linked by intervening planets,
with material from an outer belt delivered to replenish the inner belt
\citep[e.g.][]{2010ApJ...713..816N,2012MNRAS.420.2990B}. From a study of many disks,
\citet{2011ApJ...730L..29M} concluded that the warmer of the two temperatures was
typically $\sim$190K, regardless of whether the host star was Sun-like or an A-type. They
argue that the common warm dust temperatures may be a signature of sublimating comets
passed in from outer regions, or asteroid belt analogues formed just interior to the
system's ``snow line''.

Yet a third interpretation, where the two components are linked by grain dynamics (as
opposed to planetesimal dynamics), relies on Poynting-Robertson (PR) drag
\citep[e.g.][]{1979Icar...40....1B}. In this case grains ``leak'' inwards from the
planetesimal belt, but are depleted by collisions with other grains as they drift inwards
\citep{2005A&A...433.1007W,2014arXiv1404.3271V}. A steady state is reached that balances
the rates at which particles fill the region interior to the parent belt and are removed
by collisions. This process does not lead to very large levels of dust inside the parent
belt, but recent work coupled with increased mid-IR sensitivity has lead to the
conclusion that PR drag makes an important contribution in some systems
\citep{2011A&A...527A..57R,2012A&A...537A.110L,2014arXiv1404.6144S}.

Clearly, such interpretations present interesting possibilities for discerning planetary
system structure and the dynamics of such systems. The origin and evolution of the
putative warm components is of particular interest, since dust in the habitable zone may
impact a future space mission to directly image and characterise exo-Earths
\citep[e.g.][]{2006ApJ...652.1674B,2012PASP..124..799R}. For example,
\citet{2013MNRAS.433.2334K} show how the known population of warm bright debris disks
detected at 12$\mu$m can be extrapolated to fainter levels by assuming that those disks
are independent of any outer cool belts (i.e. evolve in situ). If the warm belts are
replenished by comet delivery from elsewhere, such an extrapolation cannot be made.

\begin{figure}
  \begin{center}
    \hspace{-0.5cm} \includegraphics[width=0.5\textwidth]{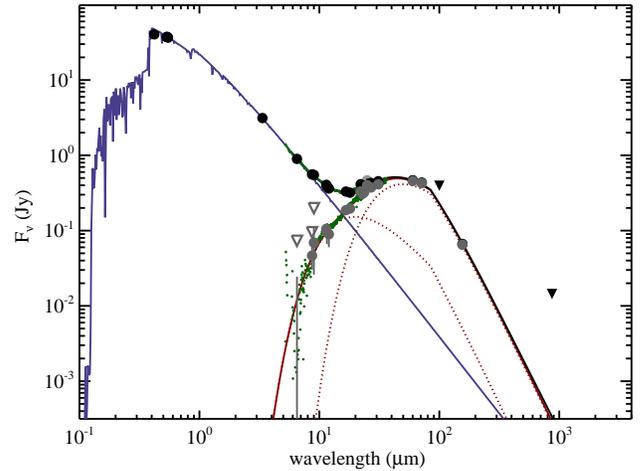}
    \caption{SED for $\eta$ Telescopii. Dots show detections and triangles are upper
      limits. Black symbols are raw photometry and brown symbols are star-subtracted
      (i.e. are disk photometry). The maroon line and dots show the observed and
      star-subtracted IRS spectra. The blue line shows the stellar photosphere model, the
      red lines the disk emission, and the black line the star+disk emission. The dotted
      lines are the two temperature components needed to fit the disk spectrum. The disk
      spectrum is sufficiently broad that a single temperature blackbody is a poor fit,
      while a model with two components at temperatures of 255 and 100K with
      $\lambda_0=85$~$\mu$m and $\beta=1.4$ works well (the SED fitting is described in
      detail in section \ref{s:sed}).}\label{fig:eta-tel}
  \end{center}
\end{figure}

Taking a step back however, the interpretation of two-temperature debris disks as
physically extended or multiple discrete structures has only been tested in a few cases
because it requires dedicated high-resolution observations. For example, $\eta$ Tel shows
what clearly appears to be a two-temperature disk spectrum (Fig. \ref{fig:eta-tel}), but
to confirm that the disk indeed comprises two distinct components has required high
resolution mid-IR observations \citep{2009A&A...493..299S}.

Therefore, our goal here is to consider a third model that could undermine work that
assumes that a broad disk spectrum is always the result of extended or multiple disk
components; namely that the debris may be confined to a relatively narrow belt and the
breadth of the spectrum simply arises due to the absorption and emission properties of
the dust \citep[e.g.][]{2007ApJ...663.1103M}. It is well known that grains of different
sizes can have different temperatures at a fixed stellocentric distance. It is also known
that debris disks comprise not grains of a single size, but a distribution that extends
from $\mu$m to at least cm sizes. Therefore, the specific question we wish to address
here is whether two-temperature disks necessarily imply multiple dust components, or if
their spectra can be reproduced by plausible grain populations residing in a narrow
planetesimal belt. Though it is likely an important effect that contributes significantly
in some cases, we do not consider PR drag here. We first compile a sample of
two-temperature debris disks (sections \ref{s:sample} \& \ref{s:sed}) and then discuss
their properties (section \ref{s:res}). We then consider whether these disks can be
modelled as single belts (section \ref{s:modelling}), and discuss the models and the
origin of multiple belts (section \ref{s:interp}).

\section{Sample}\label{s:sample}

This study was initially inspired by the presence of two-temperature disks among targets
in the \emph{Herschel} DEBRIS sample. These targets are outlined by
\citet{2010MNRAS.403.1089P}, and comprise the nearest $\sim$90 main-sequence stars of A,
F, G, K, and M spectral types (i.e. $\sim$90 of each type), excluding those near the
Galactic plane. Not all DEBRIS sample stars were observed as part of our \emph{Herschel}
programme, some being observed by DUNES \citep{2013A&A...555A..11E} and some as part of a
guaranteed time programme
\citep{2010A&A...518L.130S,2010A&A...518L.133V,2012A&A...540A.125A}. With only 9
two-temperature sources in DEBRIS (as defined below), we expand the sample with more
two-temperature disks observed by \emph{Spitzer} \citep{2004ApJS..154....1W}. These were
selected from a large database of stars with \emph{Spitzer} Infra-Red Spectrograph
\citep[IRS,][]{2004ApJS..154...18H} spectra and other mid/far-IR observations, mostly
from IRAS and the Multiband Imaging Photometer for \emph{Spitzer}
\citep[MIPS,][]{2004ApJS..154...25R} but also including \emph{Herschel} and
sub-millimetre photometry where available. The resulting sample has 48 robust
two-temperature disks around stars with a range of spectral types (see Table
\ref{tab:sample}). IRS spectra are almost always necessary for two temperatures to be
detectable, so these observations set which stars are in our sample. Some sources were
specifically targeted with IRS based on known excesses (e.g. from IRAS), so the sample
for which two-temperature disks can realistically be detected is therefore biased.

Specifically, our sample includes stars from \citet{2009ApJ...699.1067M}, who selected
known debris disks for observation by IRS based on the presence of 24$\mu$m excesses.
\citet{2011ApJ...730L..29M} found that 46\% of these showed evidence for two
temperatures. The disks in this sample are probably biased towards having two
temperatures because the presence of a warm component adds extra 24$\mu$m emission
(e.g. Fig. \ref{fig:eta-tel}). Similarly, some of our stars are from
\citet{2006ApJS..166..351C}, who observed a large sample of IRAS-discovered disks with
60$\mu$m excesses, with IRS. Given that warm components are generally only visible at
wavelengths shorter than 60$\mu$m, objects observed in this programme are probably not
biased towards two temperatures. However, as we demonstrate below, two temperatures are
easier to detect when the overall disk fractional luminosity ($f \equiv L_{\rm
  disk}/L_\star$) is greater, because the excess is detectable at a higher signal to
noise ratio (S/N) over a greater range of wavelengths. Thus, while the
\citet{2006ApJS..166..351C} sample may not be biased towards disks having two
temperatures, they are biased towards detection of two temperatures. These biases are in
general unimportant for this study, though need to be considered for the statistics in
section \ref{ss:stats}.

For an analysis of a much larger sample of stars observed with IRS see
\citet{2014ApJS..211...25C}. Our sample is not meant to be complete, but to provide a
sufficient number of sources for us to test whether the two-temperature disks arise from
single or multiple belts. Our method for deriving the properties of two-temperature disks
is different to \citet{2014ApJS..211...25C}, though we arrive at the same broad trends.

\section{SED Modelling}\label{s:sed}

Because a disk spectrum must be modelled at least once to determine whether multiple
temperature components exist, sample selection is closely linked to SED modelling, which
we now outline. For all systems, photometry ranging from optical to sub-mm wavelengths is
compiled from a wide variety of sources, including all-sky surveys such as Hipparcos
\citep{1997ESASP1200.....P,2000A&A...355L..27H}, 2MASS \citep{2003tmc..book.....C}, AKARI
\citep{2010A&A...514A...1I}, WISE \citep{2010AJ....140.1868W}, and IRAS
\citep{1990IRASF.C......0M}. References for far-IR photometry are given in Table
\ref{tab:sample} and a compiled list of the (sub-)mm photometry is given in Table
\ref{tab:submm}.

Data from the \emph{Spitzer} mission is a crucial component, with IRS spectra needed in
almost all cases to reveal two-temperature disks. These are generally obtained from the
Cornell CASSIS database \citep{2011ApJS..196....8L}. However, the CASSIS database only
provides extractions for low resolution stare-mode observations, and in some cases only
high resolution map-mode data were taken (e.g. HD 39060=$\beta$ Pic). In these cases, or
where previously published spectra were readily available
\citep{2006ApJS..166..351C,2007ApJ...666..466C,2009ApJ...701.1367C,2013ApJ...763..118S}
we preferred these over the CASSIS extractions.

The IRS instrument is split into several modules, and the spectral extractions from the
different modules must be aligned in relative terms to produce smooth self-consistent
spectra. Previous works have generally aligned the modules simply using a handful of data
in the region where the spectra overlap
\citep[e.g.][]{2009ApJ...705...89L,2014ApJS..211...25C}. We took a slightly different
approach, fitting the entire spectrum with a function and allowing the absolute values of
all but one module (LL1) to vary as part of the fit, thus forcing the spectrum to be
smooth across all modules and ensuring that any issues near the edges of each module did
not strongly affect the results. For the fitted function, we used the sum of two power
laws and one blackbody, the rationale being that the first power law accounts for the
stellar Rayleigh-Jeans tail, and that the blackbody and second power law account for
excess emission, which may look like either or a combination of both
\citep[e.g.][]{2009ApJ...699.1067M,2011ApJ...730L..29M}. We found this method to work
well and produce spectra comparable with previous methods.

Once aligned, the IRS spectra are split into 7 photometric ``bands''. Because the
absolute value of the spectrum will not necessarily agree with other photometry
\citep[the absolute level varies at the $\sim$10\% level, e.g.][]{2009ApJ...705...89L},
the spectrum is normalised so that the shortest band agrees with synthetic photometry of
the best fitting stellar photosphere model. The IRS bands are subsequently treated
identically to other photometry. Because the quality of the spectral extractions vary, we
found it necessary to add a 2\% systematic uncertainty to all spectra to avoid spurious
excesses. In most cases this uncertainty dominates, so the formal uncertainty is larger
than would be expected from looking at the point to point scatter in the spectrum.

Photometry shortward of about 10$\mu$m is used to model the stellar photospheric
emission. This wavelength is varied from star to star depending on the temperature of the
excess, ensuring both the best photospheric fit and that the excess does not affect this
fit. For each star the best fitting model from a grid of PHOENIX AMES-Cond models
\citep{2005ESASP.576..565B} is found by a combination of brute force grids and least
squares fitting. For the few stars found to be over 10,000K we use models from
\citet{2003IAUS..210P.A20C}, which span a wider range of effective temperatures. The
remaining IR photometry is used to find the best fitting disk model. We first subtracted
synthetic photometry of the photosphere model from the observed fluxes to derive disk
fluxes, with uncertainties derived from the photosphere and IR observation (including
systematic uncertainties) added in quadrature. The disk parameters are then found via
least squares minimisation, for which we use the modified blackbody,
\begin{equation}\label{eq:bb}
  F_\nu = n B_\nu(T_{\rm disk}) X_\lambda^{-1}
\end{equation}
where $n$ sets the overall level of dust with temperature $T_{\rm disk}$, and
\begin{equation}
  X_\lambda = \left\{ \begin{array}{ll} 
      1 & \lambda < \lambda_0 \\
      \left(\lambda/\lambda_0 \right)^\beta & \lambda > \lambda_0 \\
    \end{array}
  \right. .
\end{equation}
The blackbody function has units of Jy sr$^{-1}$ so $n$ is proportional to the surface
area of dust in the disk. The fractional luminosity $f_{\rm disk} = L_{\rm disk}/L_\star$
of a given disk is therefore proportional to $n T_{\rm disk}^4$ (but also depends on
$X_\lambda$).

The physical origin of this formalism comes from the inability of grains to emit
efficiently at wavelengths longer than their physical size. Therefore, $\lambda_0$ is
somehow related to grain sizes in the disk. It does not necessarily provide a direct
measure of grain size however, because the observed emission comprises contributions from
a size distribution of grains (which is related to $\beta$).

After fitting a single blackbody, each disk spectrum is inspected for goodness of fit. In
most cases where it is necessary, the need for a second temperature component is
clear. However, we found that a formal criterion (such as $\chi^2$) can be a poor
indicator because there can be other reasons for a poor model fit that are unrelated to
the number of temperature components.

\begin{figure}
  \begin{center}
    \hspace{-0.5cm} \includegraphics[width=0.5\textwidth]{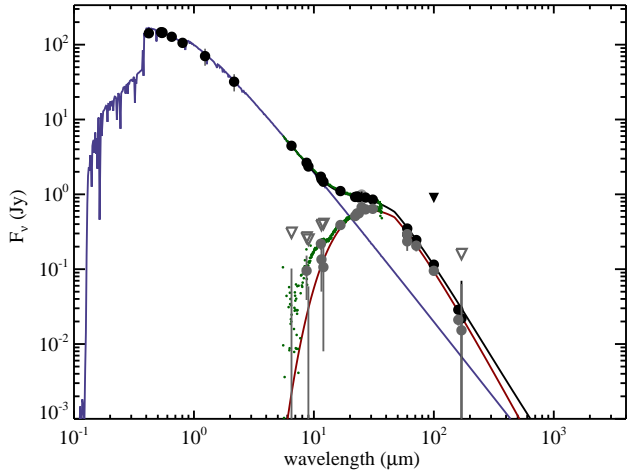}
    \caption{SED for $\zeta$ Lep. Symbols are as described in Fig. \ref{fig:eta-tel}. It
      is unclear whether the disk should be fitted as two temperature components, or
      whether the extra IRS emission near 10 $\mu$m above the disk model (green dots
      above the red line) is due to non-continuum silicate emission.}\label{fig:zeta-lep}
  \end{center}
\end{figure}

For example, $\zeta$ Lep shows evidence for extra emission above a blackbody around
10$\mu$m (Fig. \ref{fig:zeta-lep}), which may be due to a silicate feature over the
continuum, meaning that adding a second temperature component is not well justified based
on the SED \citep[mid-IR imaging suggests that the disk is somewhat
extended,][]{2007ApJ...655L.109M}. In some cases the issue may be a discontinuity in the
IRS spectrum near 15 $\mu$m, which is at the join between two different IRS modules and
can cause a dip similar to that seen for $\zeta$ Lep \citep{2009ApJ...701.1367C}. In
general, we err on the side of caution and include the two-temperature disks that appear
to be the most robust. The only targets in our sample with strong silicate features are
$\beta$ Pictoris and $\eta$ Corvi, for which the presence of two temperatures is clear
and corroborated by other studies (see end of this section).

While we could allow a separate $\lambda_0$ and $\beta$ for the two components, these
parameters are poorly constrained for the warmer component as the emission beyond the
mid-IR is almost always dominated by the cooler component. We therefore
fix $\lambda_0$ and $\beta$ to be the same for both components, and there are six model
parameters to fit.

For objects that we do model with two temperatures, there remain degeneracies between the
six parameters that are not necessarily well described by the covariance matrix that
results from the least squares fitting. The most important is that disk temperature and
normalisation are strongly correlated at constant total disk luminosity by the
Stefan-Boltzmann law. To estimate the parameters and their uncertainties in a more robust
way we use an ensemble Markov chain Monte-Carlo method \citep{goodweare}\footnote{As
  implemented in the python \texttt{emcee} package \citep{2012arXiv1202.3665F}.} using
$e^{-\chi^2/2}$ as our likelihood function. Chains with an ensemble of 200 ``walkers''
are initialised with parameters that vary randomly $\pm$1\% from the $\chi^2$ fitting
results, and then run for 50 steps as a burn-in phase to eliminate any dependence on the
initial state. This number of steps is sufficient to ensure the initial conditions do not
influence the results, being at least ten times the autocorrelation length
\citep{goodweare}. The final distributions of parameters are created from a further 20
steps, resulting in 4000 samples from which we derive the probability distributions of
each parameter.

Practically, rather than fit the component temperatures individually, we fit the
temperature and normalisation of the cool component, and $\lambda_0$ and $\beta$ where
sufficient photometry exists, and the ratio of warm to cool component temperatures
\begin{equation}
 \mathcal{R}_T = T_{\rm warm}/T_{\rm cool}
\end{equation}
and the ratio of warm to cool component normalisations \rn. The ratio of fractional
luminosities is
\begin{equation}
  \mathcal{R}_f = f_{\rm warm}/f_{\rm cool} \, .
\end{equation}
Here, \rf~is the preferable quantity to work with because \rn~and \rt~are strongly
correlated by the Stefan-Boltzmann law, but any reasonable fit must produce a disk with
roughly the same luminosity. We derive values and uncertainties by fitting a Gaussian to
the marginalised distributions for each parameter.

\begin{figure}
  \begin{center}
    \hspace{-0.5cm} \includegraphics[width=0.5\textwidth]{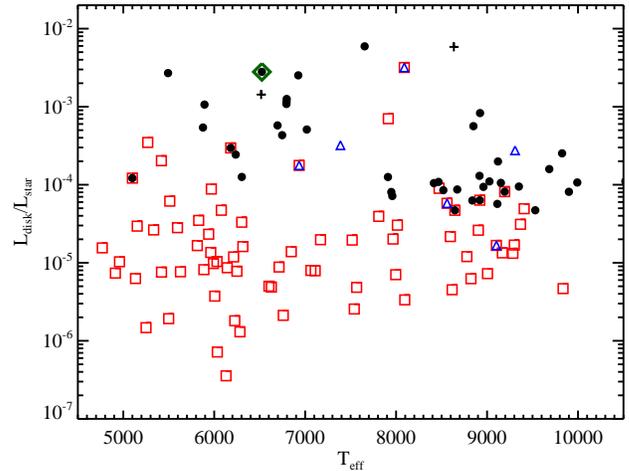}
    \caption{Sample of two-temperature disks (dots and triangles) and disks in the
      \emph{Herschel} DEBRIS sample (squares). The nine dots and triangles enclosed by
      squares are DEBRIS stars with robust two-temperature disks. All disks are shown at
      their total fractional luminosity. The dot enclosed by a diamond is HD 181327,
      which \citet{2012A&A...539A..17L} show can be modelled with dust in a single narrow
      ring. Triangles note two-temperature systems found independently to have two disk
      components (see end of section \ref{s:sed}). The two plus symbols show the single
      temperature disks, HD 191089 and HR 4796A, where two temperatures could easily have
      been detected but were not (see section \ref{ss:1t}).}\label{fig:teff-ldisk}
  \end{center}
\end{figure}

We retain two-temperature disks as those where the temperatures of the warm and cool
components are significantly different (i.e. $\mathcal{R}_T > 3\sigma_{\mathcal{R}_T}$),
and where the normalisation of the warm component is significantly different than zero
(i.e. $\mathcal{R}_n > 3 \sigma_{\mathcal{R}_n}$, all disks considered have significant
cool components). The result of this process is a sample of 48 robust two-temperature
debris disks. The targets are listed in Table \ref{tab:sample} and the SEDs available in
the online material. The overall fractional luminosities of these disks are shown in Fig.
\ref{fig:teff-ldisk}. Also shown are all disks for targets in the unbiased DEBRIS
sample. The samples in this plot should not be used to conclude that two-temperature
disks are typically brighter than other disks, or more common among bright disks, as
significant biases exist among the two-temperature sample (see section \ref{s:sample} for
a discussion of sample statistics).

Among our sample we also note systems where observations with sufficient spatial
resolution have been able to show that two distinct disk components exist. These systems
are Vega and Fomalhaut \citep{2013ApJ...763..118S}, $\eta$~Crv
\citep{2005ApJ...620..492W,2009A&A...503..265S,2014ApJ...784..148D}, HR~8799
\citep{2009ApJ...705..314S,2014ApJ...780...97M}, and $\eta$~Telescopii
\citep{2009A&A...493..299S}. We also include $\beta$ Pictoris in this list, where the
disk is seen to extend over a wide range of stellocentric radii
\citep[e.g.][]{1984Sci...226.1421S,2005Natur.433..133T}.

\section{Results}\label{s:res}

\begin{figure}
  \begin{center}
    \hspace{-0.5cm} \includegraphics[width=0.5\textwidth]{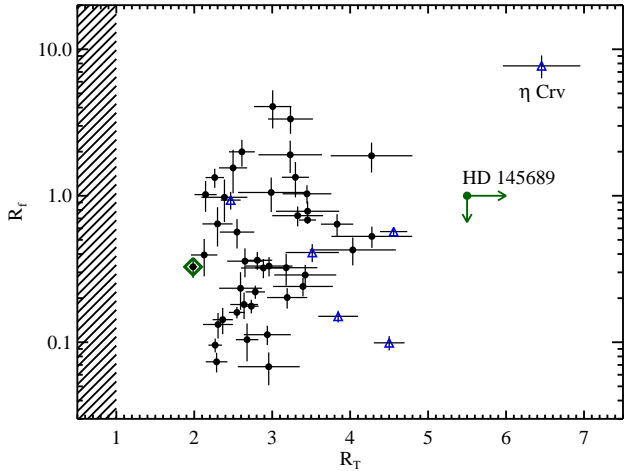}
    \caption{Two-temperature disks and uncertainties. Blue triangles note disks known two
      have multiple disk components from imaging and/or interferometry and the green
      diamond encloses HD~181327. The dot and arrow symbol marks HD~145689, host to an M9
      companion, and for which \rt~is a lower limit (and \rf~an upper
      limit).}\label{fig:rfrt}
  \end{center}
\end{figure}

A simple way to present two-temperature disks is the ratio of temperatures \rt~and
fractional luminosities \rf, as shown in Fig. \ref{fig:rfrt}. It is immediately clear
that most two-temperature disks have fairly similar temperature ratios of 2-4, but with a
range of fractional luminosity ratios. A clear outlier is $\eta$ Crv, which has the
largest temperature ratio, and is one of several systems known from detailed observations
to have two physically distinct dust belts.

The source at intermediate \rt~is HD~145689, which has an unconstrained cool component
temperature and is therefore plotted as a lower limit in \rt. It is not formally part of
our final two-temperature disk sample, but is mentioned here as a potentially interesting
two temperature system found during our sample selection. As a probable outlier, we found
it remarkable as the host of an M9 brown dwarf companion at 6\farcs7
\citep{2010A&A...521L..54H}. This source was proposed to be a $\sim$40Myr old Argus star,
and the disk model shown by \citet{2011ApJ...732...61Z} also has two temperature
components. At 52pc \citep{2007A&A...474..653V} the minimum companion separation is about
350AU. The disk radii implied by the two temperature fit of about 200 and $<$35K are 7
and $>$180AU respectively, though because these are estimates assuming blackbody grains,
the distances are probably $\sim$3 times larger
\citep{2012ApJ...745..147R,2013MNRAS.428.1263B}. Therefore, if the 70$\mu$m excess is
associated with the star or companion, this system has three possible configurations
(assuming that the larger than average temperature ratio is indeed indicative of multiple
belts). The primary may have two well-separated belts and the companion either orbits
between or beyond the two belts. The third possibility is that the cool dust component
actually orbits the companion, but is heated by the primary. HD~145689 is clearly an
intriguing system, and proof of concept that Fig. \ref{fig:rfrt} is a potentially
powerful diagnostic for finding interesting planetary systems.

Aside from HD~145689, there is a simple demarcation between $\eta$ Crv and the rest of
our sample in Fig. \ref{fig:rfrt}. However, the sources at low \rt~are not all
single-belt systems, as some that are otherwise known to have multiple dust populations
are shown as blue triangles. These sit amongst the general population, perhaps adding
strength to the standard assumption that multiple temperatures can always be interpreted
as multiple dust belts.

Using the modelling results from \citet{2013ApJ...775...55B} yields a very similar
version of Fig. \ref{fig:rfrt}, with HD~145689 again a clear outlier (their sample did
not include $\eta$ Crv). A comparison using the modelling results of
\citet{2014ApJS..211...25C} yields a very different plot, with \rt~covering a wide range
and as high as 13 (the range of \rf~is similar). Their modelling finds two-temperature
disks where we do not, and consistently hotter warm components. We suspect the origin of
these differences lies with a difference between the 2MASS and \emph{Spitzer} absolute
calibrations, which we discuss further in section \ref{ss:1t}.

However, the young F6 star HD 181327 (the dot enclosed by a green diamond) is known to be
concentrated in a relatively narrow single belt
\citep{2006ApJ...650..414S}. \citet{2012A&A...539A..17L} show that while the spectrum
cannot be modelled with a single blackbody, allowing the composition of a narrow belt to
vary can result in a good fit to the disk spectrum. Therefore, HD 181327 shows that not
all two-temperature disks necessarily originate from two distinct components.

\subsection{Sensitivity to two temperatures}\label{ss:sens}

We now consider the sensitivity to two-temperature disks in the parameter space shown in
Fig. \ref{fig:rfrt}. Clearly, disks whose components have very different levels of
emission will be hard to identify as having two components. Similarly, a warm component
with a temperature approaching that of the star will be hard to detect photometrically
(i.e. these are usually detected with interferometry), so stars with very large
temperature ratios will also be hard to detect. Of course, sufficiently small temperature
ratios will also be impossible to discern as comprising multiple temperatures. Combined,
these criteria mean that there should be bounds on all sides of Fig. \ref{fig:rfrt} that
limit the range of two-temperature disks that can be discovered. The fainter the disk
overall, the lower the S/N of all measurements will be in general, and therefore the less
parameter space these bounds will cover.

\begin{figure}
  \begin{center}
    \hspace{-0.5cm} \includegraphics[width=0.5\textwidth]{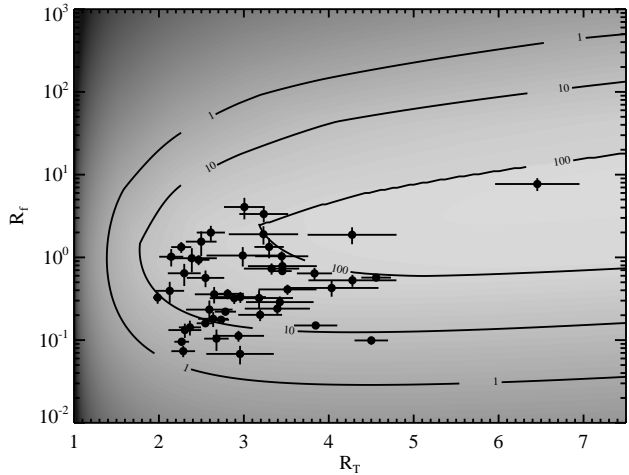}
    \caption{Simple model of sensitivity to two temperature excesses, where lighter grey
      corresponds to greater sensitivity. Contours shows $\chi^2_{\rm red}$ from fitting
      a single temperature blackbody to two-temperature disks with the parameters at each
      point in the parameter space. Disks near the $\chi^2_{\rm red}=1$ contour are well
      described by a single blackbody, and therefore two temperature disks with these
      properties cannot be discovered given observations with typical S/N. The lowest
      \rf~and \rt~two-temperature disks lie near the edge of detectability, suggesting
      that disks with $\mathcal{R}_T \lesssim 2$ and $\mathcal{R}_f \lesssim 0.1$ do
      exist but were not detected here. Similarly, two temperature disks near
      $\mathcal{R}_T \sim 5-6 $ and $\mathcal{R}_f \sim 1-10$ are easily detected, so the
      gap between most disks and $\eta$ Crv is real. Disks with $10 \lesssim
      \mathcal{R}_f \lesssim 100$ and $\mathcal{R}_T \gtrsim 2$ are detectable but were
      not seen, so must be rare.}\label{fig:sens}
  \end{center}
\end{figure}

To quantify this picture in a little more detail, Fig. \ref{fig:sens} shows a simple
approximation of the regions in which a ``typical'' set of observations could detect
two-temperature emission. At each point in the parameter space covered, we generated two
pure blackbodies, with a fixed 70K cool component, varying \rt~to set the warm component
temperature, and then ``observed'' them with synthetic photometry at seven IRS
(4-35$\mu$m), three MIPS (24-160$\mu$m), and two SCUBA (450-850$\mu$m) bands. We
peak-normalised each total spectrum to unity, and estimated the uncertainties as $0.01
F_\nu^{1/4}$, which corresponds to 1\% uncertainty at the peak, and 32\% (i.e. a
3$\sigma$ upper limit) for measurements two orders of magnitude fainter. We additionally
set the minimum uncertainty to be 5\%. While this prescription simplifies the realities
of photometry collected from different instruments with many different observing
strategies, it represents the limits reasonably well (e.g.  see Figs. \ref{fig:eta-tel}
and \ref{fig:zeta-lep}). To each spectrum we fit a single blackbody and computed the sum
of squared deviations per degree of freedom ($\chi^2_{\rm red}$), which are the contours
shown in Fig. \ref{fig:sens}.

In darker regions where $\chi^2_{\rm red}$ is low (i.e. $<$1) a single blackbody is a
good model of the emission, and a two-temperature disk cannot be confidently detected. In
lighter regions where $\chi^2_{\rm red}$ is higher, two-temperature disks are easier to
detect. We have also shown the two-temperature disks from Fig. \ref{fig:rfrt} (except
HD~145689), which shows that the simple detection simulation is reasonable in the sense
that no two-temperature disks lie where they should not be
detectable. Fig. \ref{fig:sens} shows that the most simple criterion for detecting two
temperature disks is that one component does not dominate over the other, and that their
temperatures are not too similar.

Two-temperature disks are harder to detect in overall fainter disks, simply because their
measurements are typically less precise. For the same photometric precision, lowering the
overall disk brightness by a factor of three results in uncertainties that are three
times larger. In Fig. \ref{fig:sens}, the $\chi^2_{\rm red}$ contours would therefore be
divided by a factor of nine and disks with \rf~lower than about 0.2 become harder to
detect. We return to the effect of this disk luminosity bias when looking at spectral
type trends below.

Our simulation does not include a limit at large temperature ratios because these are
limited not by the disk properties, but by difficulties in distinguishing the warm
component from the star. Because disks are rarely cooler than $\sim$30K, and disks hotter
than $\sim$500K become increasingly difficult to detect with photometry, the practical
upper limit on detectable temperature ratios is in the 15-20 range (depending on the
temperature of the cool component).

The conclusions from this analysis are i) that the lowest \rf~and \rt~disks lie near the
sensitivity limits, meaning that disks with lower \rf~and \rt~probably exist, but were
not detectable here, ii) that two-temperature disks are harder to detect when the overall
disk luminosity is lower, iii) that disks with $10 \lesssim \mathcal{R}_f \lesssim 100$
and $\mathcal{R}_T \gtrsim 2$ are detectable but were not seen, so must be rare, and iv)
the gap between most disks and $\eta$ Crv is in a region where two-temperature disks are
most easily detected, so the gap is real and the $\eta$ Crv disk is a rare outlier among
two-temperature disks.

\subsection{Trends with spectral type}\label{ss:trends}

\begin{figure*}
  \begin{center}
    \hspace{-0.5cm} \includegraphics[width=0.5\textwidth]{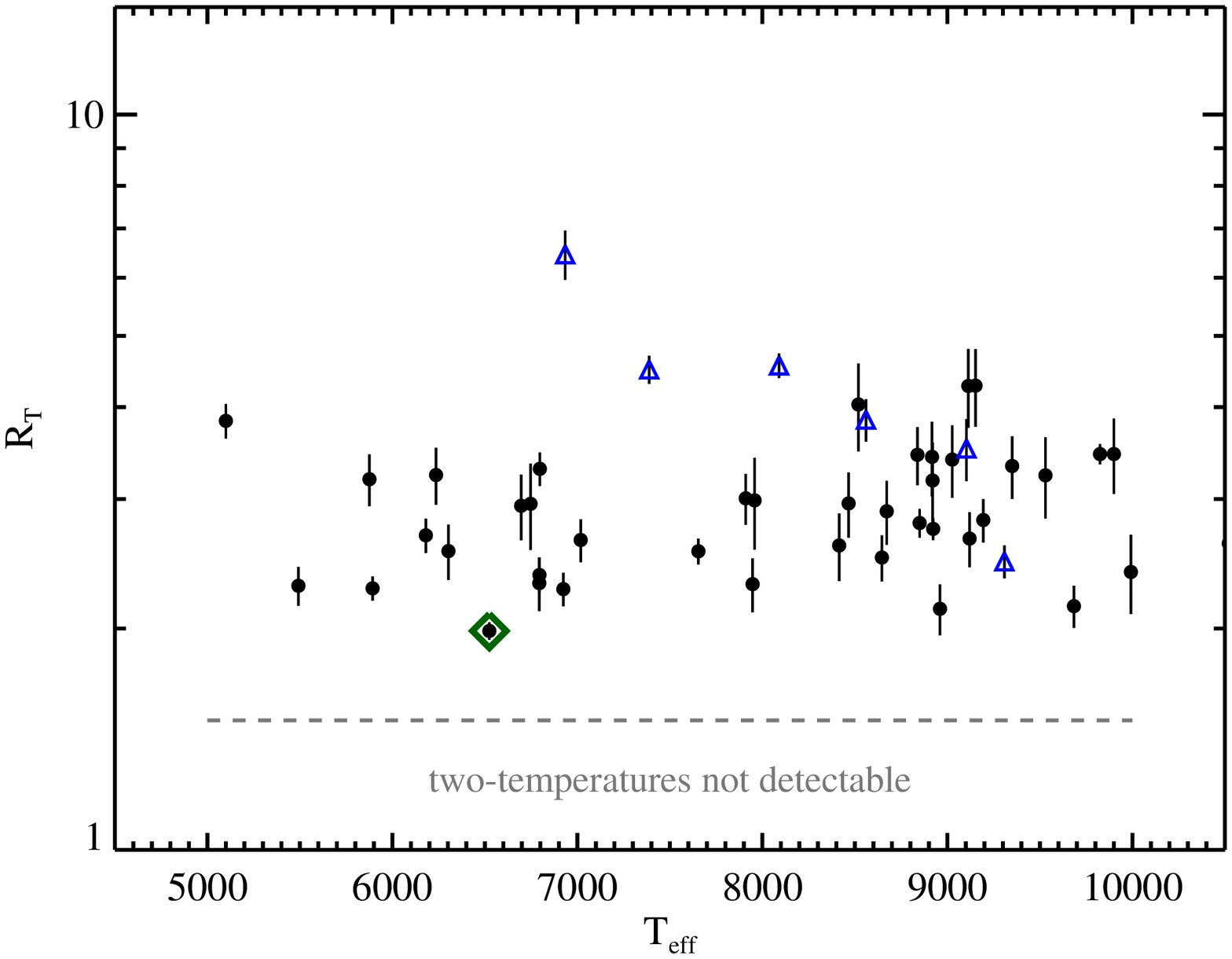}
    \includegraphics[width=0.5\textwidth]{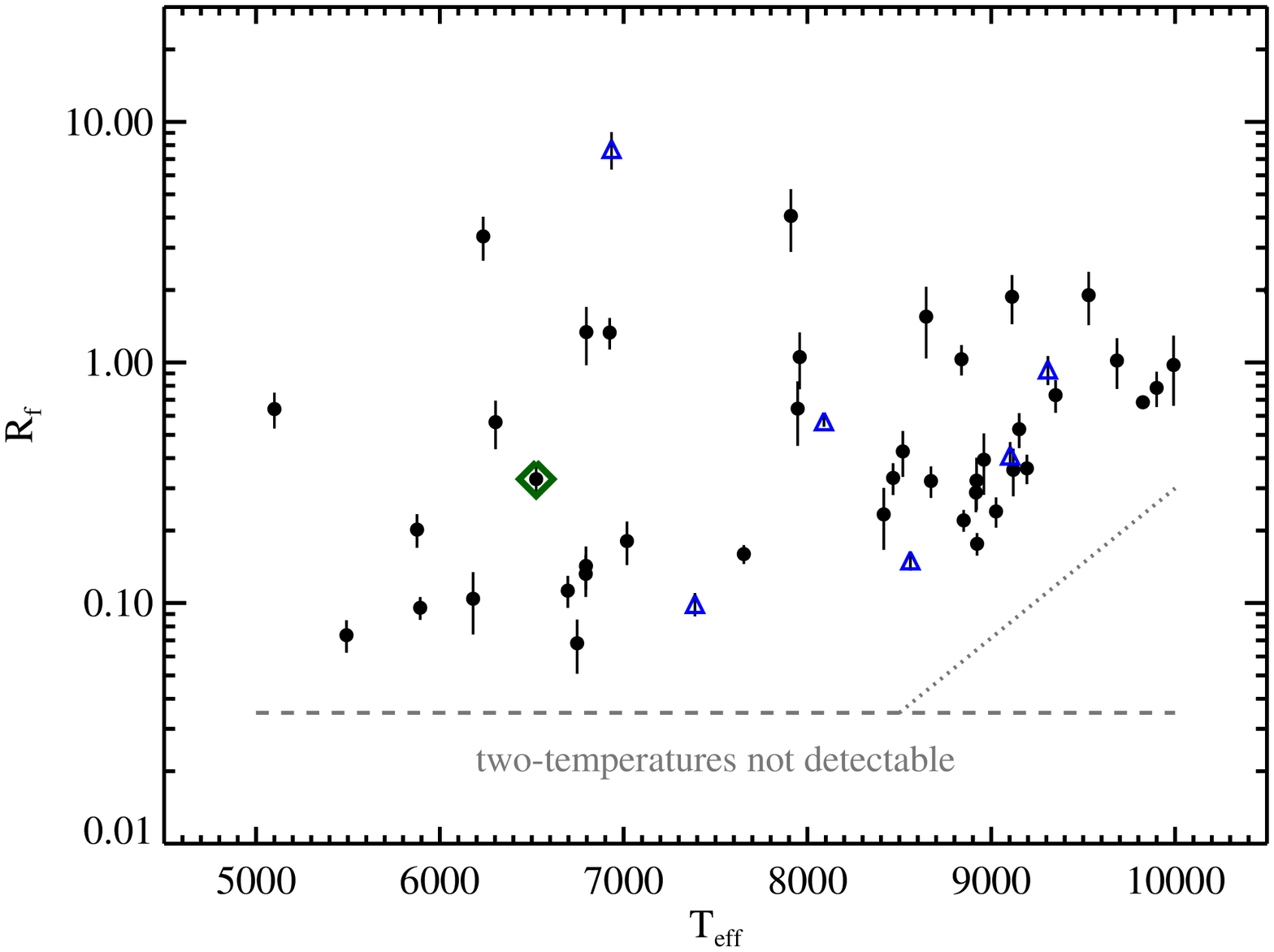}\\
    \hspace{-0.5cm} \includegraphics[width=0.5\textwidth]{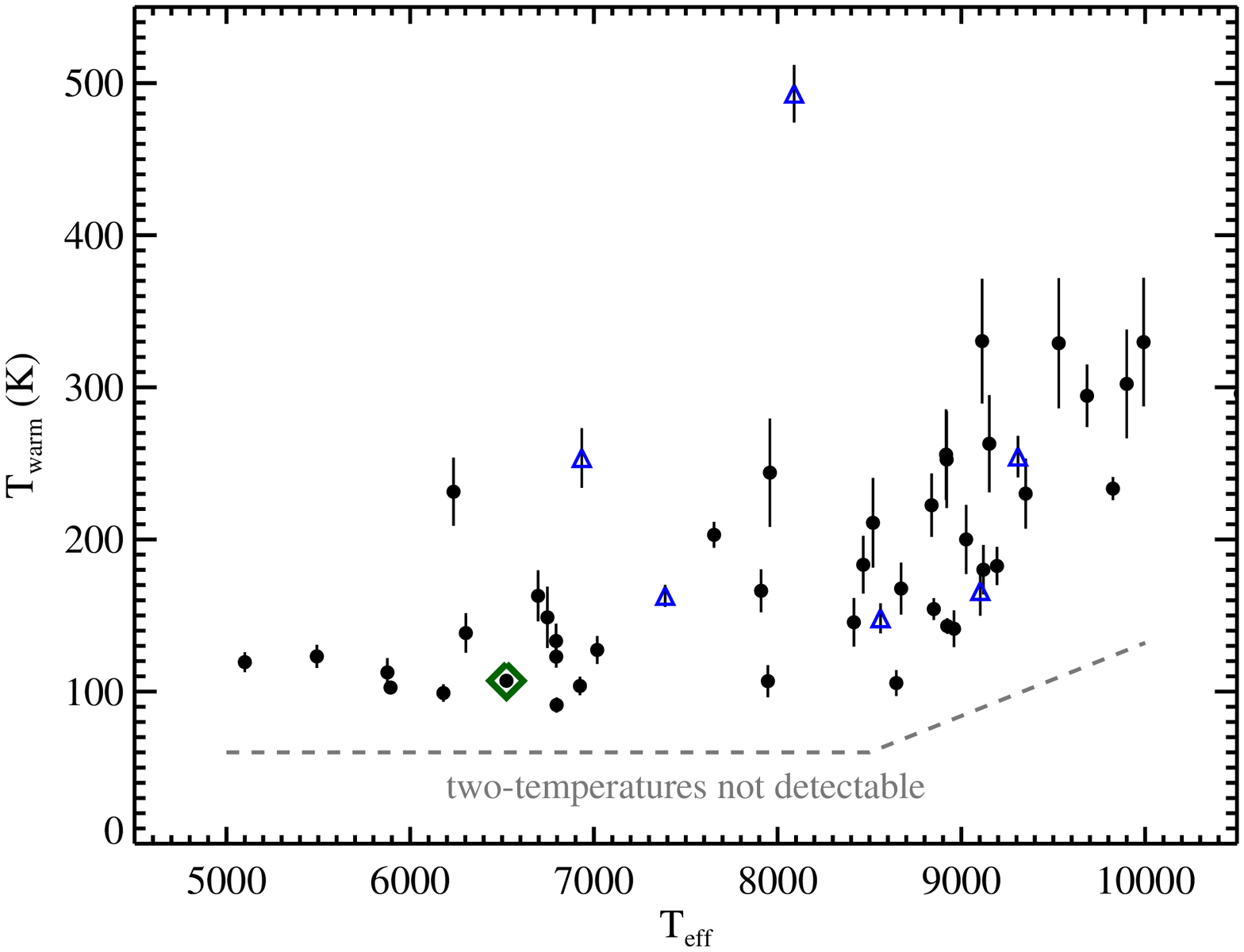}
    \includegraphics[width=0.5\textwidth]{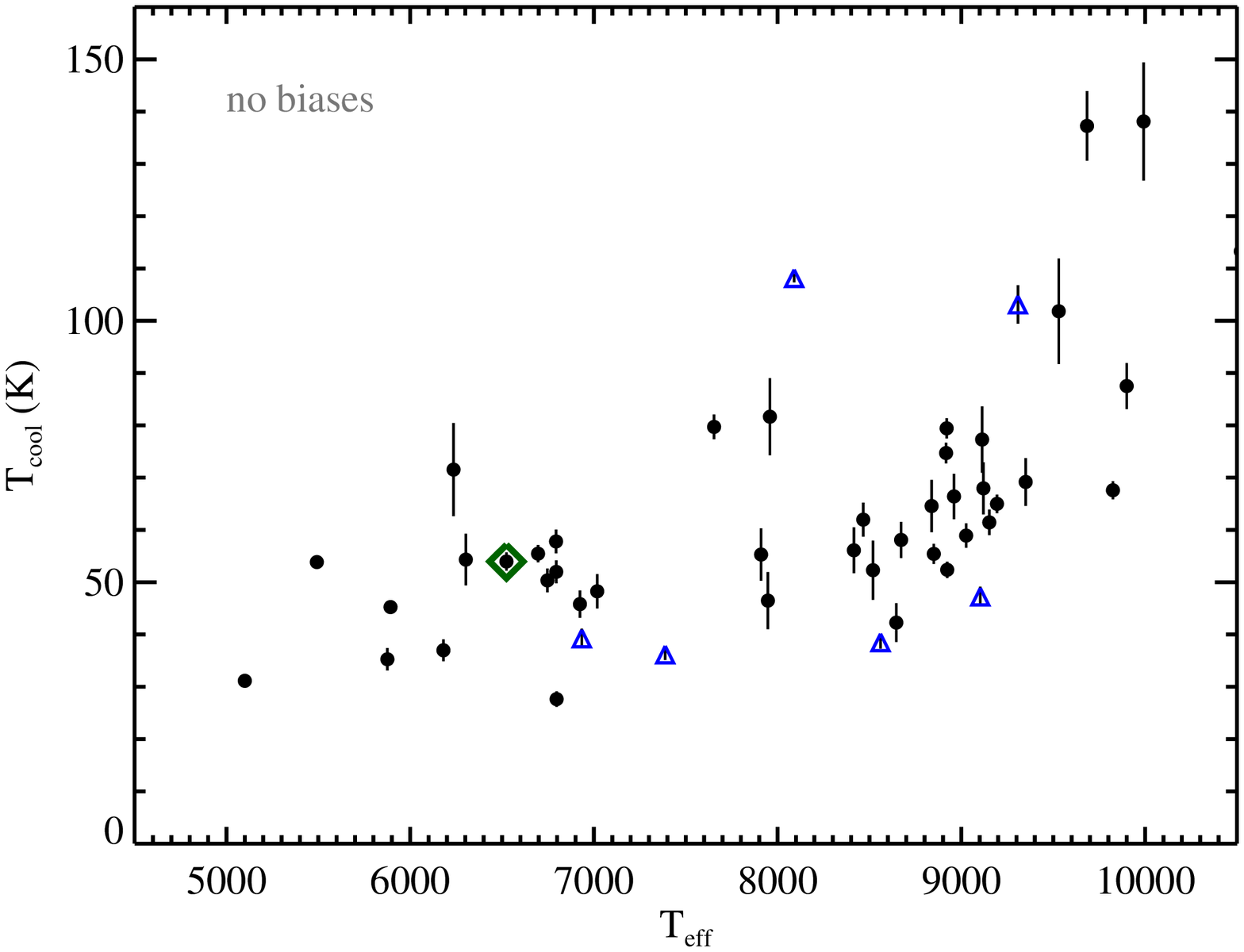}
    \caption{Relations between $\mathcal{R}_T$, $\mathcal{R}_f$, $T_{\rm warm}$, $T_{\rm
        cool}$ and stellar effective temperature. Blue triangles note disks known two
      have multiple disk components from imaging and/or interferometry and the green
      diamond encloses HD~181327. Biases derived in Fig. \ref{fig:sens} (see section
      \ref{ss:sens}) are shown by dashed lines, with an additional possible bias arising
      from disks being fainter overall around hotter stars shown by the dotted
      line. There are no biases that affect the upper envelope of points in these
      plots.}\label{fig:trends}
  \end{center}
\end{figure*}

Fig. \ref{fig:trends} shows how some of the derived disk parameters vary with stellar
effective temperature, as well as approximate biases due to the sensitivity to two
temperatures described above. \citet{2014ApJS..211...25C} analysed a much larger sample
of stars observed with IRS, and found similar trends, though they did not present the
results in terms of the ratios we use here.

The temperature of the cool component increases with $T_{\rm eff}$, as might be expected
due to increasing stellar luminosity if all disks have similar characteristic sizes
\citep[see][for further discussion of this trend]{2013ApJ...775...55B}. Because \rt~is
generally similar for all stars, the warm component temperatures show the same
trend. This conclusion is in contrast to \citet{2011ApJ...730L..29M}, who found that the
warm component temperature was generally constant, regardless of spectral type
\citep[also see][]{2014ApJS..211...25C}. However, plotting their temperatures shows a
probable correlation, with all Sun-like stars having warm components $<$220K, but 15 out
of 24 A-type stars having warm components $>$220K. Indeed, their A-type sample contains
stars that range from B8 to A7 and there is also a trend among these for the hotter stars
to have warmer warm excesses. A possible bias exists here however, because for fixed
sensitivity to \rt, increasing cool belt temperatures mean that only warm belts with
increased temperatures can be detected. A predicted detection line of $\mathcal{R}_T
\gtrsim 2$ is shown in the lower left panel of Fig. \ref{fig:trends}, based on the lower
envelope of temperatures in the lower right panel. The lowest warm temperatures around
the hottest stars lie farther above the detection line than would be expected, so it
seems likely that the trend towards warmer warm belts around hotter stars is real. There
is no such bias for the top envelope of points in any of the panels, so the observed
increase in the warmest warm component temperatures with stellar temperature also argues
that the trend is real. Our interpretation of both sets of two-temperature fitting
results is therefore different to the \citet{2011ApJ...730L..29M} conclusion of common
warm dust temperatures, instead finding a probable trend for warmer dust around hotter
stars.

While the maximum \rf~appears to be constant, the minimum \rf~appears to increase with
$T_{\rm eff}$, and the hottest stars tend to have similarly luminous warm and cool
components. This trend may be a bias however, as the hotter stars in our sample tend to
have lower overall fractional luminosities (Fig. \ref{fig:teff-ldisk}), for which
detecting low \rf~disks is more difficult (as indicated by the dotted line in the upper
right panel).

\subsection{Disks with single temperature spectra}\label{ss:1t}

As we have emphasised, to be sure that a debris disk comprises multiple components
requires at least one of those components to be resolved. For example, disks with
well-resolved outer belts such as Fomalhaut and Vega show unresolved emission closer to
the star at $\sim$10 au, strong evidence for multiple belts that in these cases is
consistent with their two-temperature SEDs \citep{2013ApJ...763..118S}.

There are however, also disks that show both warm and cool emission from imaging but
appear to have single temperature disk spectra. For example, $\alpha$ CrB is resolved in
both mid-IR and far-IR imaging, suggesting that the disk has either two belts, or a disk
that extends over a wide range of radii, yet the SED is well fit by a single modified
blackbody \citep{2010ApJ...723.1418M,2012MNRAS.426.2115K}. An intermediate class also
exists, where the disk spectra are sufficiently complicated by spectral features that a
poor single temperature fit does not immediately suggest that a two-temperature model
would be better (e.g. Fig. \ref{fig:zeta-lep}). As another example, the HD~113766A disk
spectrum can be modelled moderately well as a single component, but when considered in
light of mid-IR interferometry, the combined photometry shows that a two-belt model is a
better interpretation \citep{2013A&A...551A.134O}.

\begin{figure*}
  \begin{center}
    \hspace{-0.5cm} \includegraphics[width=0.5\textwidth]{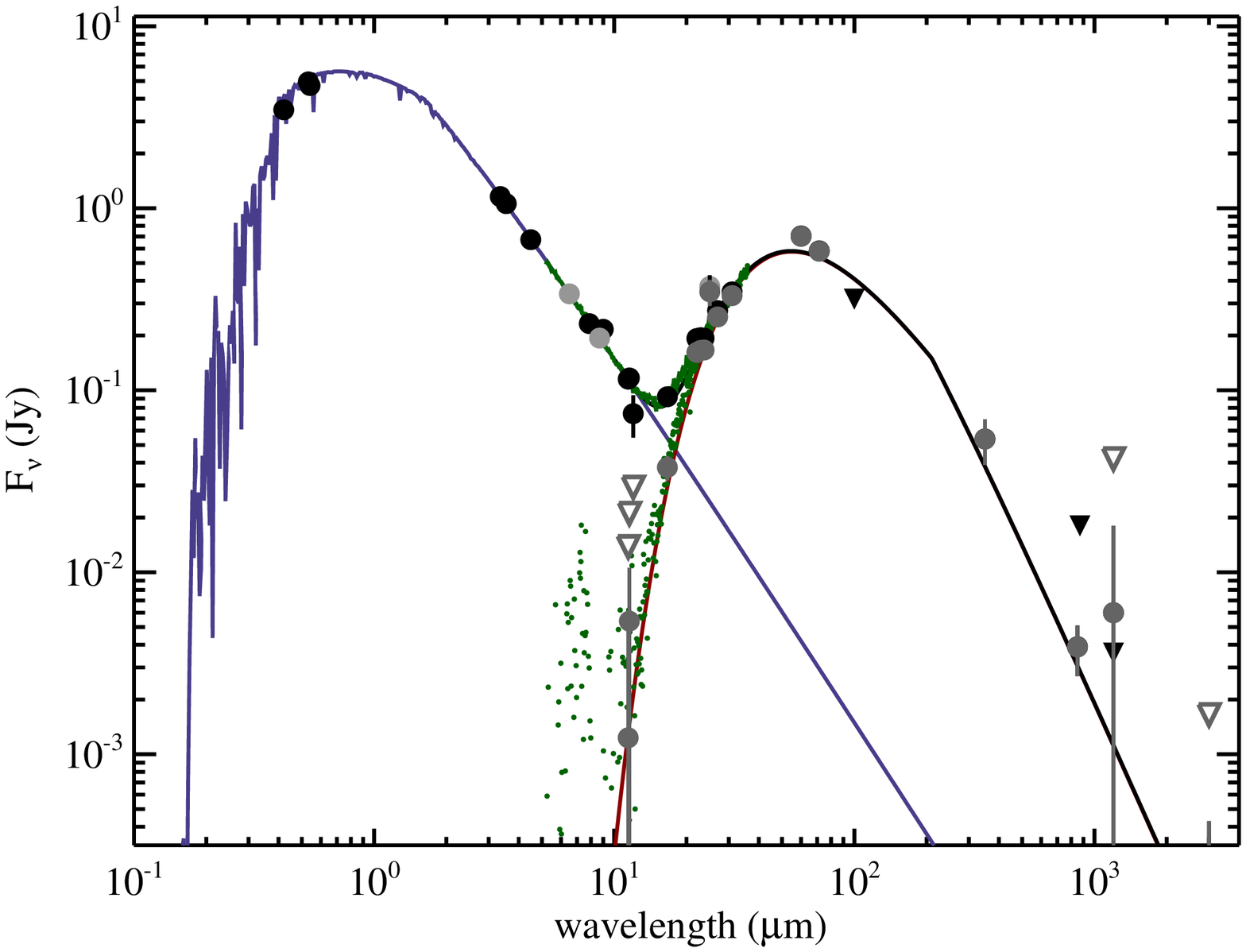}
    \includegraphics[width=0.5\textwidth]{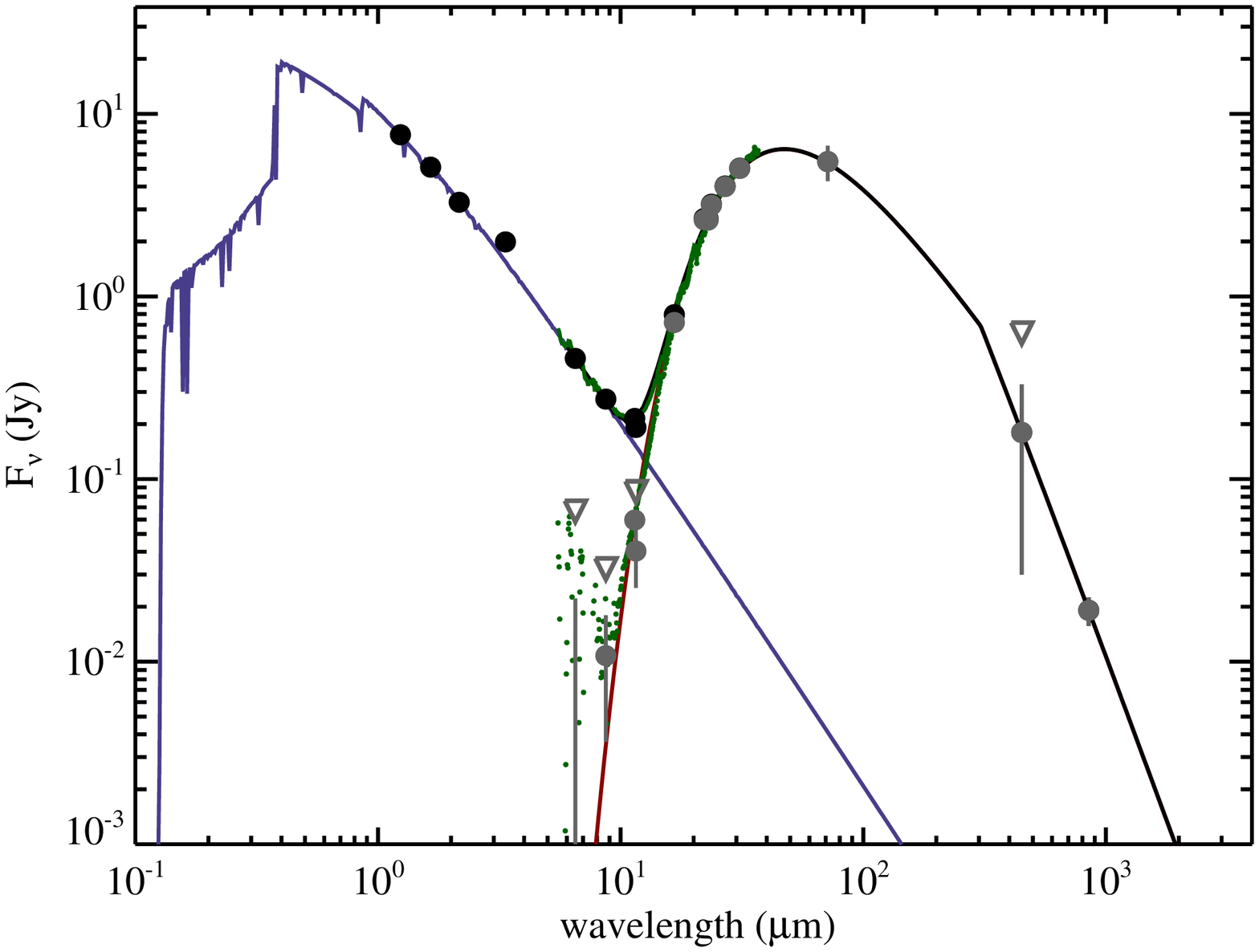}
    \caption{SEDs for the single-temperature systems HD~191089 (left) and HR 4796A
      (right). Black dots and triangles show photometry and upper limits, and grey dots
      and triangles show star-subtracted photometry and upper limits. The green line and
      dots show the observed and star-subtracted IRS spectra. The blue line shows the
      stellar photosphere model, the red lines the individual and total disk emission,
      and the black line the star+disk emission.}\label{fig:seds2}
  \end{center}
\end{figure*}

An additional question is whether there are disks where two-temperature behaviour was not
detected where it could have been, or stated another way; do all disks have two
temperatures? While it does not have two temperatures, a disk such as that around
HD~113766A is not particularly well suited for this test because it has strong spectral
features. Two that are well suited are HD~191089 and HR 4796A, as shown in Fig.
\ref{fig:seds2}. Both disks are extremely bright ($f=0.001-0.006$) and given the high S/N
are very well modelled by modified blackbodies at about 100K. From the range of \rt~and
\rf~at the effective temperatures of 6500 and 8630K, Fig. \ref{fig:trends} shows that a
typical warm component would have $T_{\rm warm}=200-400$K and $f= 10^{-4}$ to $10^{-2}$,
which would have been easily detected. Fig. \ref{fig:sens} suggests that warm disks in
this temperature range would still be detectable if they were up to an order of magnitude
fainter.

For HD~191089, \citet{2011MNRAS.410....2C} concluded from mid-IR imaging that the inner
regions were truly depleted, suggesting that in addition to appearing as a single
temperature disk, that the emission does actually come from a single belt. Our result of
a single temperature is in contrast to the two temperature model of
\citet{2014ApJS..211...25C}, which they strongly prefer. However, this appears to be an
artefact of their analysis, which decreases the MIPS 70$\mu$m uncertainties by a factor
of 70. Our inclusion of longer wavelength photometry shows that the second temperature
component is not justified.

Detailed modelling of HR 4796A requires two spatially distinct dust belts
\citep{1999A&A...348..557A,2005ApJ...618..385W}. Again our model is in disagreement with
\citet{2014ApJS..211...25C}, who find a warm temperature component at 231K. The main
difference between our methods is that we tie our IRS spectra to the photosphere, whereas
\citet{2014ApJS..211...25C} tie it to MIPS 24$\mu$m photometry, which requires an
accurate relative calibration between the photosphere models and MIPS data. Inspection of
the distribution of 13 and 24$\mu$m observed/star flux ratios from their Table 2 provides
a possible resolution; near unity their distributions have means of 1.02 and 1.03
respectively, suggesting that the photospheres are on average underestimated relative to
IRS. A possible origin of this discrepancy is that the 2MASS photometric system is about
2\% fainter than that used by MIPS \citep{2008AJ....135.2245R}, which if not corrected
for will lead to slightly fainter photospheres and the inference of warm disk components
where the evidence is marginal.

\citet{2014ApJS..211...25C} find that there are many other examples of single temperature
disks. Therefore, not all stars have strong evidence for two-temperature disks even when
they could have been detected, and single temperature disks may or may not actually
have multiple belts.

\subsection{Statistics}\label{ss:stats}

Given that many two-temperature disks are known to exist, it is desirable to make
an estimate of how common the phenomenon is. Ultimately we are biased by the overall set
of stars that were observed with IRS, since among nearby stars these were typically those
already known to host bright disks, and we are therefore biased towards detecting
two-temperatures in general. This bias means that any simple estimate of the two-temperature
occurrence rate will very likely be an overestimate.

We first consider stars in the unbiased DEBRIS sample (i.e. including stars observed by
DUNES and with guaranteed time). Only one of these was observed by the
\citet{2009ApJ...699.1067M} programme (HD~110411), meaning that objects in this sample
observed with IRS are unlikely to be strongly biased towards having two temperatures (see
section \ref{s:sample}). Our sample of two-temperature disks has 9 DEBRIS stars (6 A-type, 2
F-type, 1 K-type), while the DEBRIS sample itself has 83 A-type, 94 F-type, 89 G-type, 91
K-type, and 89 M-type primary stars. Of these, 21, 17, 9, 6, and 1 were observed with IRS
and have detected disks respectively, meaning that the raw fractions of disks that have
two-temperatures are 6/21, 2/17, 0/9, 1/6, and 0/1.

Fig. \ref{fig:teff-ldisk} shows how our overall sample of two-temperature disks compares
to the DEBRIS sample. The volume-limited DEBRIS sample includes relatively few bright
disks (as these are rare), and only the brightest DEBRIS disks are seen to have two
temperatures. However, the fainter disks may have two-temperature components that could
not be detected. A lack of sensitivity may therefore account for only a few (3/32)
two-temperature disks among FGK stars in DEBRIS. With the above caveat about a bias
towards two temperatures among IRS-observed disks, this fraction suggests that
two-temperature disks are fairly common around FGK stars. For A-types, there is still
little overlap in the two-temperature and DEBRIS stars in Fig. \ref{fig:teff-ldisk}, but
nearly 30\% of A-star disks are seen to have two temperatures. Further,
Fig. \ref{fig:trends} shows that these disks tend to have higher \rf, meaning that it is
likely that some of the remaining 70\% of DEBRIS A-type disks should have detectable
two-temperature behaviour that was not seen. Therefore, this largely qualitative look at
two-temperature disks among DEBRIS stars suggests that the phenomenon could be relatively
common, at a level of a few tens of percent.

To approach this issue from another angle, we consider the brightest disks observed with
IRS by \citet{2006ApJS..166..351C}, those with overall fractional luminosities above
$10^{-3}$. These six disks are all around stars younger than $\sim$20 Myr old, and are
sufficiently bright that two-temperature disks similar to others in our sample should
have been easily detectable. These are HD~95086, HD~110058, HD~113766, HD~146897,
HD~181327, and HD~191089. Of these, HD~95086 and HD~181327 show two temperatures
\citep{2012A&A...539A..17L,2013ApJ...775L..51M}, while the other four do not. HD~113766
has a silicate feature that makes SED fitting complex, but is inferred to have two
spatially distinct components \citep{2013A&A...551A.134O}. Therefore, one third of this
small number of stars with disks show two temperatures.

This rough estimate is lower than the 66\% found by \citet{2014ApJS..211...25C}, and in
closer agreement with 46 and 33\% found by \citet{2011ApJ...730L..29M} and
\citet{2013ApJ...775...55B}. There seem to be three possible reasons for this
difference. The first two are related to the way \citet{2014ApJS..211...25C} model their
disk spectra, which may result in detection of more two-temperature disks as described
above in relation to HD~191089 and HR~4796A. A third possible reason is the difference in
samples, because \citet{2014ApJS..211...25C} include many young stars. They found that
younger systems are more likely to have two-temperature disks, which may increase their
detection rate of two-temperature disks, particularly if many of these young stars were
observed based on previous disk detections.

Another question is therefore whether the frequency of two-temperature disks changes with
age, or whether they are just easier to detect due to brighter disks at younger ages. For
example, if the warm components are related to ongoing terrestrial planet formation and
are independent of the outer cool components, then two-temperature disks would only be
expected to appear around stars younger than $\sim$100 Myr. We can therefore compare our
rough two-temperature occurrence rates from DEBRIS and the younger
\citet{2006ApJS..166..351C} sample. Though our power to distinguish them is limited,
there is no evidence among these samples that the fraction of two-temperature disks
changes with age. Though they did not consider this possibility, Fig. 9 from
\citet{2014ApJS..211...25C} suggests that the frequency of two-temperature disks is
generally fairly constant, but could be higher for systems younger than $\sim$10 Myr. We
consider this issue in more detail in section \ref{ss:evol}.

To summarise, the results do not rule out two-temperature behaviour for most disks,
because most are too faint for it to be detected reliably. We have also shown that there
are examples of single temperature disks where two temperatures could have been
detected. Based on this discussion it seems that two-temperature disks are certainly not
rare, but neither are they ubiquitous.

\section{Theoretical grain models}\label{s:modelling}

Having studied the observational properties of two-temperature disks, we now explore how
two-temperature emission might appear from a dust belt at a single radial distance $r$
from the central star. These ideas largely rely on the differences in temperature that
grains of different sizes can have at a single stellocentric radius, which we consider
first. We then move on to numerical models, whose spectra in general depend on the host
star, the grain size distribution, and the optical properties of the grains.

\subsection{Grain temperatures}\label{ss:temps}

\begin{figure}
  \begin{center}
    \hspace{-0.5cm} \includegraphics[width=0.5\textwidth]{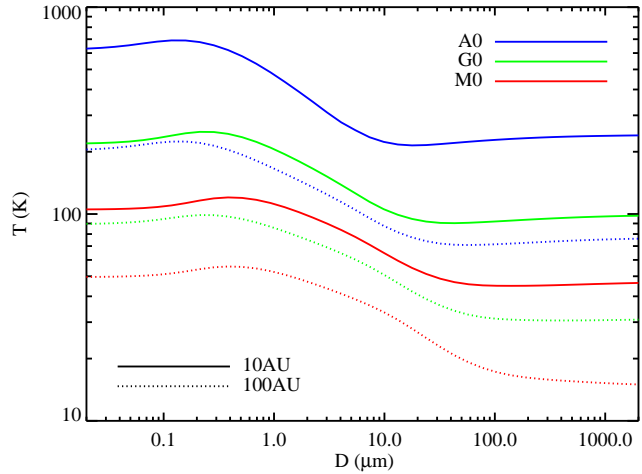}
    \caption{Grain temperatures as a function of diameter for dust around A, G, and
      M-type stars at 10 and 100AU. The temperature ratios for small and large grains in
      this figure vary from 2.2 to 3.3. The grain model is described in section
      \ref{ss:moddesc}.}\label{fig:temps}
  \end{center}
\end{figure}

Fundamentally, two-temperature emission could arise from single dust belts because the
temperature of dust grains depends on their size, as shown in Fig.
\ref{fig:temps}. There are two temperature regimes separated by a transition region; the
cooler blackbody regime is where grains are large and absorb and emit efficiently at all
wavelengths, and the ``small'' regime is where grains absorb and emit inefficiently at
all wavelengths. The exact location of these regimes depends on the spectral type of the
star and the radial location of the dust, but is a fairly weak function of these
parameters. Fig. \ref{fig:temps} shows that while the temperatures vary considerably
with spectral type and radial distance, the ratio of temperatures in the small ($T_{\rm
  sm}$) and blackbody ($T_{\rm BB}$) grain regimes is fairly constant at around 2-3.

To understand this ratio theoretically we balance the energy absorbed by a dust particle
of diameter $D$ over area $\pi D^2/4$ and emitted from an area $\pi D^2$ at stellocentric
distance $r$
\begin{equation}\label{eq:tdisk}
  \frac{R_\star^2}{4 r^2} \int_0^\infty Q_{\rm abs} B_\nu(T_\star) d\lambda = 
  \int_0^\infty Q_{\rm abs} B_\nu(T_{\rm dust}) d\lambda \, .
\end{equation}
For a blackbody particle that is perfectly absorbing and emitting $Q_{\rm abs}=1$ and the
blackbody dust temperature is
\begin{equation}
  T_{\rm BB}^4 = \frac{T_\star^4 R_\star^2}{4 r^2} = \frac{L_\star}{16 \sigma_{\rm K} \pi r^2}
\end{equation}
where $\sigma_{\rm K}$ is the Stefan-Boltzmann constant.

If we now consider a grain that is small relative to the peak wavelengths of star and
disk emission, the absorption and emission efficiency is $Q_{\rm abs} \propto
\lambda^{-n}$ at wavelengths that contribute significantly to the integrals in
Eq. (\ref{eq:tdisk}). For fixed $n$ these integrals are $\propto T^{4+n}$, so
Eq. (\ref{eq:tdisk}) can be rewritten as
\begin{equation}\label{eq:tsmtbb}
  \frac{T_{\rm sm}}{T_{\rm BB}} = \left( \frac{T_\star}{T_{\rm BB}} \right)^{n/(4+n)} \, .
\end{equation}
For typical values of $n=1-2$ \citep[e.g.][]{1989IAUS..135..285H}, $T_\star=6000$K, and
$T_{\rm BB}=70$K, equation (\ref{eq:tsmtbb}) yields $T_{\rm sm}/T_{\rm
  BB}=2.4-4.4$, which is in good agreement with the ratios found in Fig. \ref{fig:temps}.

These values are therefore representative of the temperature ratios \rt~that are
achievable in a two-temperature debris disk with dust in a single belt. While the
observed ratios could be smaller, if for example the smallest grains present in the disk
were $\sim$10$\mu$m in size and therefore grains of $T_{\rm sm}$ nonexistent, they cannot
be larger than allowed by the grain properties. If we take $T_{\rm sm}/T_{\rm BB}=5$ as
an approximate maximum allowed value, then $\eta$ Crv with $\mathcal{R}_T=6.4$ is the
only source for which the conclusion of two spatially distinct belts would be well
founded based purely on these simple temperature considerations.

\begin{figure}
  \begin{center}
    \hspace{-0.5cm} \includegraphics[width=0.5\textwidth]{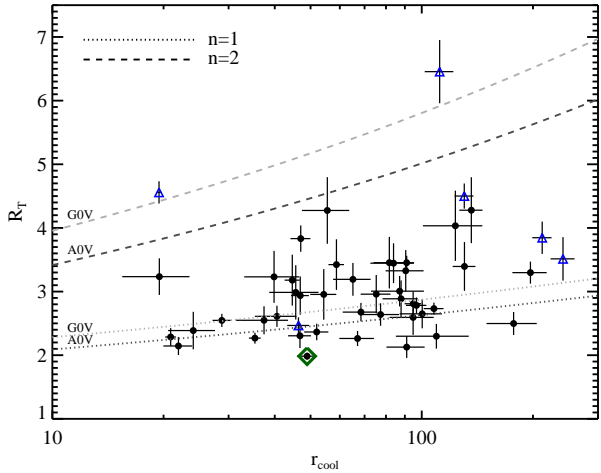}
    \caption{\rt~as a function of $r_{\rm cool}$. A weak positive correlation is
      predicted by Eq. (\ref{eq:tsmtbb}), shown for A0V and G0V stars for $n=1$ and
      2.}\label{fig:rcoldrt}
  \end{center}
\end{figure}

Eq. (\ref{eq:tsmtbb}) also predicts a weakly increasing temperature ratio with disk
radius (via $T_{\rm BB} \propto \sqrt{r}$). Fig. \ref{fig:rcoldrt} shows \rt~against disk
radius in search of this trend, with lines showing the predicted correlation. No clear
trend is visible, and the scatter in the points is larger than the variation expected for
$n=1$. The expected trend for $n=2$ is larger, but as comparison with Fig.
\ref{fig:temps} shows, $n=2$ tends to overestimate values for \rt~so is probably too
extreme to be representative. Inspection of this plot as a function of spectral type
shows no trends, though the relatively small difference between lines of different
spectral types shows that the expected differences are small. Therefore, this comparison
shows that a possible \rt~dependence on $r_{\rm cool}$ is not a good diagnostic of
two-temperature disks that may arise from the range of grain temperatures in a single
belt.

There is one more important aspect to be explored before turning to a more complex model,
which is the minimum grain size. The conclusion that any disk with $\mathcal{R}_T
\lesssim 5$ can be produced by a single dust belt relies on grains smaller than
$\sim$1$\mu$m being able to survive in the disk. However, it is thought that such small
grains are blown out by radiation pressure on dynamical timescales for Sun-like and
earlier-type stars. Adopting the relation for the blowout size in microns
\begin{equation}
  D_{\rm bl} = 0.8 (L_\star / M_\star) (2700/\rho) ,
\end{equation}
where $\rho$ is density in kg m$^{-3}$ and the luminosity and mass are in Solar units,
implies that while Sun-like stars can retain grains that are small enough to achieve
$T_{\rm sm}$, A0-type stars with typical blowout sizes of 10$\mu$m do not. Therefore, a
conclusion from Fig. \ref{fig:temps} is that early A-type stars only retain grains that
are all roughly the temperature of a blackbody. Therefore, if the calculated blowout size
is representative of the minimum grain size, early A-stars should not show
two-temperature spectra from a single dust belt. Another conclusion is that the minimum
grain size for Sun-like stars is sufficiently small that two-temperature spectra from a
single belt are possible.

\subsection{Numerical models}\label{ss:realgrain}

Having considered the possible range of grain temperatures, we now consider what is
probable given realistic size distributions and compositions. The size distribution can
be reasonably approximated by a power-law with a single slope parameter, while the
optical properties require a handful of parameters that describe the material
composition.

\subsubsection{Model description}\label{ss:moddesc}

In what follows we use a fairly standard grain composition model
\citep{1999A&A...348..557A}. In this model all grains have the same basic properties, and
are some mix of crystalline/amorphous silicates and organics with some porosity, where
the vacuum arising due to the porosity can be filled to some degree with crystalline or
amorphous water ice. Grain properties are calculated using Mie or Rayleigh-Gans theory or
geometric optics in the appropriate regimes \citep{1993ApJ...402..441L}, using optical
properties derived by \citet{1997A&A...323..566L,1998A&A...331..291L}. The refractive
indices of each material are mixed according to Maxwell-Garnett effective-medium theory,
which is not the only choice, but was preferred because it allows the grains to be
treated as a silicate core with a mantle of organics (i.e. the components are not treated
equally when mixed). This grain model is not the only possible approach, and for example
has problems reproducing observations when confronted with debris disks observed both in
thermal and scattered light
\citep[e.g.][]{2010AJ....140.1051K,2012A&A...540A.125A}. However, it has succeeded in
many instances when modelling of mid to far-IR disk spectra was required
\citep[e.g.][]{1999A&A...348..557A,2002MNRAS.334..589W,2012A&A...539A..17L}, and provides
a unified approach that has proven a useful tool for creating models more realistic and
complex than simple blackbodies.

Even with the various assumptions that were made in creating this grain model, there are
several more to be made. As outlined above, the minimum grain size and the size
distribution are important in setting the disk spectrum. We must also specify the mixture
of silicates, organics, vacuum (via porosity), and water ice (if the porosity is
non-zero), and whether these are crystalline or not. Formally, $q_{\rm si}$ sets the
fraction of total silicate+organic volume occupied by silicates, $p$ the porosity
fraction, and $q_{\rm H_2O}$ the fraction of vacuum filled with ice. We set the blowout
size as the size at which the radiation to gravitational force ratio parameter $\beta$ is
0.5.

For the composition, we use two different models that illustrate how the disk spectra can
vary. While previous authors \citep[e.g.][]{1999A&A...348..557A} have used compositions
at different extremes of what is possible, for example interstellar medium-like and
comet-like to explore how disk spectra can vary, we found that crystalline comet-like
compositions produced disks with strong spectral features that are not seen among our
sample. Our two compositions are therefore not so extreme. We default to moderately
porous grains with 1/3 amorphous silicates and 2/3 organics, and a small amount of water
ice (i.e. $p=0.5$, $q_{\rm si}=1/3$, $q_{\rm H_2O}=0.05$), which we call our ``rocky''
composition. For a second model, we increase $q_{\rm H_2O}$ to 0.85, which we call the
``icy'' composition.

Above, we considered the possible range of temperatures that could be present in a dust
belt. However, how different sizes contribute to the overall emission is set by the size
distribution, so unless it allows both the coolest and warmest grains to contribute
roughly equally to the emission, the presence of multiple dust temperatures will not
result in a two-temperature disk spectrum. One common way of describing the number of
objects between diameters $D$ and $D + dD$ in the size distribution is
\begin{equation}
  n(D) dD = K D^{2-3q} dD,
\end{equation}
where $K$ sets the normalisation and $q$ the steepness of the distribution (the origin of
$q$ being in the mass distribution $n(M)dM \propto M^{-q}dM$). When $q>1.67$ the surface
area is dominated by the smallest particles. A standard value for $q$ under the
assumption of an infinite size distribution with strength-independent size is 1.83
\citep{1969JGR....74.2531D}. However, the slope $q$ varies depending on the
size-dependent strength of objects \citep{2003Icar..164..334O}, with typical values
expected to be around 1.8-1.9 for objects that dominate the observable emission
\citep[e.g.][]{2012ApJ...754...74G}.

\subsubsection{Model results}\label{ss:modres}

We now show disk spectra for single-belt models with a range of size distributions and
our two compositions. We first show why changing the size distribution has important
consequences for the disk spectrum, and then how these models compare to the observed
sample of two-temperature disks for a range of size distributions and our two compositions.

\begin{figure}
  \begin{center}
    \hspace{-0.5cm} \includegraphics[width=0.5\textwidth]{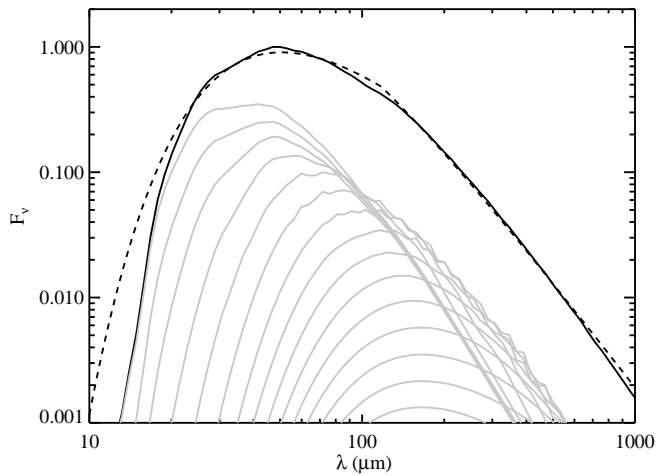}
    \caption{Disk spectrum for $r=100$ AU around a Sun-like star with our ``rocky''
      composition and $q=1.9$, $D_{\rm bl}=3\mu$m. The maximum object size is 1km. The
      solid black line shows the total spectrum, solid grey lines show the contribution
      from 15 logarithmically-spaced size bins (from 3$\mu$m to 40mm with a factor of
      1.96 spacing). Pebbles larger than 40mm lie off the bottom of the plot. The dashed
      line shows a 106K blackbody with $\lambda_0=123\mu$m and
      $\beta=0.88$.}\label{fig:rg_eg_1t}
  \end{center}
\end{figure}

We begin this part of the analysis with Fig. \ref{fig:rg_eg_1t}, which shows a typical
disk spectrum using the ``rocky'' composition with $q=1.9$ at 100AU from a Sun-like star
with the solid black line. The resulting spectrum is well matched by a modified blackbody
beyond 20 $\mu$m (dashed line). No two-temperature behaviour is present and if anything,
the blackbody model has too much warm emission rather than too little. What is also
clearly visible is that the emission is made up of a range of temperature components
(grey lines), to the extent that it is perhaps remarkable that the overall spectrum is
well described by a modified blackbody. The size distribution of $q=1.9$ means that the
spectrum contains a reasonable balance of these temperature components and the overall
spectrum has a temperature that is somewhere between $T_{\rm sm}$ and $T_{\rm BB}$.

\begin{figure}
  \begin{center}
    \hspace{-0.5cm} \includegraphics[width=0.5\textwidth]{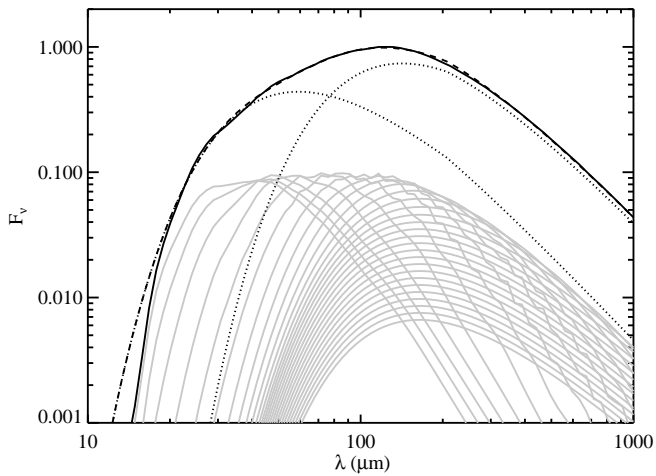}
    \caption{Same as Fig. \ref{fig:rg_eg_1t}, but with $q=1.73$. The model has
      temperatures of 36 and 86 (dotted lines, with $\mathcal{R}_T=2.6$) and
      $\mathcal{R}_f=1.8$, with $\lambda_0=206\mu$m and $\beta=0.35$. The flatter size
      distribution means that large objects, even those with 1 km diameters, contribute
      significantly to the overall spectrum.}\label{fig:rg_eg_2t}
  \end{center}
\end{figure}

To create a disk spectrum that looks more like it has two temperatures will require
different relative numbers of small and large grains. Decreasing the contribution from
large grains will only push the spectrum further towards something that looks like a
modified blackbody at $T_{\rm sm}$. However, increasing the number of large grains can
result in a more even contribution and a broader spectrum, as shown in Fig.
\ref{fig:rg_eg_2t}. This figure shows the result if $q=1.73$, which clearly shows a
two-temperature spectrum. The two-blackbody fit has $\mathcal{R}_T=2.6$ and
$\mathcal{R}_f=1.8$, which lies amongst the disks shown in Fig. \ref{fig:rfrt}, though
with a relatively high \rf. Some experimentation shows that two-temperatures are required
to fit our grain model SEDs for Sun-like stars when $q \lesssim 1.8$, and that similar
results can be obtained using either our rocky or icy compositions. As noted above, this
scenario does not work for early A-type stars, as long as we retain the assumption that
sub-blowout size grains do not make a significant contribution to the disk spectrum.

\begin{figure*}
  \begin{center}
    \hspace{-0.5cm} \includegraphics[width=1\textwidth]{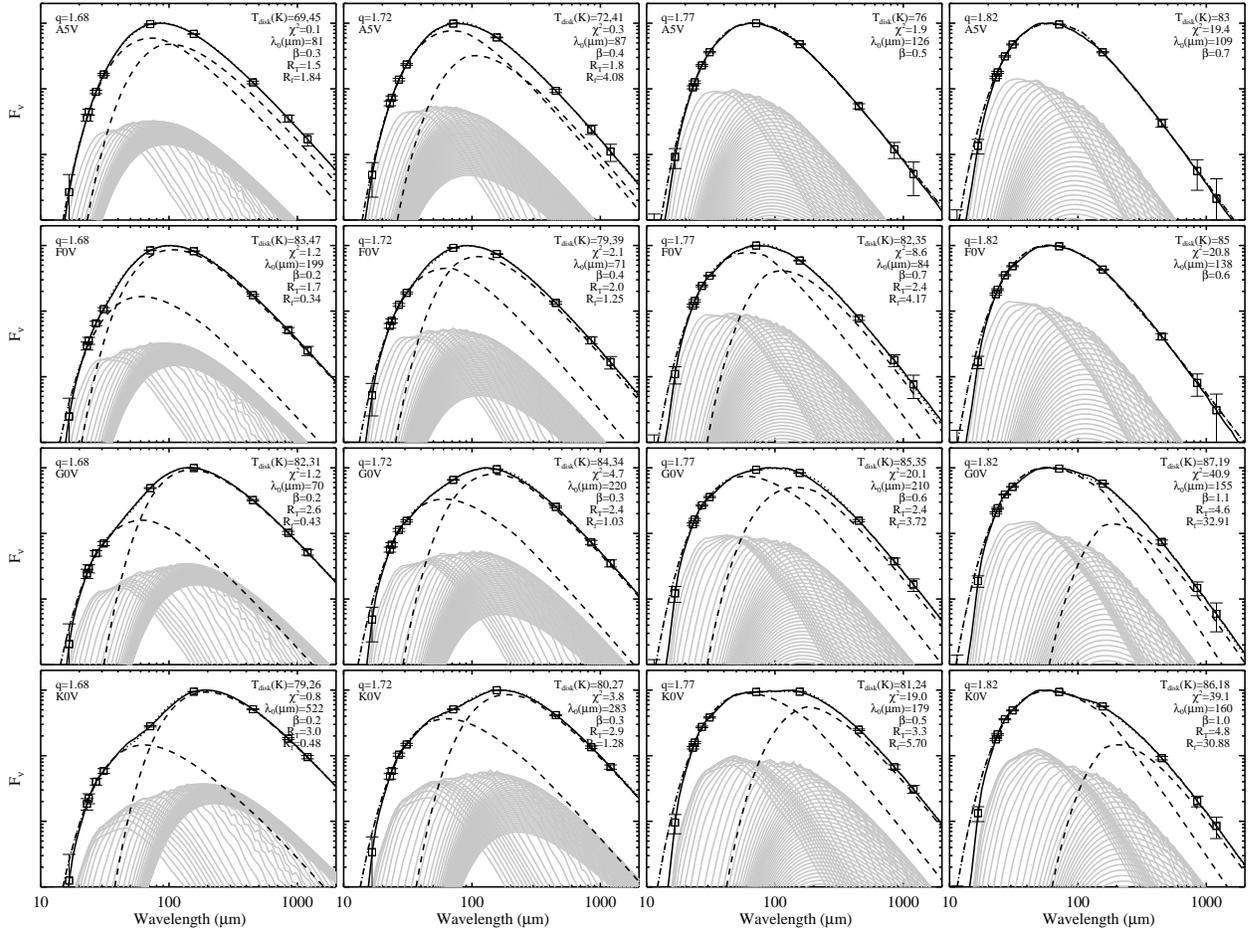}
    \caption{Example two-temperature disk spectra for a range of spectral types and size
      distribution slopes (noted in the top left of each panel), assuming that the
      minimum grain size is the blowout size and with our ``rocky'' composition ($q_{\rm
        H_2O}=0.05$). All disks are at 100 au. Each panel shows the contribution of
      grains of different sizes (grey lines) and the total spectrum (solid line). The
      best-fit two-temperature blackbody model is shown (dotted line), as is each
      component (dashed lines). The synthetic photometry used to fit the model are shown
      as squares. Each legend shows the disk temperatures, best fit $\chi^2$,
      $\lambda_0$, $\beta$, \rt, and \rf.}\label{fig:seds}
  \end{center}
\end{figure*}

To illustrate the range of behaviour and the trends that arise from these relatively
steep size distributions, Fig. \ref{fig:seds} shows models with our rocky composition
($q_{\rm H_2O}=0.05$) for a range of size distributions ($q=1.68$, 1.72, 1.77, and 1.82)
and spectral types (A5, F0, G0, and K0) with fitted blackbody models. We use the same
assumptions for fitting blackbody models as in Fig. \ref{fig:sens}, but now fit two
temperature components. The second fitted component is not shown if it has the same
parameters as the first. Some trends are clear; $\mathcal{R}_f$ tends to increase with
the steepness of the size distribution, which can be understood simply as a result of the
decreasing contribution of the large grains. The typical $\mathcal{R}_T$ increases to
later spectral types, which is the result of smaller blowout sizes and therefore a wider
range of grain temperatures (e.g. Fig. \ref{fig:temps}).

\begin{figure*}
  \begin{center}
    \hspace{-0.5cm} \includegraphics[width=0.5\textwidth]{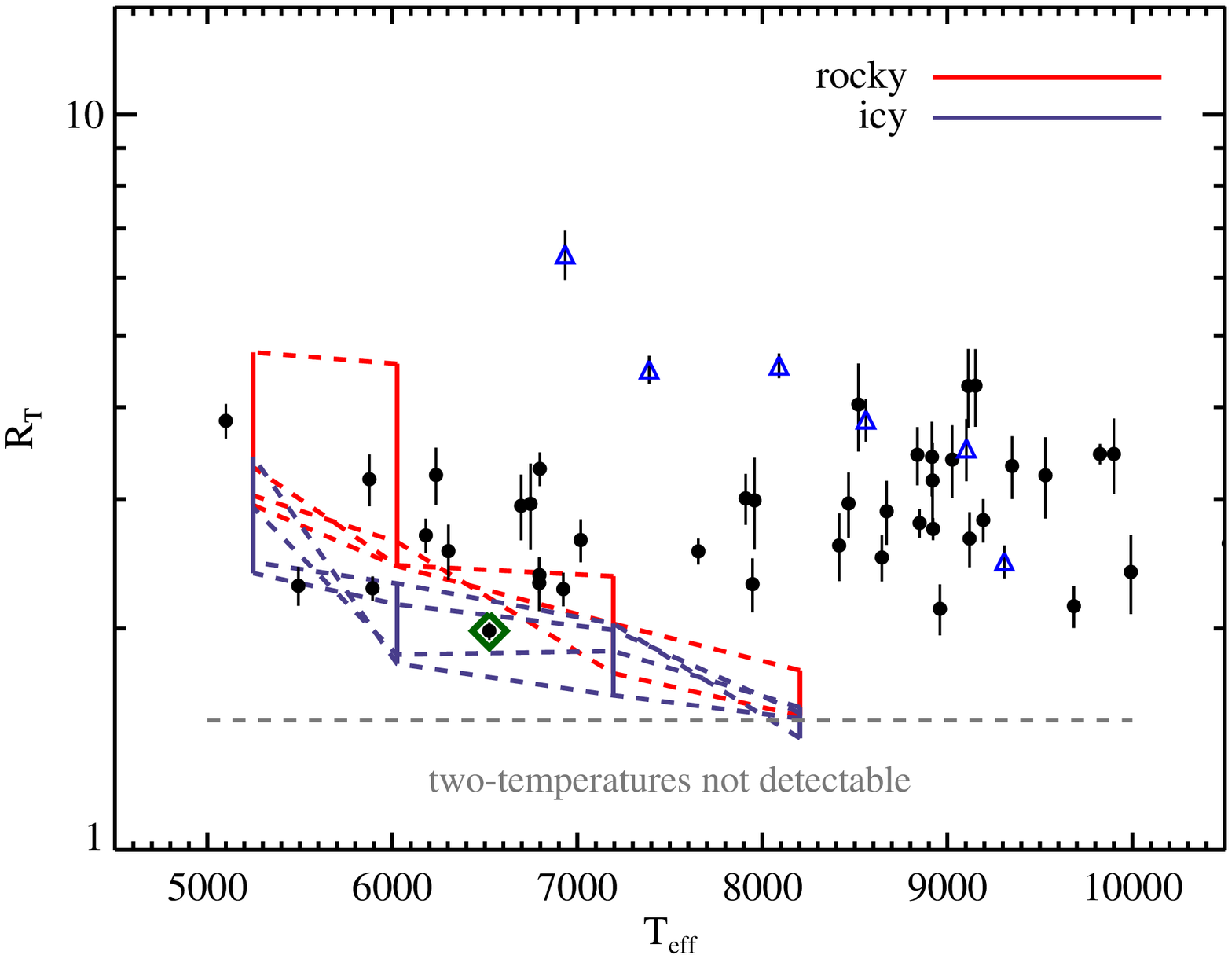}
    \includegraphics[width=0.5\textwidth]{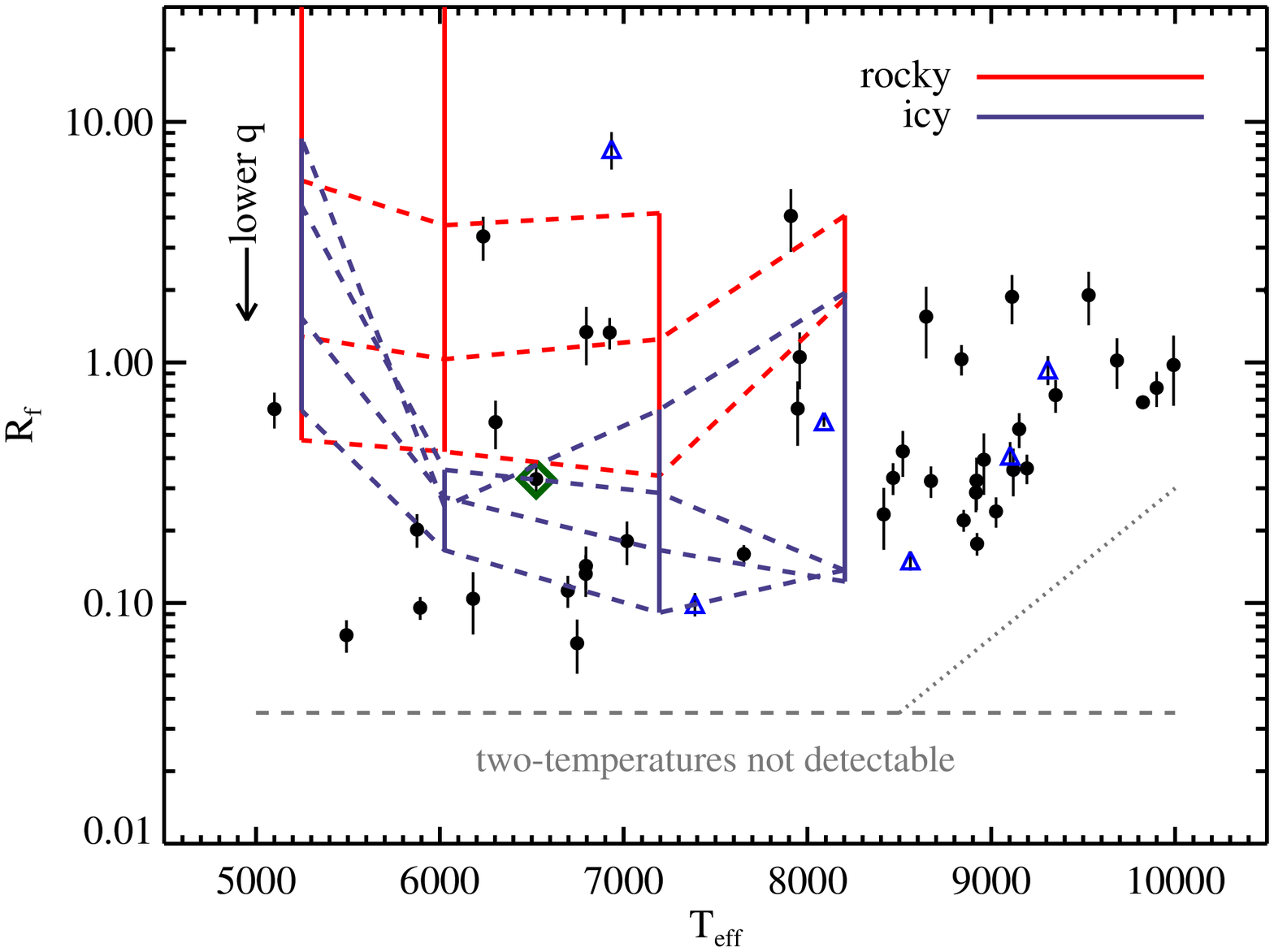}\\
    \caption{Sample of two-temperature debris disks (same as the top panels of Fig.
      \ref{fig:trends}), with additional lines showing parameter space covered by the
      models from size distribution models from Fig. \ref{fig:seds} and section
      \ref{ss:modres}. The two different sets of lines show the ``rocky'' and ``icy''
      compositions we considered. Each vertex represents a model that resulted in a
      two-temperature disk. Vertical solid lines connect models at constant $T_{\rm eff}$
      and near-horizontal dashed lines connect models of constant $q$ (with values of
      1.68, 1.72, 1.77, and 1.82 as shown in Fig. \ref{fig:seds}). The main trends are
      that earlier spectral types have lower \rt, and smaller $q$ results in lower
      \rf.}\label{fig:trends2}
  \end{center}
\end{figure*}

Fig. \ref{fig:trends2} shows the temperature and fractional luminosity ratios against
stellar temperature for our observed sample (i.e. the same as the top panels of
Fig. \ref{fig:trends}), with the addition of the range of parameters found from the grid
of ``rocky'' composition models in Fig. \ref{fig:seds} plotted as red lines. Each vertex
corresponds to a model that required a two-temperature blackbody fit. For these grids,
vertical solid lines show the effect of varying the size distribution slope $q$ at
constant spectral type, and near-horizontal dashed lines show the effect of varying the
spectral type at constant size distribution slope. The trends described above for
Fig. \ref{fig:seds} are apparent; models with flatter size distributions tends to have
lower \rf, and \rt~generally decreases as stellar effective temperature increases but
depends only weakly on $q$. For this composition, the models have lower \rt~than most
observed disks, and only reproduce the disks with higher \rf.

We now introduce the second ``icy'' composition ($q_{\rm H_2O}=0.85$), plotted as blue
lines in Fig. \ref{fig:trends2}. These models have been calculated over the same grid of
size distribution slopes and spectral types. A set of SEDs analogous to those in
Fig. \ref{fig:seds} looks qualitatively similar, and shows the same trends. This icier
composition covers a somewhat different range of \rt~and \rf, resulting in
two-temperature disks with lower values of both ratios. The effect of changing $q$ gives
a similar change as the difference between the two different compositions in logarithmic
\rf, suggesting that these can be equally important effects. In general however, flatter
size distributions are required to reproduce the lowest observed values of \rf. Both
models have trouble producing \rt~as large as those observed, particularly for earlier
spectral types due to larger blowout sizes. For the few models where \rt~is in reasonable
agreement, \rf~is relatively high ($\gtrsim$1).

\begin{figure}
  \begin{center}
    \hspace{-0.5cm} \includegraphics[width=0.5\textwidth]{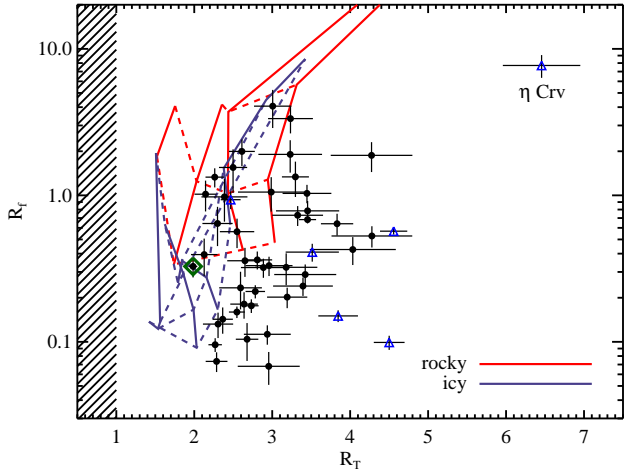}
    \caption{Sample of two-temperature debris disks (same as Fig. \ref{fig:rfrt}), with
      lines showing parameter space covered by the models from size distribution models
      from Fig. \ref{fig:seds} and section \ref{ss:modres}. The two different sets of
      lines show the ``rocky'' and ``icy'' compositions we considered. Each vertex
      represents a model that resulted in a two-temperature disk. Solid lines connect
      models at constant $T_{\rm eff}$ and dashed lines connect models of constant
      $q$.}\label{fig:trends3}
  \end{center}
\end{figure}

To emphasise the \rt~vs. \rf~parameter space covered, Fig. \ref{fig:trends3} shows the
model grids compared to our two-temperature sample. As is expected from the variations in
Fig. \ref{fig:trends2}, the models do not cover this space in a simple linear fashion,
but do show that only disks with relatively low \rt~are reproduced by narrow belt
models. At low \rf, model disks with \rt~larger than about 2 are not seen, increasing to
about 3 at higher \rf.

As noted above, the main conclusion from Fig. \ref{fig:trends2} is that \rt~decreases to
more luminous stars; their blowout sizes increase with luminosity and they do not show
two-temperature spectra for spectral types earlier than about A5. Based on the overlap
between the observed two-temperature and model disks in Figs. \ref{fig:trends2} and
\ref{fig:trends3}, Sun-like stars with relatively small \rt~may be described by our
single belt model. The single-belt system HD~181327 lies within our model grids with a
range of size distributions depending on the composition ($q \approx 1.67-1.82$) compared
to the results of \citet{2012A&A...539A..17L}, who found $q \approx 1.8$ but used a
carefully tuned mix of three material components. That such complex single-belt models
can be successfully constructed in individual cases suggests that our two simple models
can be taken as a general indication of where plausible models lie. This parameter space
can no doubt be expanded somewhat with more complex prescriptions. We return to tests of
this single belt model below.

\section{Discussion}\label{s:interp}

In the preceding sections, we have shown the properties of a sample of two-temperature
debris disks. These disks typically have warm/cool component temperature ratios \rt~of
2-4, and warm/cool component fractional luminosity ratios \rf~below ten
(Fig. \ref{fig:rfrt}). Warm components are detected with \rf~down to about 0.1, and those
fainter than this level become difficult to detect (Fig. \ref{fig:sens}). Biases in the
sample mean that the frequency of the two-temperature phenomenon is hard to estimate, but
it appears fairly common, at the tens of percent level.

We then explored how two-temperature disk spectra may arise from narrow planetesimal
belts, rather than two distinct belts as is generally assumed. The motivation comes from
the known variation of dust temperature with size (Fig. \ref{fig:temps}), and the
possibility that a single belt with properly modelled grain emission might provide a
simpler explanation for two-temperature disks than the assumption of multiple belts. We
found that a single belt can appear to have two temperatures, but that the specific
parameters depend on the size distribution and grain composition assumed. A weakness of
this model is that the maximum \rt~produced is 2-3, smaller than seen for many
systems. In addition, the single belt model only works for disks around Sun-like stars,
because more luminous stars remove small grains via radiation forces, and the range of
grain temperatures across the size distribution is then smaller than observed.

\subsection{Testing the single belt model}

The single belt scenario ultimately relies on the range of temperatures that grains can
have at a single stellocentric distance, which range from those of blackbodies at $T_{\rm
  BB}$, to those of very small grains at $T_{\rm sm}$. To reproduce the low \rf~values
seen for some systems requires relatively flat size distributions, where the two
temperature components correspond to these extremes, and in particular the cool component
is dominated by emission from objects that behave like blackbodies. A prediction of this
scenario is therefore that the true radius of the belt should correspond to that
predicted by the blackbody temperature of the cool component, which can be tested in
cases where two-temperature disks have been spatially resolved. For example,
\citet{2013ApJ...776..111M} found for four two-temperature disks that the resolved size
was 2-3 times larger than that predicted by a blackbody. Similar conclusions were reached
by \citet{2013MNRAS.428.1263B} for two resolved disks. Therefore, comparison of predicted
and resolved disk sizes does not appear to support the single belt scenario when
relatively flat size distributions are required. However, most debris disks are not
resolved, so this test is inconclusive in general.

\begin{figure}
  \begin{center}
    \hspace{-0.5cm} \includegraphics[width=0.5\textwidth]{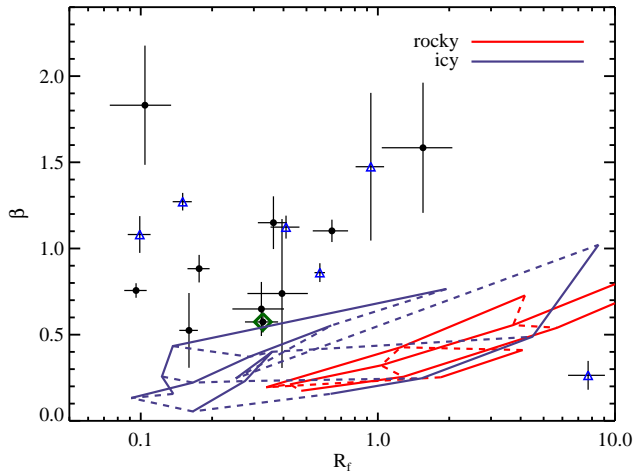}
    \caption{Two-temperature debris disks where the far-IR/sub-mm slope $\beta$ is
      constrained, with lines showing parameter space covered by the models from size
      distribution models from Fig. \ref{fig:seds} and section \ref{ss:modres}. The two
      different sets of lines show the ``rocky'' and ``icy'' compositions we
      considered. Each vertex represents a model that resulted in a two-temperature
      disk. Solid lines connect models at constant $T_{\rm eff}$ and dashed lines connect
      models of constant $q$. The observed $\beta$ are generally larger than expected for
      models with similar \rf.}\label{fig:beta-rf}
  \end{center}
\end{figure}

In cases where the cool belts are cool enough for the predicted blackbody size to be in
rough agreement with the resolved size, or where the resolved size is unknown, an
alternative test can be made. If the cool component is dominated by emission from grains
at the blackbody temperature, then the spectral slope of the far-IR and millimetre
emission should also appear similar to a blackbody (i.e. $\beta \approx 0$). This
property can be seen in Fig. \ref{fig:seds}, where $\beta$ is closer to zero for disks
with flatter size distributions (left column). In these spectra $\beta$ does not reach
zero because the cool component contains some emission from grains large enough to have
near-blackbody temperatures, but small enough to emit inefficiently at sub-mm
wavelengths.

To illustrate this point, Figure \ref{fig:beta-rf} shows two-temperature disks from our
sample where $\beta$ is constrained, which includes a range of host spectral types. The
plot also includes the grids of models described in section \ref{ss:modres}. The size
distribution slope $q$ affects both $\beta$ and \rf, so this plot tests whether the
$\beta$ predicted for a given \rf~is similar to that observed. The models are again shown
as lines of constant $T_{\rm eff}$ (solid) and $q$ (dashed), and as in
Fig. \ref{fig:trends3} the models do not cover this space in a simple linear
fashion. Overall however, these models predict lower $\beta$ for lower \rf, and for the
compositions used here consistently lie below the observed disks, with the exception of
$\eta$ Crv, which lies below the models. Different or more complex grain models could be
consistent with disks that lie near the model lines, with HD~181327 being a specific
example. Overall however, the single belt model again appears inconsistent with the
observed disk properties in the few cases where it can be tested. This test can only be
made for relatively few disks because sufficiently sensitive observations at far-IR/mm
wavelengths are required, and these are difficult to obtain. This difficulty leads to a
bias, in that disks with lower $\beta$ are more easily detected at long wavelengths
(i.e. those closer to pure blackbodies), and strengthens the conclusion that the observed
two-temperature disks have larger $\beta$ than expected from the single belt model.

Disks that lie close to HD~181327 in Fig. \ref{fig:beta-rf} may be the best place to look
for two-temperature disks arising from single belts. The four lying along a locus with
similar slope to the models are HD~39060 ($\beta$ Pic), HD~32297, HD~110411 ($\rho$ Vir),
and HD~161868 ($\gamma$ Oph). Of these, $\beta$ Pic has a well studied and complex disk
structure, that extends over a range of radii
\citep[e.g.][]{1984Sci...226.1421S,2005Natur.433..133T,2014arXiv1404.1380D}. HD 32297 was
modelled as two belts by \citet{2013ApJ...772...17D}, and while they find that the inner
component could not be accounted for by their models of the outer component, this
putative inner component has yet to be confirmed. HD~110411 and HD~161868 are have
relatively little spatial disk information and no resolved detection of an inner
component \citep{2010ApJ...723.1418M,2013MNRAS.428.1263B}. Therefore, three of these four
disks represent worthy targets for future high resolution imaging that test for the
presence of inner disk components.

In summary, our models show that in some cases two-temperature disks can arise from
single belts. As long as the minimum grain size is set by radiation pressure,
two-temperature disks around A-type stars probably arise from multiple belts. In
addition, a few two-temperature disks have been confirmed to have multiple belts by high
resolution observations, and these comprise both A-type and Sun-like stars. For Sun-like
stars, single belt models, particularly those with relatively flat size distributions,
can produce two temperature disks, and this model is not conclusively ruled out because
not all disks are resolved and/or detected at far-IR/mm wavelengths. Where observations
exist however, this model is disfavoured. In addition, the flatter size distributions are
steeper than those expected from collisional models and inferred from detailed modelling
of well characterised systems. Therefore, in general, the assumption that two temperature
disks have multiple belts should not be made without considering the properties of those
disks and their host stars, but it seems likely that the bulk of two-temperature disks do
arise from multiple belts.

\subsection{Evolution of multiple belts}\label{ss:evol}
 
We now consider whether our results shed light on the origin of multiple belts, which in
general appear to be the origin of two-temperature disks. A possible constraint could
come from the expected collisional evolution. For example, if two-temperature disks arise
from a single belt and the material compositions do not change, no significant evolution
of \rf~or \rt~would be expected over time because the observed emission always comes from
material in the same location. However, this similarity may also be expected if the warm
belts are made of material delivered from the outer belt, perhaps scattered by planets
\citep[e.g.][]{2007ApJ...658..569W,2012MNRAS.420.2990B}, in which case the brightness of
the inner belt is reasonably connected to that of the outer one.

On the other hand, if two temperatures arise from two independent belts (i.e. as in the
Solar system), the two belts are expected to collisionally evolve at different rates. We
can estimate the results of differential evolution by assuming that two belts at
different radii began their evolution at the same time, soon after the debris disk
emerged from the gaseous protoplanetary disk. The collision rate in the disk depends
strongly on orbital radius, and for an equal number of objects is higher at smaller radii
due both to greater relative velocities and a smaller enclosed volume. The brightness of
a belt will start to decay when the largest objects start to collide, which will take
longer for the outer belt. Therefore, for two belts that have the same initial
brightness, the inner one will start to decay first, and the outer belt will follow
later, and all other things being equal, in the long term the brightness difference
between two belts is set by the difference in their radii.

\citet{2007ApJ...658..569W} estimate that the maximum fractional luminosity of an
individual belt is
\begin{equation}\label{eq:fmax}
  f_{\rm max} = 1.6 \times 10^{-4} r^{7/3} M_\star^{-5/6} L_\star^{-0.5} t^{-1} \, ,
\end{equation}
where the variables are disk radius (in au), stellar mass and luminosity (in Solar
units), and system age (in Myr). This fractional luminosity applies to all disks because
the brightness decay rate is proportional to the disk mass (and hence the
brightness). Thus, all disks tend to the same brightness level once the most massive
objects have started to collide. In disks with relatively low initial masses the largest
objects take longer to start to collide and decay, and until that time the fractional
luminosity lies below the level given by equation (\ref{eq:fmax}).

This theory was applied to systems with dust at au scales, such as $\eta$ Crv and
HD~69830 \citep{2007ApJ...658..569W}. These disks were found to lie well above $f_{\rm
  max}$ and were deemed ``transient'', in that they could not be described by this model
of collisional evolution. Here however, the warm dust components lie roughly in the range
of a few to 10 au, so for the typical $<$Gyr ages of objects in our sample the observed
fractional luminosities of both the warm and cool components are comparable with $f_{\rm
  max}$, rather than significantly above it. This agreement suggests that the warm
components, if interpreted as distinct belts, are undergoing the collisional evolution
expected within the framework of this model, but are not brighter than expected (with the
notable exception of $\eta$ Crv). This inference in turn suggests that we may see the
differential evolution of warm/cool component brightnesses described above. However,
given the radii inferred for the warm components and their correspondingly long collision
timescales, which may be similar to their ages \citep[e.g.][]{2013ApJ...768...25G}, the
non-detection of such differential evolution would not rule out the two-belt scenario.

To consider the expected evolution of \rf, we assume a typical \rt~of 3, so the radii are
a factor of 9 different. Initially, both belts will be very bright, having just emerged
from the protoplanetary disk phase, so if both belts are assumed to be near to radially
optically thick then \rf~will be of order unity \citep[but could of course be different,
for example if the outer disk is shadowed by the inner one,][]{2014MNRAS.tmp...88K}. The
brightness of the inner belts is expected to be currently decaying, though this decay may
have only begun recently. The outer belts, at roughly ten times greater distances are not
expected to be decreasing in brightness significantly due to much longer collision
timescales (roughly a factor 10$^4$). Therefore, the basic expectation is that the ratio
of warm/cool belt brightnesses will start somewhere near unity and decrease over
time. Given sufficient time, the difference in belt radii implies this ratio would
eventually reach a value of
\begin{equation}
  \mathcal{R_{\rm f,max}} = f_{\rm max,warm}/f_{\rm max,cool} = \mathcal{R}_T^{-14/3},
\end{equation}
or approximately $10^{-2}$ to $10^{-3}$. However, the evolution is sufficiently slow that
this limit will not be reached for the $\lesssim$Gyr ages within our sample.

\begin{figure}
  \begin{center}
    \hspace{-0.5cm} \includegraphics[width=0.5\textwidth]{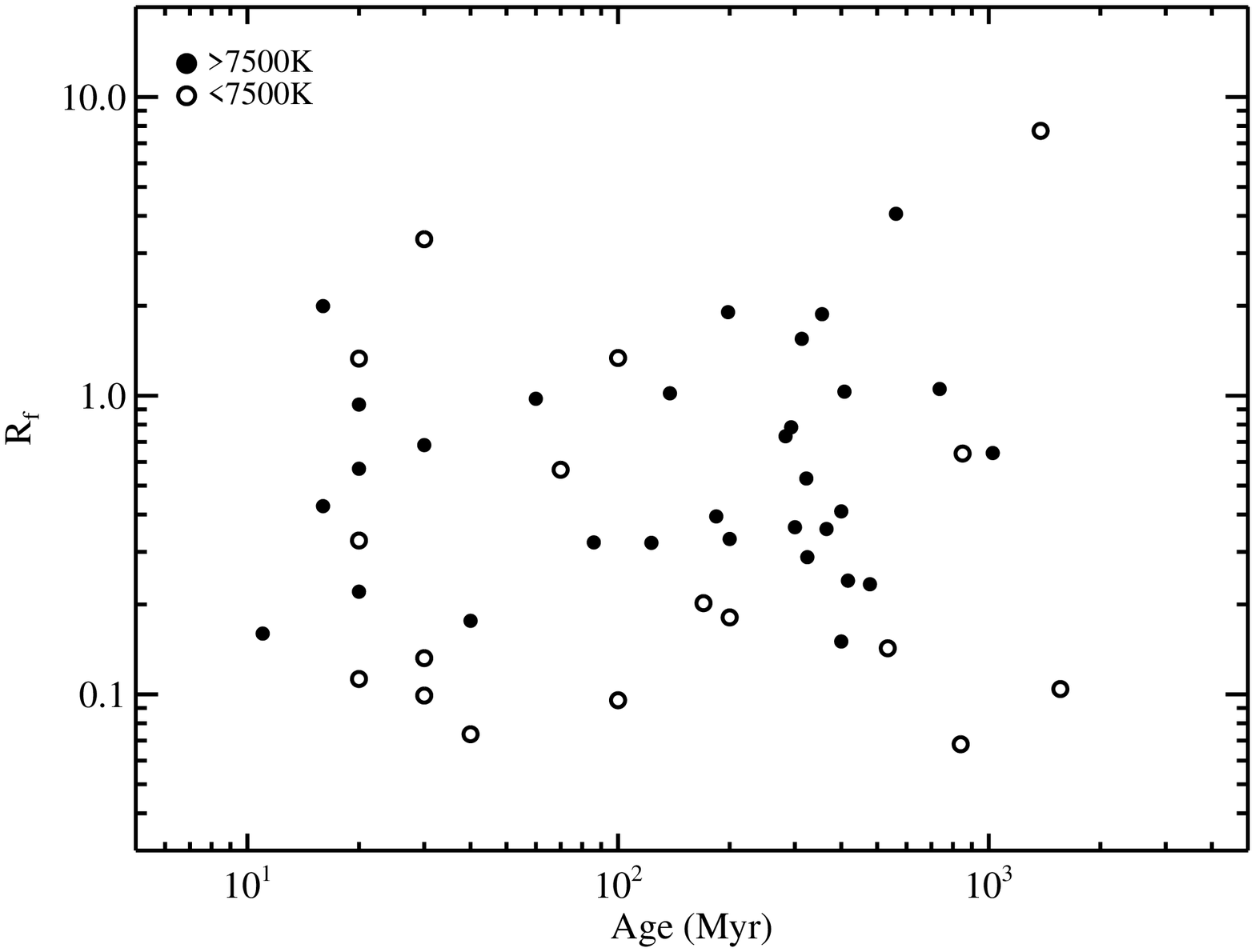}
    \caption{Dependence of \rf~on stellar age. No trends are visible, but the $\sim$10 au
      radii of the warm components and their relatively slow collisional evolution means
      that the collisional two-belt scenario is not ruled out.}\label{fig:age}
  \end{center}
\end{figure}

Fig. \ref{fig:age} shows the evolution of \rf~with time for our sample. We use ages from
\citet{2004A&A...426..601D,2013ApJ...763..118S,2014ApJS..211...25C}, but adjust the age
of the $\beta$ Pictoris moving group to 20 Myr \citep{2014MNRAS.438L..11B}, and the age
of HD~61005 to 40 Myr \citep{2013MNRAS.431.1005D}. These ages are of course very
uncertain and there are many disagreements in the literature, but there is a reasonable
age distinction between moving group/association stars and older field stars
\citep[listed by][]{2014ApJS..211...25C}, so the ages should at least be
representative. No significant evolution is seen, though this lack of evolution does not
strongly rule out the hypothesis that the two temperatures correspond to two distinct
belts. An additional expectation is that the ages of two-temperature systems would be
biased towards young ages, because the warm components should decay to undetectable
levels more rapidly than the cool components. Our sample is inevitably biased towards
younger ages due to young disks being brighter, so any conclusions drawn based on the
relative youth of our sample would not be definitive. While \citet{2014ApJS..211...25C}
find that two-temperature disks are more likely to be found around younger stars, this
tendency is not strong and 50\% of their two-temperature disks are $\gtrsim$100 Myr old.

In summary, we cannot rule out the hypothesis that two-temperature disks arise from two
independent belts that are decaying due to collisions. The reason being that the warm
belts are at sufficiently large radial distances that their brightness is not at odds
with models of collisional evolution.

\subsection{Planetary system structure}

Our main conclusion is that most two-temperature debris disks comprise two disk
components. Consideration of collisional models shows that these components could be two
independent belts undergoing normal collisional evolution, analogous to the Solar
System's Asteroid and Edgeworth-Kuiper belts. The exception among our sample is $\eta$
Crv, whose warm component is too bright to be explained by collisional models and may
originate from material scattered from the outer belt
\citep[e.g.][]{2007ApJ...658..569W,2012ApJ...747...93L}. In other systems the inner
components may also be linked to the outer belts via inward scattering of material by
intervening planets \citep[e.g.][]{2012MNRAS.420.2990B}. While such a scenario is not
required to explain the observed warm dust levels, such scattering no doubt occurs in
some, and perhaps all, systems.

Considering the scattering scenario, the inner belt is generally thought to originate due
to objects depositing their mass near the star after passing inside some comet
sublimation and/or disintegration radius
\citep[e.g.][]{2008Icar..195..871K,2012MNRAS.420.2990B,2012A&A...548A.104B}. Naively,
such a picture is inconsistent with the trend towards higher warm component temperatures
for more massive stars, since sublimation should occur at constant temperature. In
addition, we find that the warm components can be as cool as 100 K around Sun-like stars,
lower than the expected sublimation temperature of $\sim$150 K. However, as planetesimals
are scattered inward they will collide most often at stellocentric distances near the
innermost planet where the volume density and relative velocities are highest, and may
even be disrupted due to tidal forces given sufficiently close encounters with this
planet. In this case the scattering scenario still allows the creation of two-temperature
disks in the absence of thermal destruction of planetesimals within a few au of the star.

Circumstantial evidence that warm and cool belts are separated by planets is provided by
extra-Solar systems with two disk components and intervening planets, but does not
distinguish between the independent-belt and scattering scenarios. The two belts in the
HR~8799 system are separated by a series of massive planets
\citep{2008Sci...322.1348M,2009A&A...503..247R,2009ApJ...705..314S,2014ApJ...780...97M},
and HD~95086 has a single planet that resides in a two-temperature disk
\citep{2013ApJ...779L..26R,2013ApJ...775L..51M}, though the warm component has yet to be
confirmed by high resolution observations for the latter system.
 
The $\sim$10 au typical radii of the warm components of our two-temperature disks do not
rule out planets and debris disks at smaller distances. These tend to be much less
massive and much harder to detect
\citep[e.g.][]{2012ApJS..201...15H,2013MNRAS.433.2334K}, so it seems probable that
low-mass planets and fainter exo-Zodiacal clouds reside interior to the warm components
of the disks we have considered here.

Therefore, in either of the above scenarios, two-temperature debris disks seem to give
information on the typical scales of outer planetary systems, with the warm/cool
temperature ratios suggesting that these typically span a factor of ten in radius.

\section{Conclusions}

We have presented a study of debris disks whose emission spectra are well modelled by
dust at two temperatures. These disks are typically assumed to be a 
sign of multiple belts, so here our goal was to explore whether this emission could arise
from dust in a single belt, with the range of temperatures arising from the natural
variation in grain temperature with size.

We collected a sample of 48 nearby stars with two-temperature debris disks, and used the
ratios of warm/cool component temperatures (\rt) and fractional luminosities (\rf) as a
diagnostic of disk properties. A plot of \rf~versus \rt~shows that $\eta$ Crv is
clearly an outlier among two-temperature disks, having an unusually large warm/cool
temperature ratio. We also identified HD~145689 as a potentially interesting system,
where the M9 companion may orbit outside or between two debris disk components, or host a disk
itself.

Using a grain emission model, we test whether two-temperature disks can arise from single
belts. As long as the minimum grain size is set by radiation pressure, two-temperature
disks around A-type stars probably arise from multiple belts. In addition, a few
two-temperature disks have been confirmed to have multiple belts by high resolution
observations, and these comprise both A-type and Sun-like stars. For Sun-like stars, our
single-belt model can produce two temperature disks. Where observations allow tests to be
made this model is disfavoured, but it is not conclusively ruled out because not all
disks are resolved and/or detected at far-IR/mm wavelengths. In general therefore, the
assumption that two temperature disks have multiple belts should not be made, but it
seems likely that the bulk of two-temperature disks do arise from multiple belts. As
noted at the outset, PR drag may allow disks whose planetesimals reside in a narrow belt
to have two temperatures due to small grains extending in towards the star. Whether this
process can generically reproduce a subset of two-temperature disks is clearly worth
future effort.

Assuming the multiple belt interpretation is correct, we considered the expected
collisional evolution of two distinct belts. Aside from $\eta$ Crv, the warm components
could be independent belts undergoing normal collisional evolution, so it is possible
that two-temperature disks represent systems with analogues of the Asteroid and
Edgeworth-Kuiper belts that are separated by planets. Scattering of material from the
outer regions could still be an important, or even dominant, mechanism for creating
two-temperature debris disks, with the warm component comprising material scattered from
the cool component, again due to the presence of intervening planets. For either
scenario, the ratio of warm/cool component temperatures is indicative of the scale of
outer planetary systems, which typically span a factor of about ten in radius.

\section*{Acknowledgements}

This work was supported by the European Union through ERC grant number 279973. We thank
the referee for their comments and suggestions, Alexander Krivov for helpful discussions,
and Christine Chen and Kate Su for sharing their IRS spectra.


\appendix


\section{Sample and (sub-)mm photometry}

\begin{table*}
  \caption{Sample and results of blackbody fitting for 48 two-temperature disks, 9
    sources marked with a * are part of the DEBRIS sample
    \citep{2010MNRAS.403.1089P}. The ``Ref'' column notes papers from which far-IR
    photometry was obtained:
    1: \citet{1988iras....7.....H},
    2: \citet{1990IRASF.C......0M},
    3: \citet{2006ApJ...653..675S},
    4: \citet{2008ApJ...681.1484R},
    5: \citet{2008ApJS..179..423C},
    6: \citet{2009ApJ...705.1226B},
    7: \citet{2009ApJ...699.1067M},
    8: \citet{2009ApJ...705..314S},
    9: \citet{2010A&A...518L.130S},
    10: \citet{2010A&A...518L.132L},
    11: \citet{2010A&A...518L.133V},
    12: \citet{2011ApJ...732...61Z},
    13: \citet{2011ApJS..193....4M},
    14: \citet{PhillipsThesis},
    15: \citet{2012A&A...540A.125A},
    16: \citet{2013ApJ...768...25G},
    17: \citet{2013MNRAS.428.1263B}.}\label{tab:sample}
    \begin{tabular}{lrrrrrrrrrrrrrrl}
    \hline \hline Name & $T_\star$ & Age (Myr) & $T_{\rm warm}$ & $e_{T_{\rm warm}}$ & $T_{\rm cool}$
    & $e_{T_{\rm cool}}$ & $\lambda_0$ & $e_{\lambda_0}$ & $\beta$ & $e_\beta$ &
    $\mathcal{R}_T $ & $e_{\mathcal{R}_T}$ & $\mathcal{R}_f $ & $e_{\mathcal{R}_f}$ & Ref \\
    \hline
HD 377 & 5876 & 170 & 113 & 9 & 35 & 2 &  &  &  &  & 3.2 & 0.3 & 0.20 & 0.03 & 5 \\
HD 6798 & 9120 & 365 & 180 & 16 & 68 & 5 &  &  &  &  & 2.7 & 0.2 & 0.36 & 0.08 & 2 \\
HD 9672 & 8923 & 40 & 143 & 5 & 52 & 2 & 61 & 10 & 0.9 & 0.1 & 2.7 & 0.1 & 0.18 & 0.02 & 2 \\
HD 10647* & 6181 & 1560 & 99 & 6 & 37 & 2 & 56 & 8 & 1.8 & 0.3 & 2.7 & 0.1 & 0.10 & 0.03 & 2,6,10,16 \\
HD 10939 & 9026 & 417 & 200 & 23 & 59 & 2 &  &  &  &  & 3.4 & 0.4 & 0.24 & 0.03 & 1,7 \\
HD 13246 & 6236 & 30 & 231 & 22 & 72 & 9 &  &  &  &  & 3.2 & 0.3 & 3.34 & 0.69 & 12 \\
HD 14055* & 9197 & 300 & 183 & 13 & 65 & 2 & 192 & 21 & 1.1 & 0.2 & 2.8 & 0.2 & 0.36 & 0.05 & 2,3,14 \\
HD 15115 & 6696 & 20 & 163 & 17 & 55 & 2 &  &  &  &  & 2.9 & 0.3 & 0.11 & 0.02 & 2,13 \\
HD 15745 & 6924 & 20 & 104 & 6 & 46 & 3 &  &  &  &  & 2.3 & 0.1 & 1.33 & 0.20 & 2,13 \\
HD 16743 & 7018 & 200 & 127 & 9 & 48 & 3 &  &  &  &  & 2.6 & 0.2 & 0.18 & 0.04 & 13 \\
HD 22049* & 5100 & 850 & 119 & 7 & 31 & 1 & 73 & 11 & 1.1 & 0.1 & 3.8 & 0.2 & 0.64 & 0.11 & 2,6 \\
HD 23267 & 9992 & 60 & 330 & 42 & 138 & 11 &  &  &  &  & 2.4 & 0.3 & 0.98 & 0.32 & 7 \\
HD 25457 & 6303 & 70 & 138 & 13 & 54 & 5 &  &  &  &  & 2.5 & 0.2 & 0.56 & 0.13 & 2,5 \\
HD 30447 & 6794 & 30 & 133 & 11 & 58 & 2 &  &  &  &  & 2.3 & 0.2 & 0.13 & 0.03 & 2,13 \\
HD 31295 & 8673 & 123 & 168 & 17 & 58 & 3 &  &  &  &  & 2.9 & 0.3 & 0.32 & 0.05 & 2,3,14 \\
HD 32297 & 7654 & 11 & 203 & 9 & 80 & 2 & 237 & 105 & 0.5 & 0.2 & 2.5 & 0.1 & 0.16 & 0.01 & 1 \\
HD 38056 & 9900 & 293 & 302 & 36 & 88 & 4 &  &  &  &  & 3.5 & 0.4 & 0.78 & 0.13 & 7 \\
HD 38206 & 9825 & 30 & 233 & 8 & 68 & 2 &  &  &  &  & 3.5 & 0.1 & 0.68 & 0.04 & 2,7 \\
HD 38207 & 6795 & 534 & 123 & 7 & 52 & 2 &  &  &  &  & 2.4 & 0.1 & 0.14 & 0.03 & 5 \\
HD 39060* & 8090 & 20 & 493 & 19 & 108 & 1 & 204 & 15 & 0.9 & 0.1 & 4.6 & 0.2 & 0.57 & 0.03 & 2,11,14 \\
HD 61005 & 5492 & 40 & 123 & 8 & 54 & 1 &  &  &  &  & 2.3 & 0.1 & 0.07 & 0.01 & 1,5 \\
HD 70313 & 8466 & 200 & 183 & 19 & 62 & 3 &  &  &  &  & 3.0 & 0.3 & 0.33 & 0.05 & 2,7 \\
HD 71722 & 8917 & 324 & 256 & 30 & 75 & 2 &  &  &  &  & 3.4 & 0.4 & 0.29 & 0.05 & 7 \\
HD 79108 & 9350 & 283 & 230 & 23 & 69 & 5 &  &  &  &  & 3.3 & 0.3 & 0.73 & 0.11 & 2,7 \\
HD 80950 & 9684 & 138 & 294 & 21 & 137 & 7 &  &  &  &  & 2.1 & 0.1 & 1.02 & 0.24 & 7 \\
HD 98673 & 7958 & 737 & 244 & 36 & 82 & 7 &  &  &  &  & 3.0 & 0.4 & 1.05 & 0.28 & 7 \\
HD 107146 & 5893 & 100 & 103 & 4 & 45 & 1 & 338 & 20 & 0.8 & 0.0 & 2.3 & 0.1 & 0.10 & 0.01 & 2,5 \\
HD 109085* & 6934 & 1380 & 254 & 20 & 39 & 2 & 34 & 12 & 0.3 & 0.1 & 6.5 & 0.5 & 7.70 & 1.36 & 2 \\
HD 110411* & 8920 & 86 & 253 & 32 & 79 & 2 & 75 & 9 & 0.6 & 0.2 & 3.2 & 0.4 & 0.32 & 0.08 & 2,14,17 \\
HD 125162* & 8646 & 313 & 106 & 8 & 42 & 4 & 93 & 24 & 1.6 & 0.4 & 2.5 & 0.2 & 1.55 & 0.51 & 2,14 \\
HD 136246 & 8519 & 16 & 211 & 29 & 52 & 6 &  &  &  &  & 4.0 & 0.6 & 0.43 & 0.09 & 7 \\
HD 136482 & 10521 & 16 & 296 & 20 & 113 & 8 &  &  &  &  & 2.6 & 0.2 & 1.99 & 0.41 & 7 \\
HD 138965 & 8850 & 20 & 154 & 7 & 55 & 2 &  &  &  &  & 2.8 & 0.1 & 0.22 & 0.02 & 1,7 \\
HD 141378 & 8415 & 478 & 146 & 16 & 56 & 4 &  &  &  &  & 2.6 & 0.3 & 0.23 & 0.07 & 2,7 \\
HD 153053 & 7947 & 1025 & 107 & 11 & 46 & 5 &  &  &  &  & 2.3 & 0.2 & 0.64 & 0.19 & 1,7 \\
HD 159492 & 7910 & 562 & 166 & 14 & 55 & 5 &  &  &  &  & 3.0 & 0.2 & 4.06 & 1.18 & 2,14 \\
HD 161868 & 8960 & 184 & 141 & 12 & 66 & 4 & 162 & 95 & 0.7 & 0.4 & 2.1 & 0.2 & 0.39 & 0.11 & 2,3,14 \\
HD 172167* & 9103 & 400 & 166 & 16 & 47 & 2 & 67 & 5 & 1.1 & 0.1 & 3.5 & 0.3 & 0.41 & 0.06 & 2,9,14 \\
HD 181296 & 9308 & 20 & 254 & 14 & 103 & 4 & 87 & 13 & 1.5 & 0.4 & 2.5 & 0.1 & 0.93 & 0.13 & 2,4 \\
HD 181327 & 6524 & 20 & 107 & 3 & 54 & 2 & 59 & 4 & 0.6 & 0.1 & 2.0 & 0.1 & 0.33 & 0.05 & 2 \\
HD 182919 & 9530 & 198 & 329 & 43 & 102 & 10 &  &  &  &  & 3.2 & 0.4 & 1.90 & 0.47 & 7 \\
HD 191174 & 9113 & 355 & 330 & 41 & 77 & 6 &  &  &  &  & 4.3 & 0.5 & 1.87 & 0.43 & 7 \\
HD 192425 & 8838 & 408 & 222 & 21 & 65 & 5 &  &  &  &  & 3.4 & 0.3 & 1.03 & 0.15 & 2,7 \\
HD 205674 & 6747 & 840 & 149 & 20 & 50 & 2 &  &  &  &  & 3.0 & 0.4 & 0.07 & 0.02 & 2,13 \\
HD 216956* & 8560 & 400 & 148 & 10 & 39 & 1 & 74 & 6 & 1.3 & 0.1 & 3.8 & 0.3 & 0.15 & 0.01 & 2,14,15 \\
HD 218396 & 7388 & 30 & 163 & 7 & 36 & 1 & 67 & 7 & 1.1 & 0.1 & 4.5 & 0.2 & 0.10 & 0.01 & 8 \\
HD 221853 & 6797 & 100 & 91 & 5 & 28 & 1 &  &  &  &  & 3.3 & 0.2 & 1.34 & 0.36 & 2,13 \\
HD 225200 & 9152 & 322 & 263 & 32 & 61 & 2 &  &  &  &  & 4.3 & 0.5 & 0.53 & 0.09 & 7 \\
\hline
\end{tabular}
\end{table*}

\begin{table*}
  \caption{Sub-mm and mm photometry of targets in our sample. The 3$\sigma$ limit column
    indicates that the flux is an upper limit. Fluxes without this flag are not
    necessarily significant detections.}\label{tab:submm}
    \begin{tabular}{lrrrrrl}
    \hline \hline Name & $\lambda (\mu$m) & Instrument & Flux (mJy) & Unc (mJy) & 3$\sigma$
    flag & Reference \\
    \hline
HD 377 & 3000 & OVRO & 0.79 & 0.61 &  & {\citet{2005AJ....129.1049C}} \\
HD 377 & 1200 & IRAM & 4 & 1 &  & {\citet{2009A&A...497..409R}} \\
HD 377 & 2700 & OVRO & 0.32 & 0.8 &  & {\citet{2005AJ....129.1049C}} \\
HD 9672 & 1300 & IRAM & 13.9 & 2.48 &  & {\citet{1995Ap&SS.224..389W}} \\
HD 14055 & 850 & SCUBA & 5.5 & 1.8 &  & {\citet{2006ApJ...653.1480W}} \\
HD 15115 & 850 & SCUBA & 4.9 & 1.6 &  & {\citet{2006ApJ...653.1480W}} \\
HD 15115 & 870 & LABOCA & 15.3 &  & 1 & {\citet{2009A&A...508.1057N}} \\
HD 22049 & 1300 & IRAM & 12.7 & 3.9 &  & {\citet{1995Ap&SS.224..389W}} \\
HD 22049 & 1300 & MPIfR & 24.2 & 3.4 &  & {\citet{1991A&A...252..220C}} \\
HD 22049 & 450 & SCUBA & 225 & 10 &  & {\citet{2004MNRAS.348.1282S}} \\
HD 22049 & 850 & SCUBA & 37 & 3 &  & {\citet{2005ApJ...619L.187G}} \\
HD 22049 & 450 & SCUBA & 250 & 20 &  & {\citet{2005ApJ...619L.187G}} \\
HD 22049 & 350 & CSO & 366 & 50 &  & {\citet{2009ApJ...690.1522B}} \\
HD 22049 & 850 & SCUBA & 40 & 1.5 &  & {\citet{2004MNRAS.348.1282S}} \\
HD 25457 & 1200 & SEST & -8 & 14 &  & {\citet{2005AJ....129.1049C}} \\
HD 25457 & 870 & LABOCA & 9.9 &  & 1 & {\citet{2010A&A...518A..40N}} \\
HD 25457 & 2700 & OVRO & -1.51 & 1.33 &  & {\citet{2005AJ....129.1049C}} \\
HD 25457 & 3000 & OVRO & 0.37 & 0.62 &  & {\citet{2005AJ....129.1049C}} \\
HD 25457 & 1200 & IRAM & 2.2 &  & 1 & {\citet{2009A&A...497..409R}} \\
HD 30447 & 870 & LABOCA & 6.9 & 5 &  & {\citet{2010A&A...518A..40N}} \\
HD 32297 & 870 & LABOCA & 19.5 &  & 1 & {\citet{2010A&A...518A..40N}} \\
HD 32297 & 1300 & CARMA & 5.1 & 1.1 &  & {\citet{2008ApJ...686L..25M}} \\
HD 38207 & 1200 & SEST & -3 & 12 &  & {\citet{2005AJ....129.1049C}} \\
HD 38207 & 1200 & IRAM & 0.33 &  & 1 & {\citet{2009A&A...497..409R}} \\
HD 39060 & 1200 & SIMBA & 24.3 & 3 &  & {\citet{2003A&A...402..183L}} \\
HD 39060 & 870 & LABOCA & 63.6 & 6.7 &  & {\citet{2009A&A...508.1057N}} \\
HD 39060 & 850 & SCUBA & 58.3 & 6.5 &  & {\citet{1998Natur.392..788H}} \\
HD 39060 & 1300 & MPIfR & 24.9 & 2.6 &  & {\citet{1991A&A...252..220C}} \\
HD 61005 & 870 & LABOCA & 18 &  & 1 & {\citet{2010A&A...518A..40N}} \\
HD 61005 & 1200 & SEST & 31 & 34 &  & {\citet{2005AJ....129.1049C}} \\
HD 61005 & 350 & CSO & 95 & 12 &  & {\citet{2009A&A...497..409R}} \\
HD 107146 & 350 & CSO & 319 & 6 &  & {\citet{2009A&A...497..409R}} \\
HD 107146 & 3000 & OVRO & 1.42 & 0.23 &  & {\citet{2005AJ....129.1049C}} \\
HD 107146 & 850 & SCUBA & 20 & 3.2 &  & {\citet{2005ApJ...635..625N}} \\
HD 107146 & 880 & SMA & 36 & 1 &  & {\citet{2011ApJ...740...38H}} \\
HD 107146 & 450 & SCUBA & 130 & 40 &  & {\citet{2004ApJ...604..414W}} \\
HD 107146 & 450 & SCUBA & 130 & 12 &  & {\citet{2005ApJ...635..625N}} \\
HD 107146 & 850 & SCUBA & 20 & 4 &  & {\citet{2004ApJ...604..414W}} \\
HD 109085 & 450 & SCUBA & 58.2 & 9.8 &  & {\citet{2005ApJ...620..492W}} \\
HD 109085 & 850 & SCUBA2 & 15.5 & 1.4 &  & {\citet{2014ApJ...784..148D}} \\
HD 109085 & 850 & SCUBA & 14.3 & 1.8 &  & {\citet{2005ApJ...620..492W}} \\
HD 161868 & 870 & LABOCA & 12.8 & 5.2 &  & {\citet{2010A&A...518A..40N}} \\
HD 172167 & 850 & SCUBA & 45.7 & 5.4 &  & {\citet{1998Natur.392..788H}} \\
HD 172167 & 1300 & IRAM & 11.4 & 1.7 &  & {\citet{2002ApJ...569L.115W}} \\
HD 172167 & 3300 & IRAM & 0.39 &  & 1 & {\citet{2002ApJ...569L.115W}} \\
HD 181296 & 870 & LABOCA & 14.4 &  & 1 & {\citet{2009A&A...508.1057N}} \\
HD 181327 & 870 & LABOCA & 51.7 & 6.2 &  & {\citet{2009A&A...508.1057N}} \\
HD 181327 & 3190 & ATCA & 0.72 & 0.25 &  & {\citet{2012A&A...539A..17L}} \\
HD 192425 & 1300 & SMTO & 7.95 &  & 1 & {\citet{2003AJ....125.3334H}} \\
HD 192425 & 870 & SMTO & 33.3 &  & 1 & {\citet{2003AJ....125.3334H}} \\
HD 216956 & 450 & SCUBA & 595 & 35 &  & {\citet{2003ApJ...582.1141H}} \\
HD 216956 & 1300 & MPIfR & 21 & 2.5 &  & {\citet{1991A&A...252..220C}} \\
HD 216956 & 850 & SCUBA & 81 & 7.2 &  & {\citet{1998Natur.392..788H}} \\
HD 216956 & 850 & SCUBA & 97 & 5 &  & {\citet{2003ApJ...582.1141H}} \\
HD 218396 & 850 & SCUBA & 10.3 & 1.8 &  & {\citet{2006ApJ...653.1480W}} \\
HD 218396 & 1100 & UKT14 & 33 &  & 1 & {\citet{1996MNRAS.279..915S}} \\
\hline
\end{tabular}
\end{table*}




\onecolumn

The following pages contain flux density distributions (SEDs) for the 48-star
sample. Stars are ordered as in Table A1 left to right and down. Each panel is labelled
by the stars' HD identifier. Symbols are as in Fig. 1 in the main article.

\vspace{-1cm}

\begin{center}
\begin{tabular}{cc}
  HD 377 & HD 6798 \\
  \includegraphics[width=0.5\textwidth]{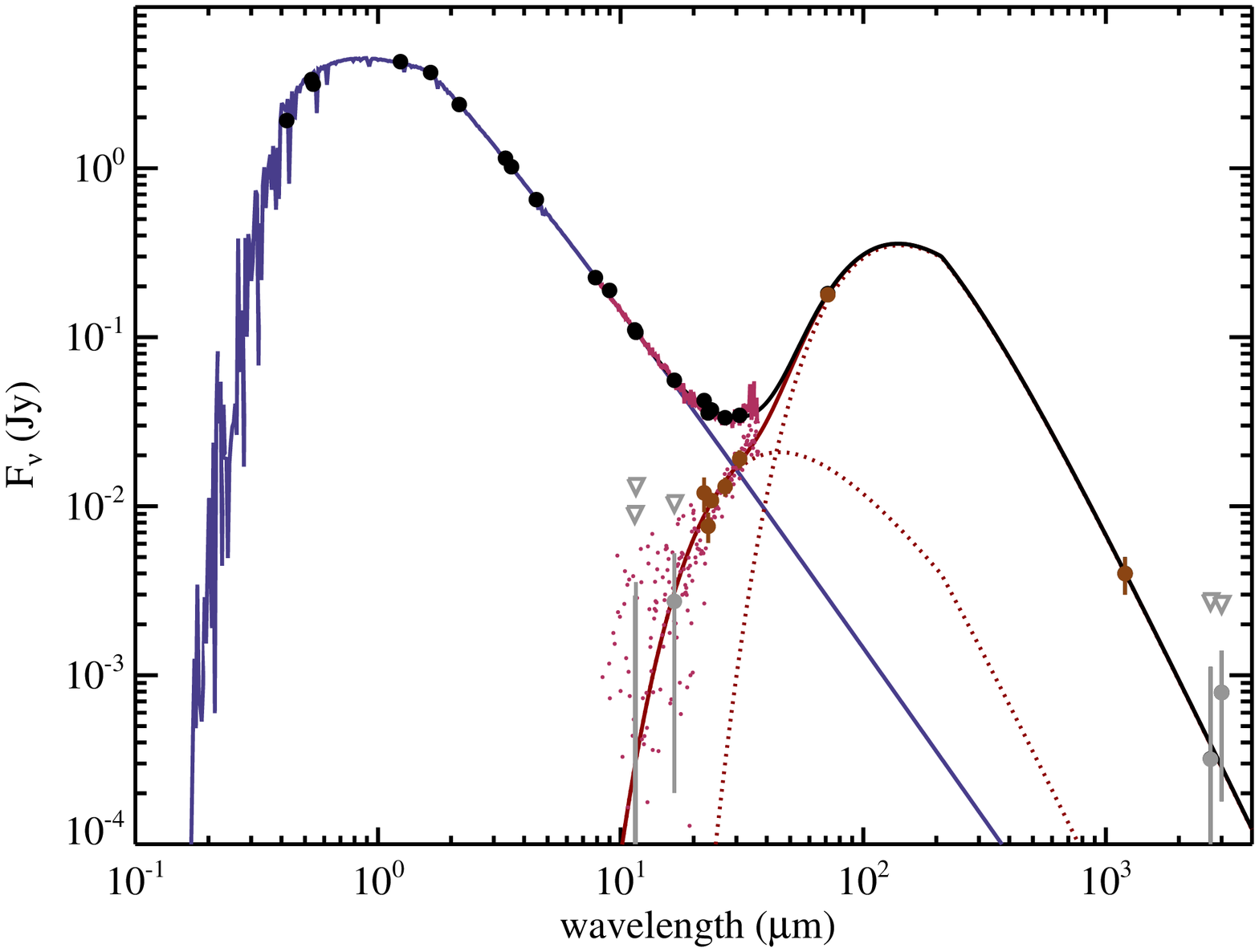}&
  \includegraphics[width=0.5\textwidth]{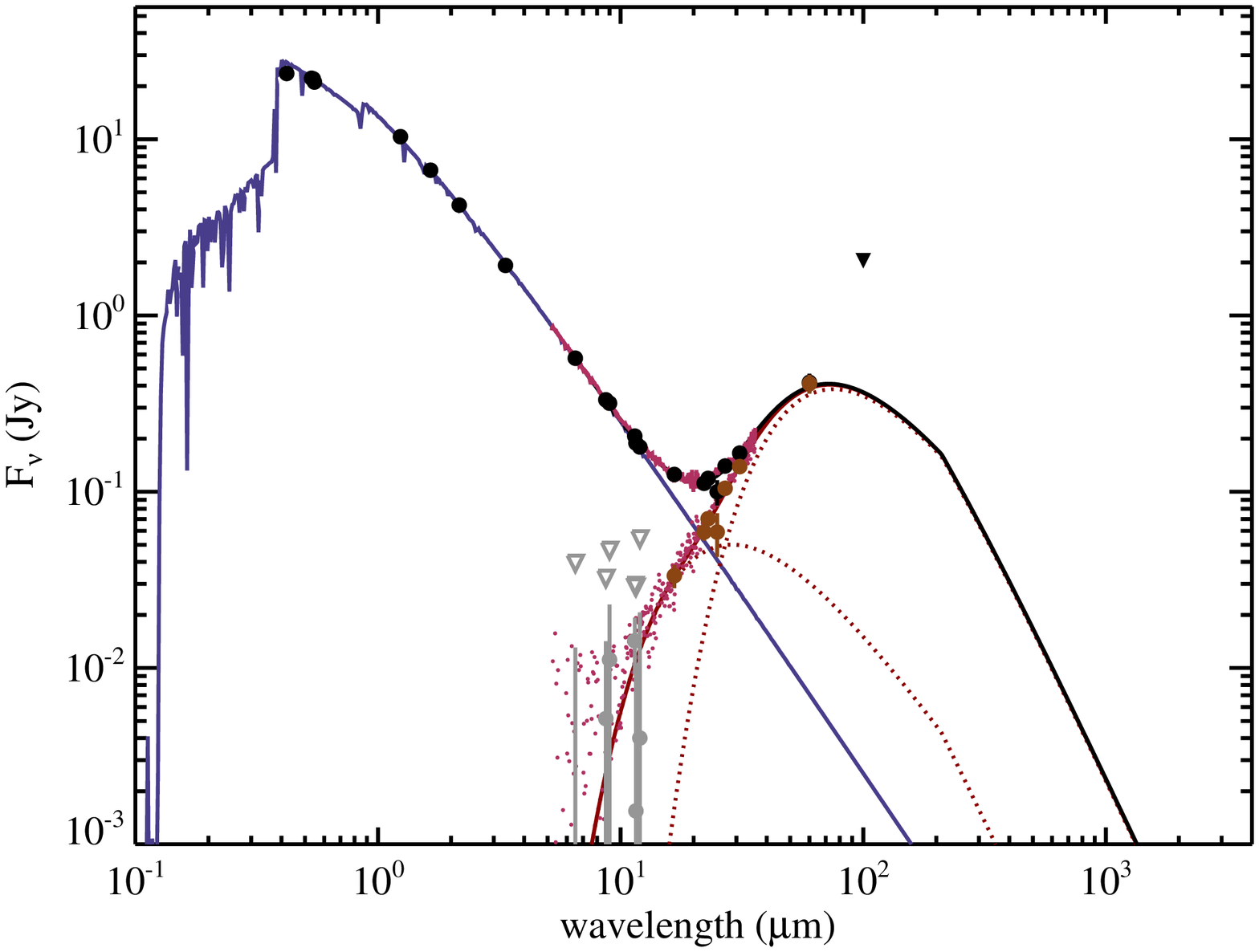}\\
  HD 9672 & HD 10647 \\
  \includegraphics[width=0.5\textwidth]{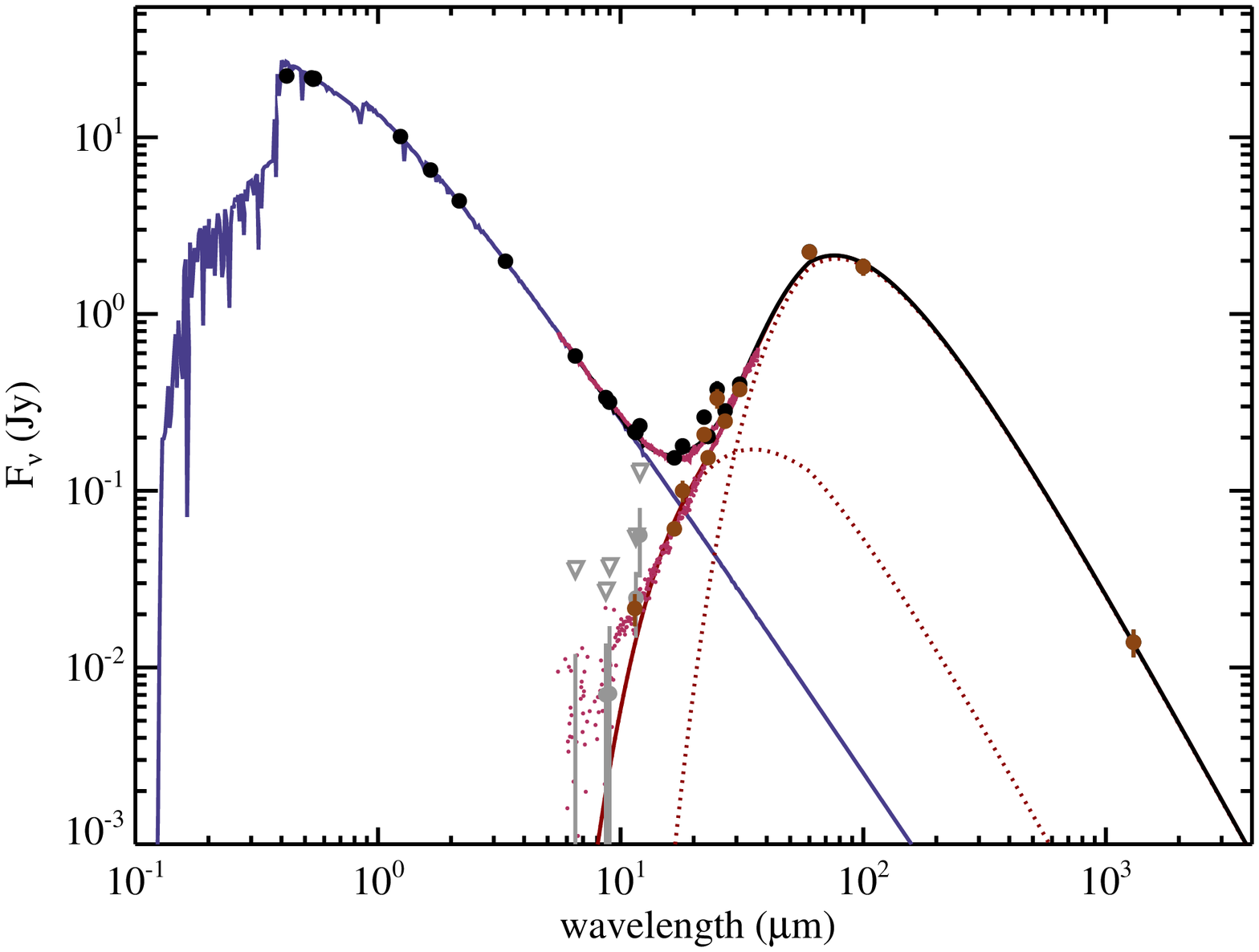}&
  \includegraphics[width=0.5\textwidth]{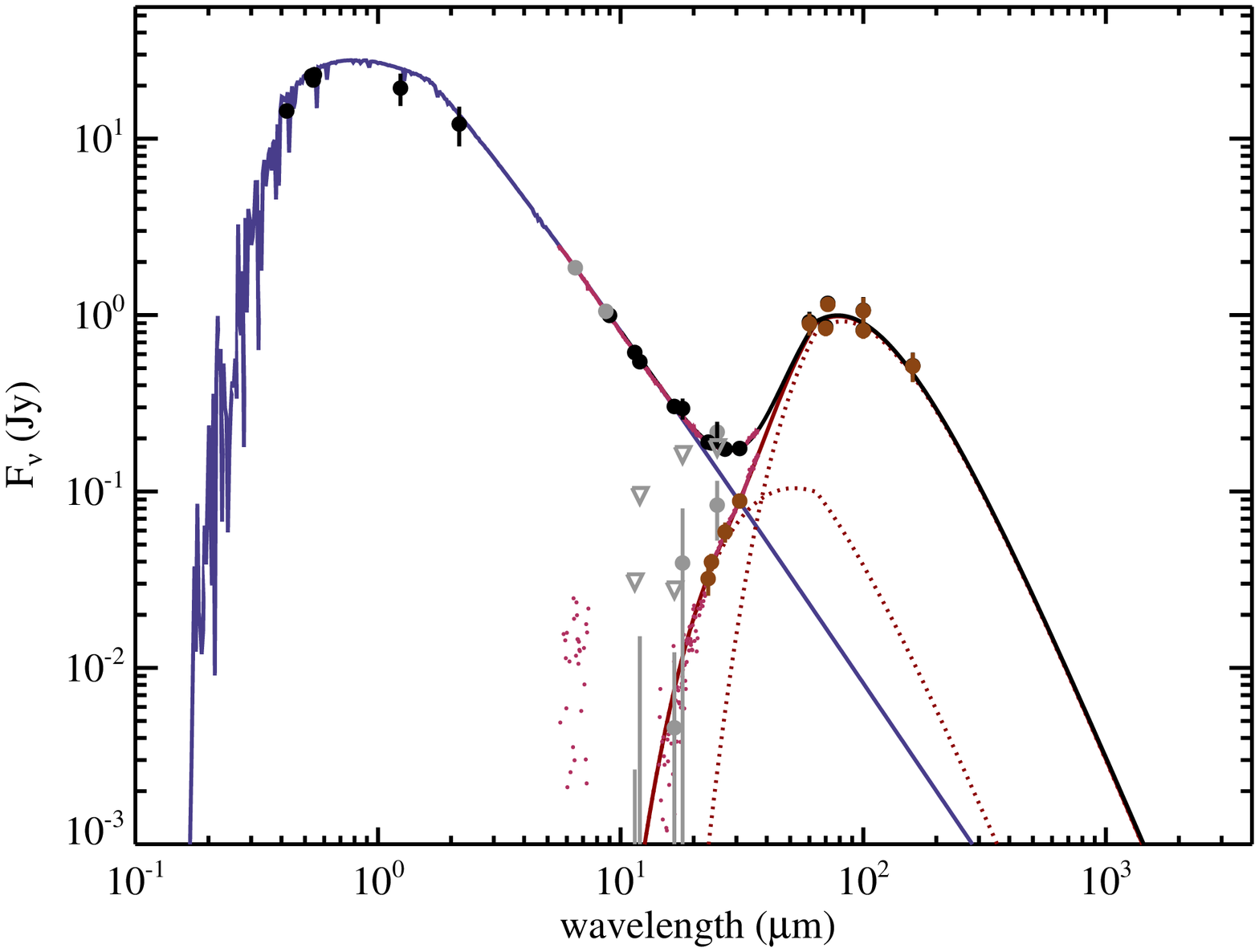}\\
  HD 10939 & HD 13246 \\
  \includegraphics[width=0.5\textwidth]{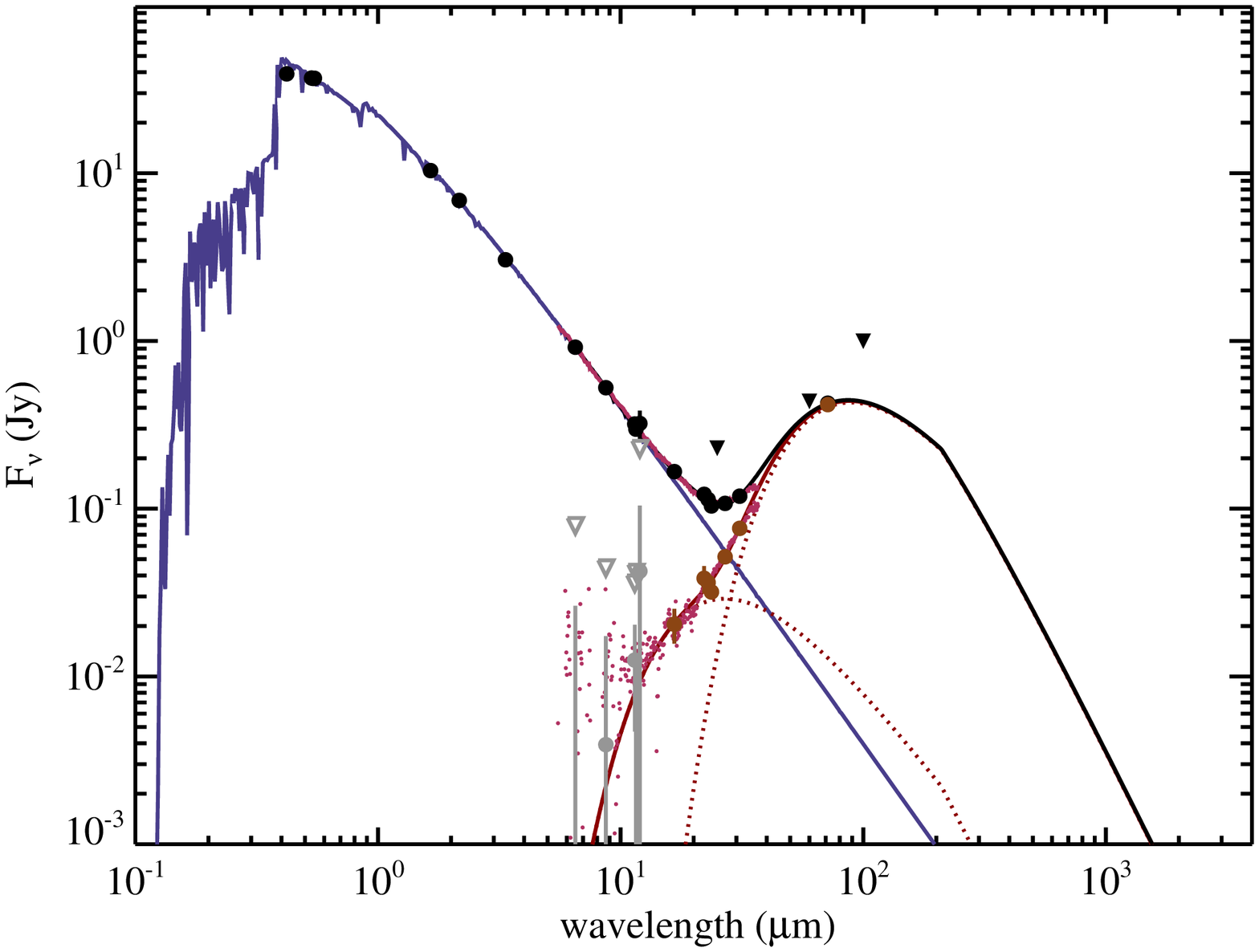}&
  \includegraphics[width=0.5\textwidth]{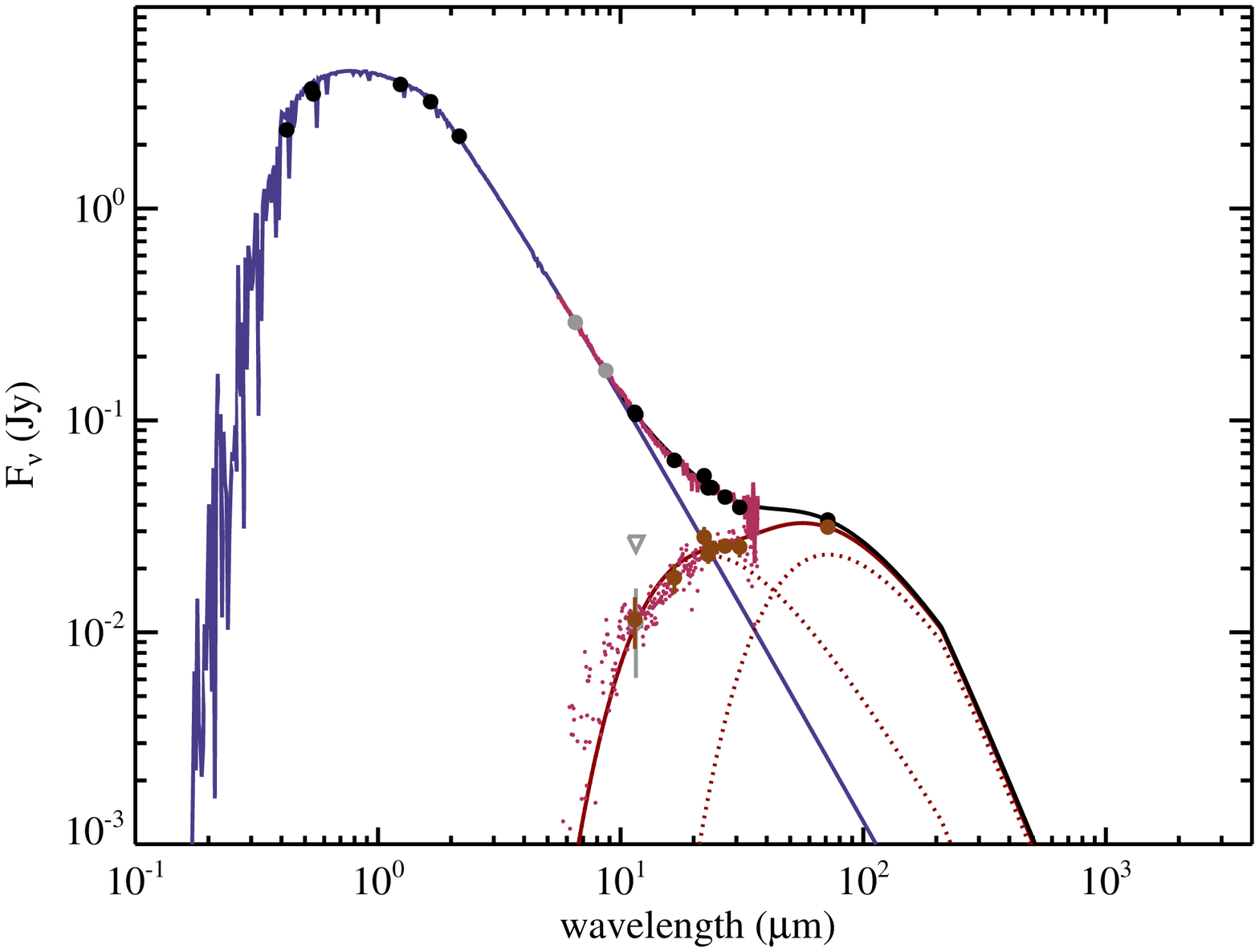}\\
\end{tabular}
\end{center}
\begin{center}
\begin{tabular}{cc}
  HD 14055 & HD 15115 \\
  \includegraphics[width=0.5\textwidth]{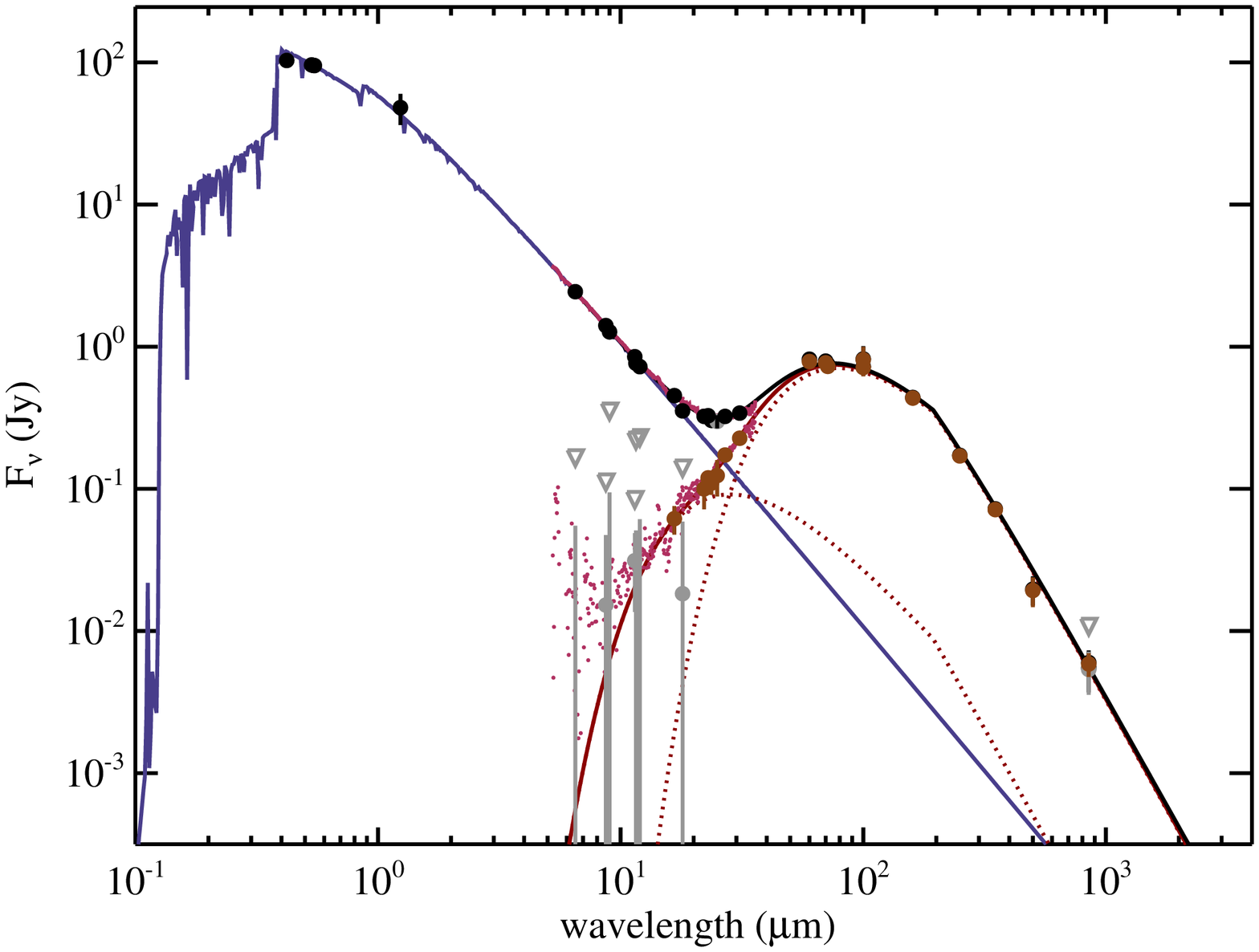}&
  \includegraphics[width=0.5\textwidth]{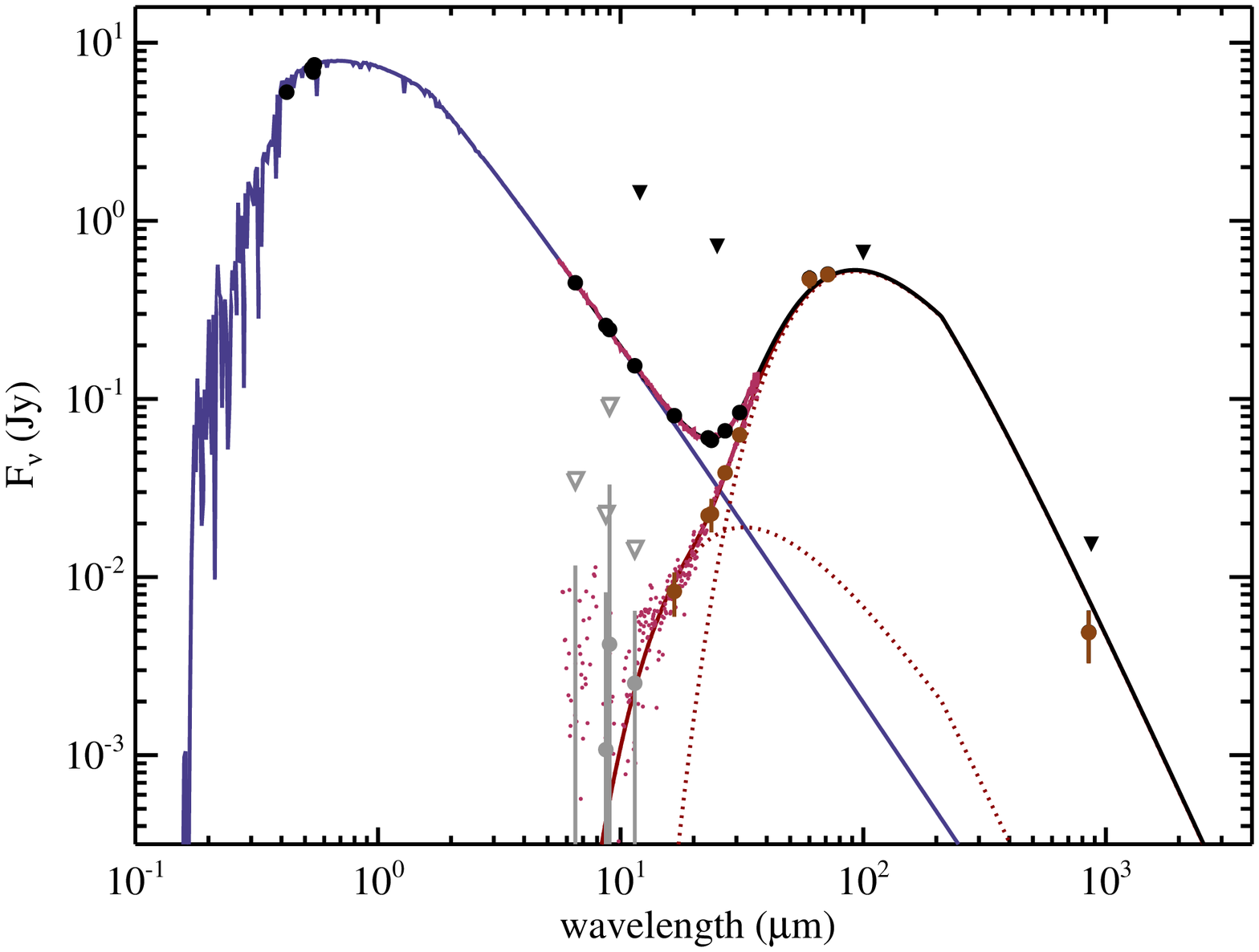}\\
  HD 15745 & HD 16743 \\
  \includegraphics[width=0.5\textwidth]{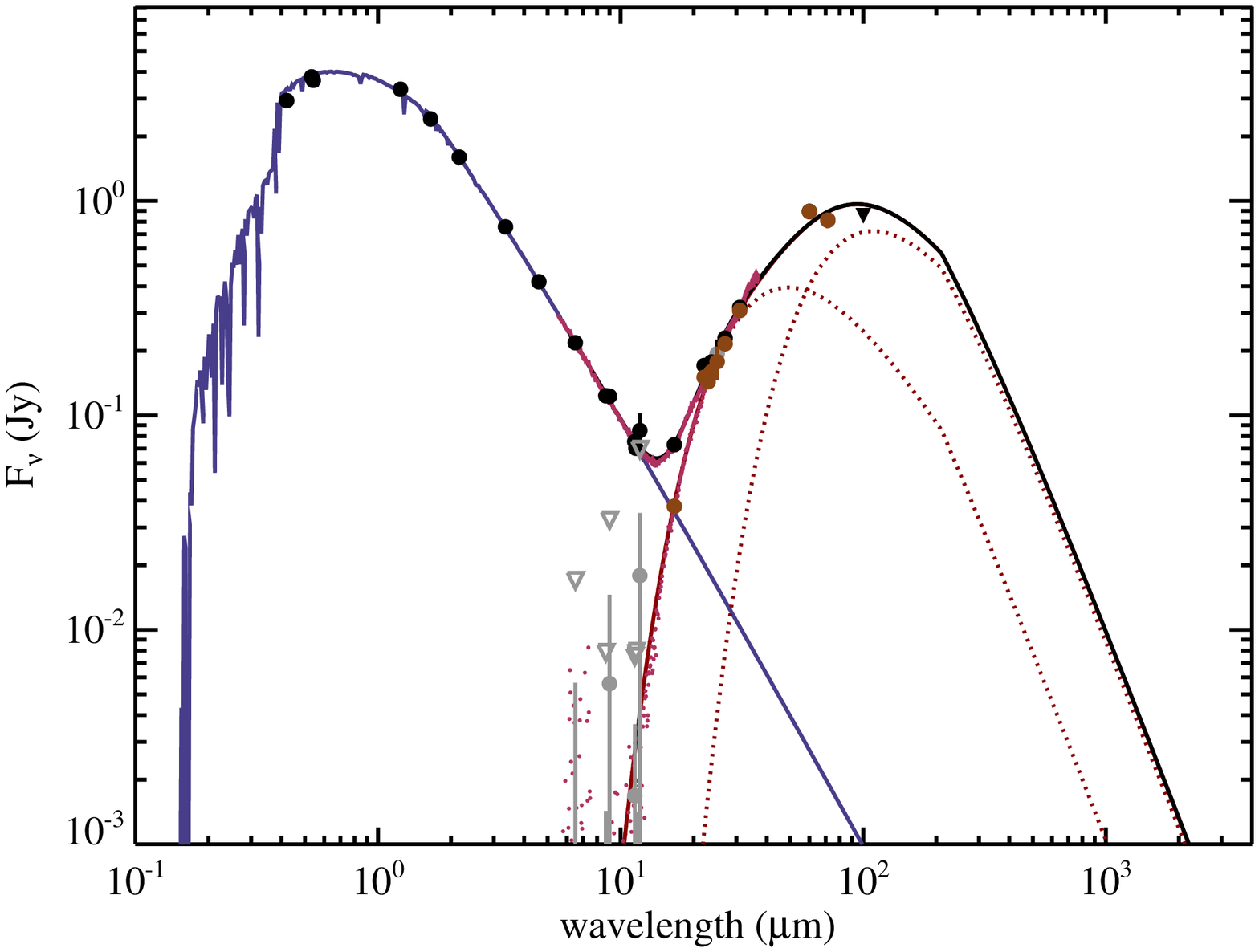}&
  \includegraphics[width=0.5\textwidth]{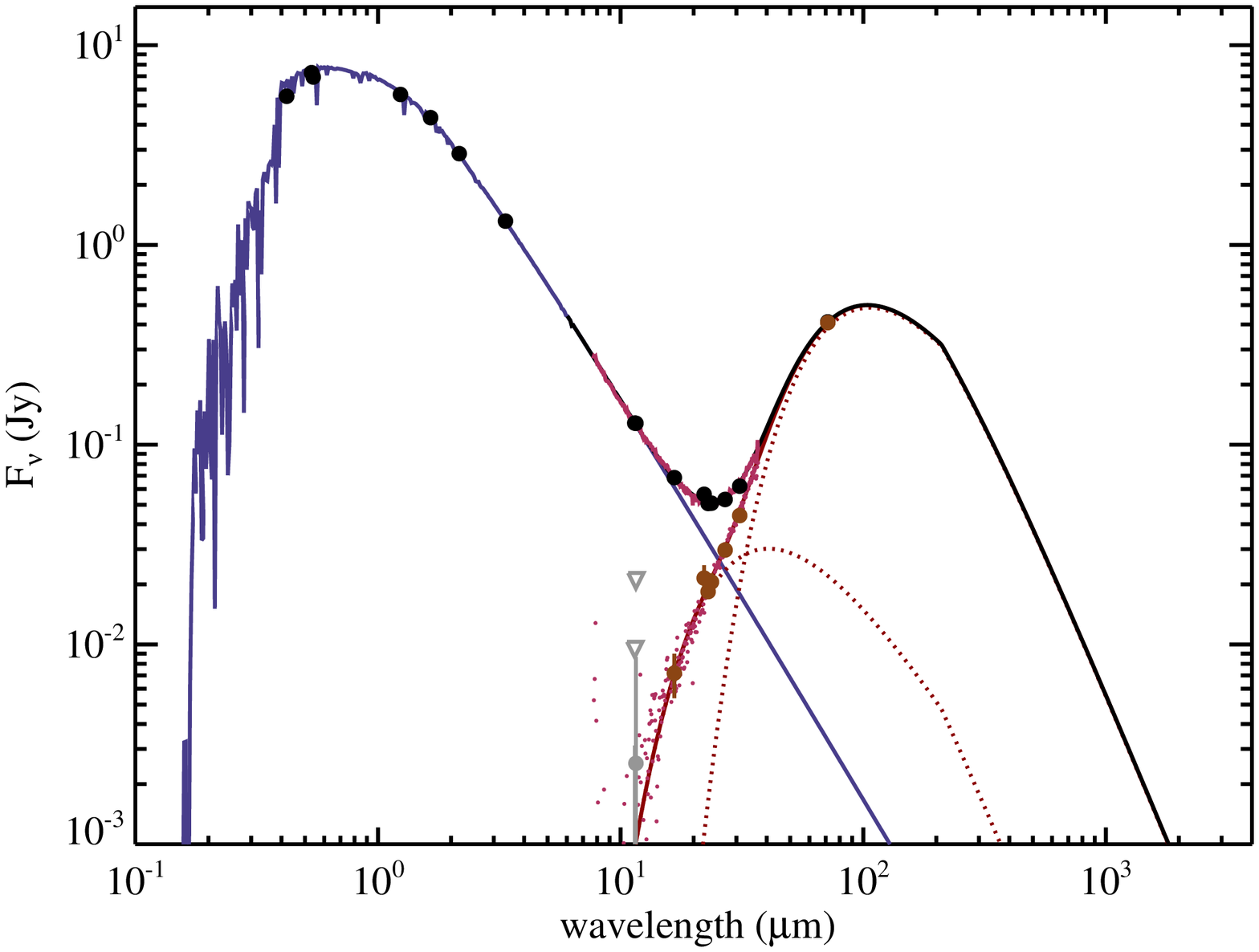}\\
  HD 22049 & HD 23267 \\
  \includegraphics[width=0.5\textwidth]{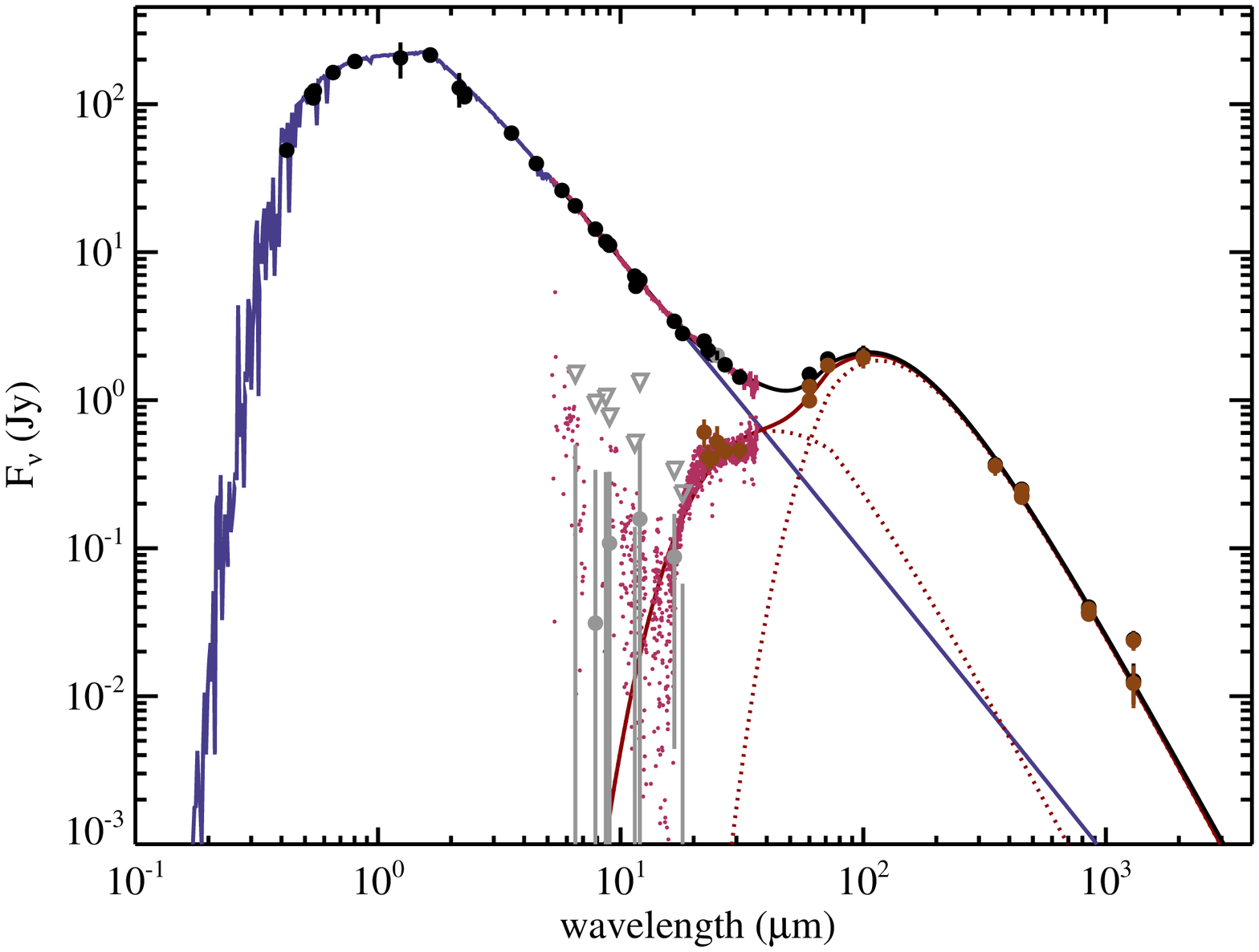}&
  \includegraphics[width=0.5\textwidth]{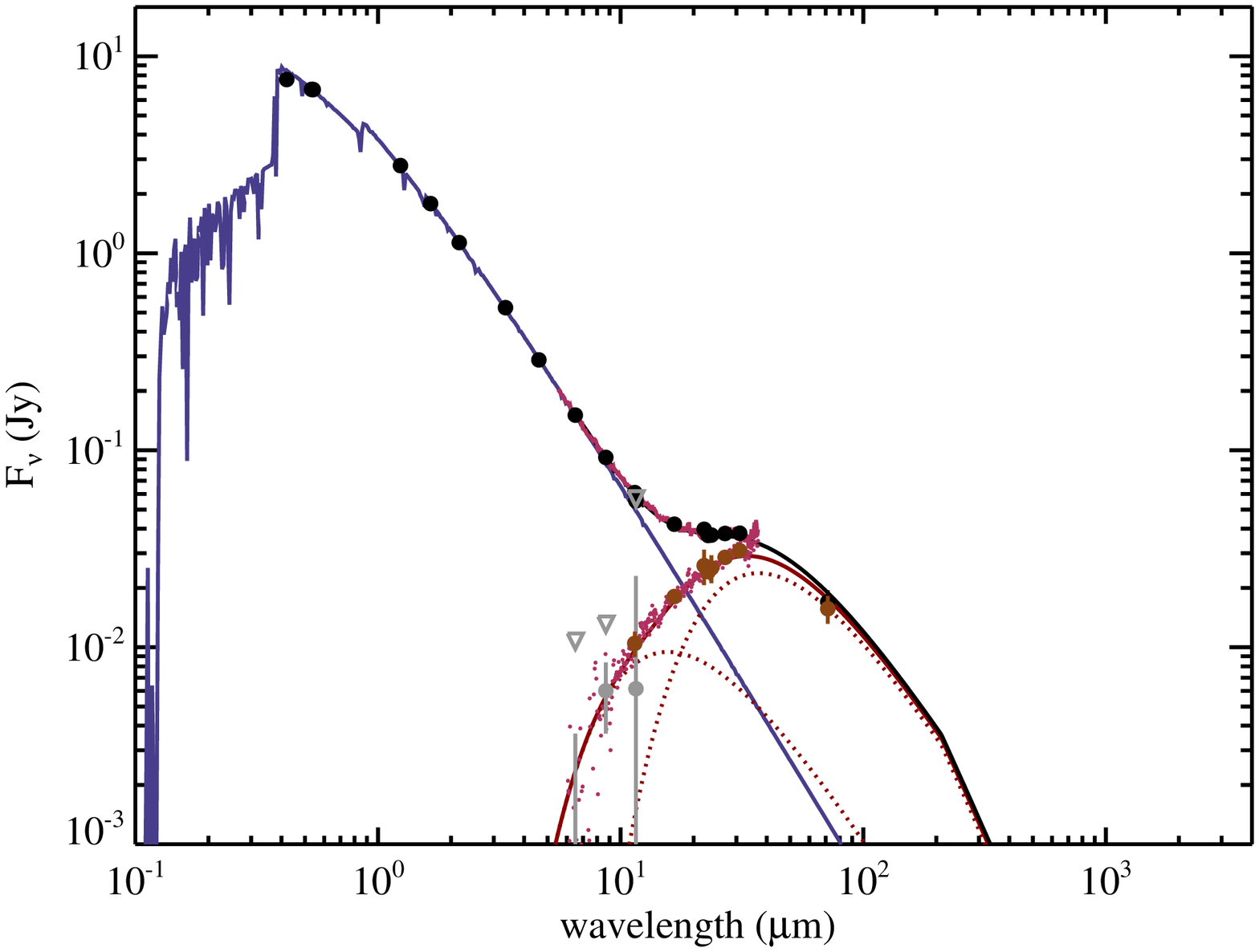}\\
\end{tabular}
\end{center}
\begin{center}
\begin{tabular}{cc}
  HD 25457 & HD 30447 \\
  \includegraphics[width=0.5\textwidth]{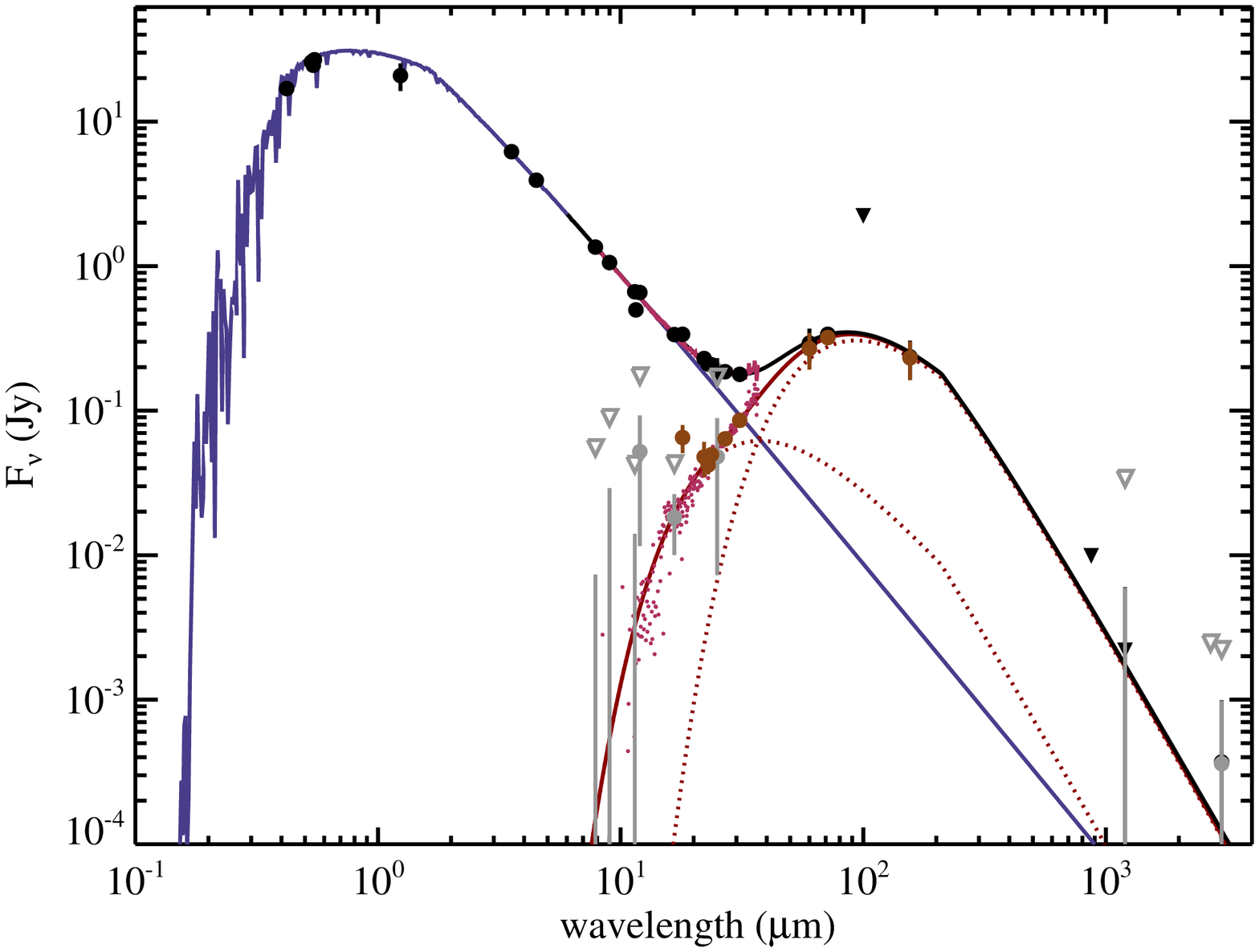}&
  \includegraphics[width=0.5\textwidth]{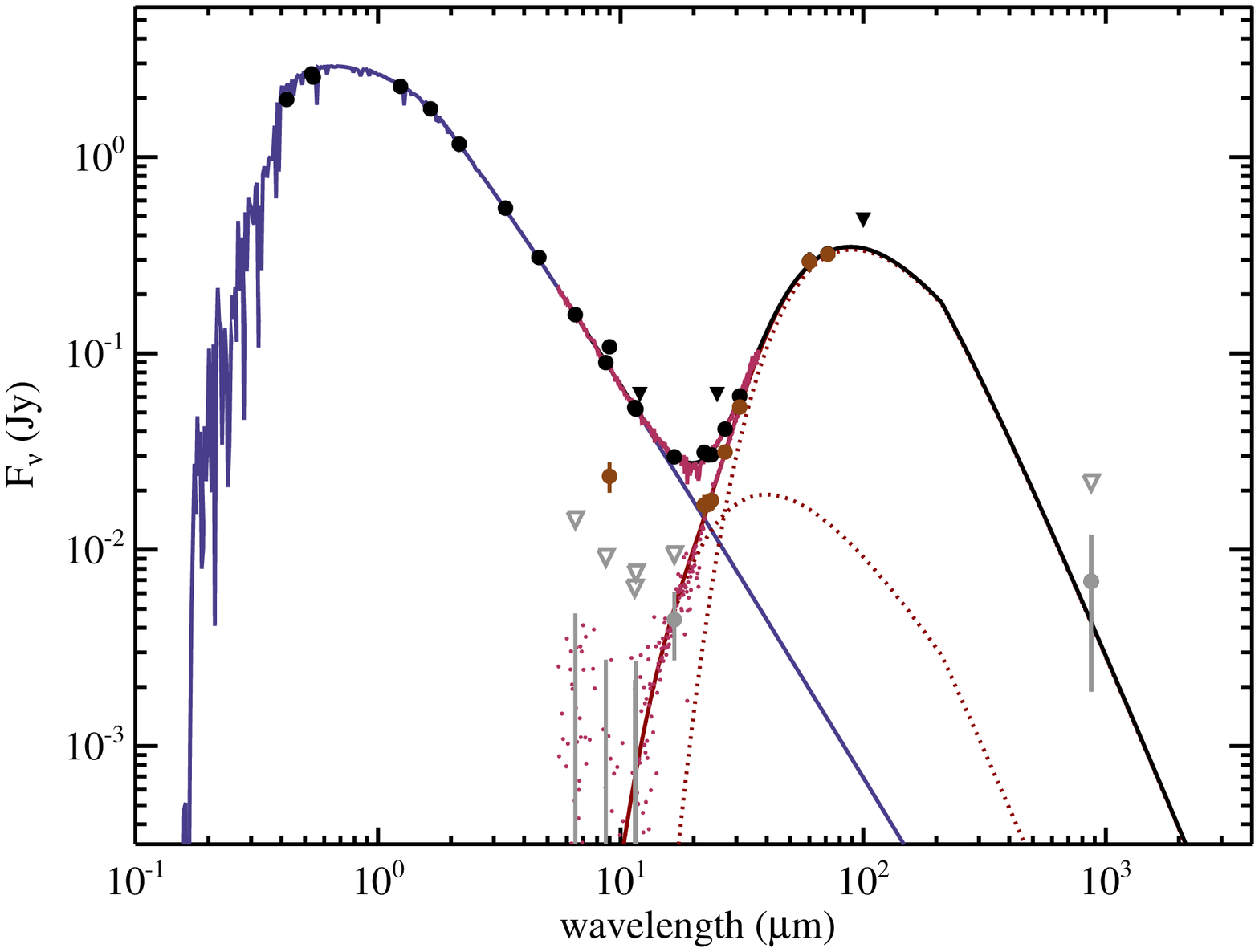}\\
  HD 31295 & HD 32297 \\
  \includegraphics[width=0.5\textwidth]{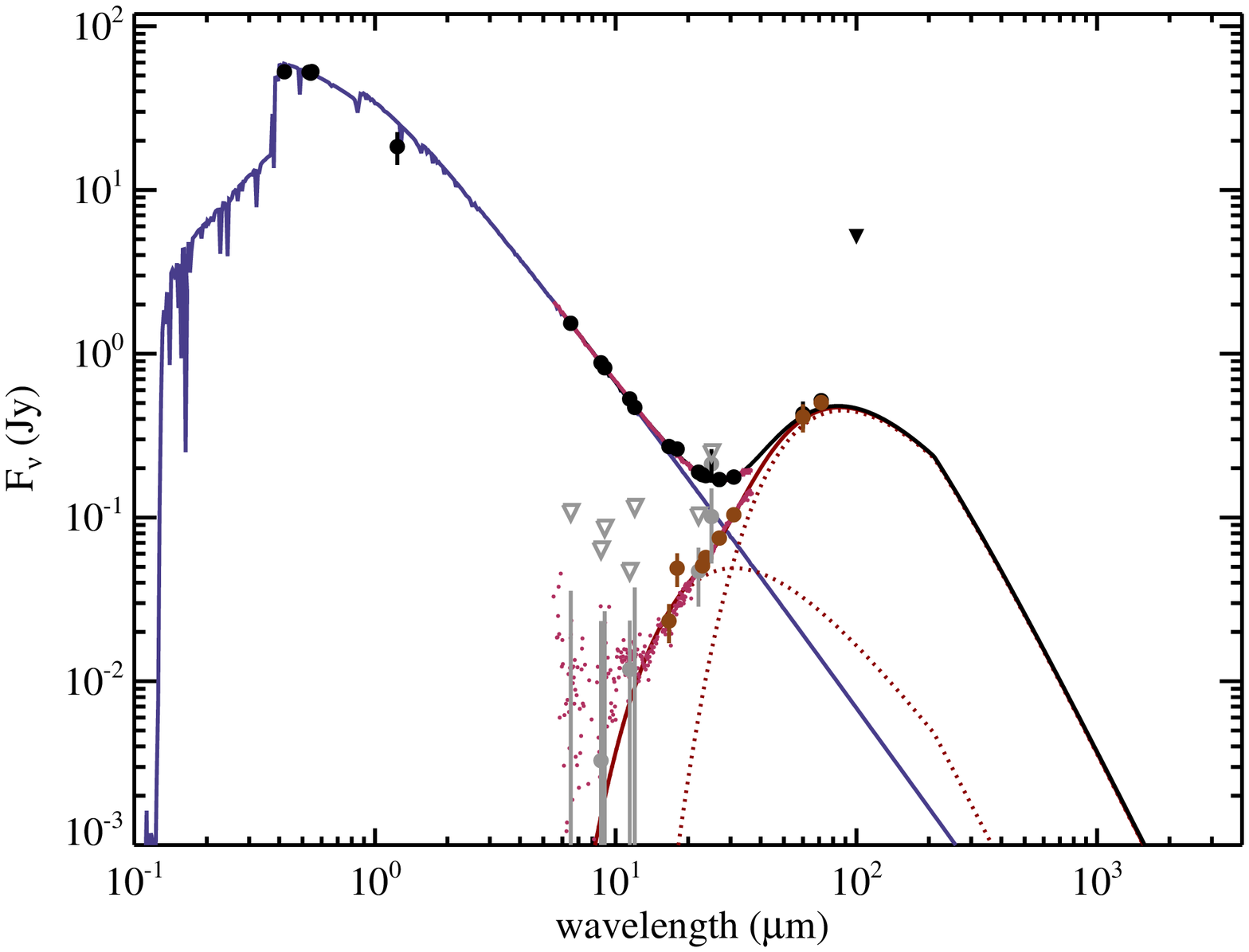}&
  \includegraphics[width=0.5\textwidth]{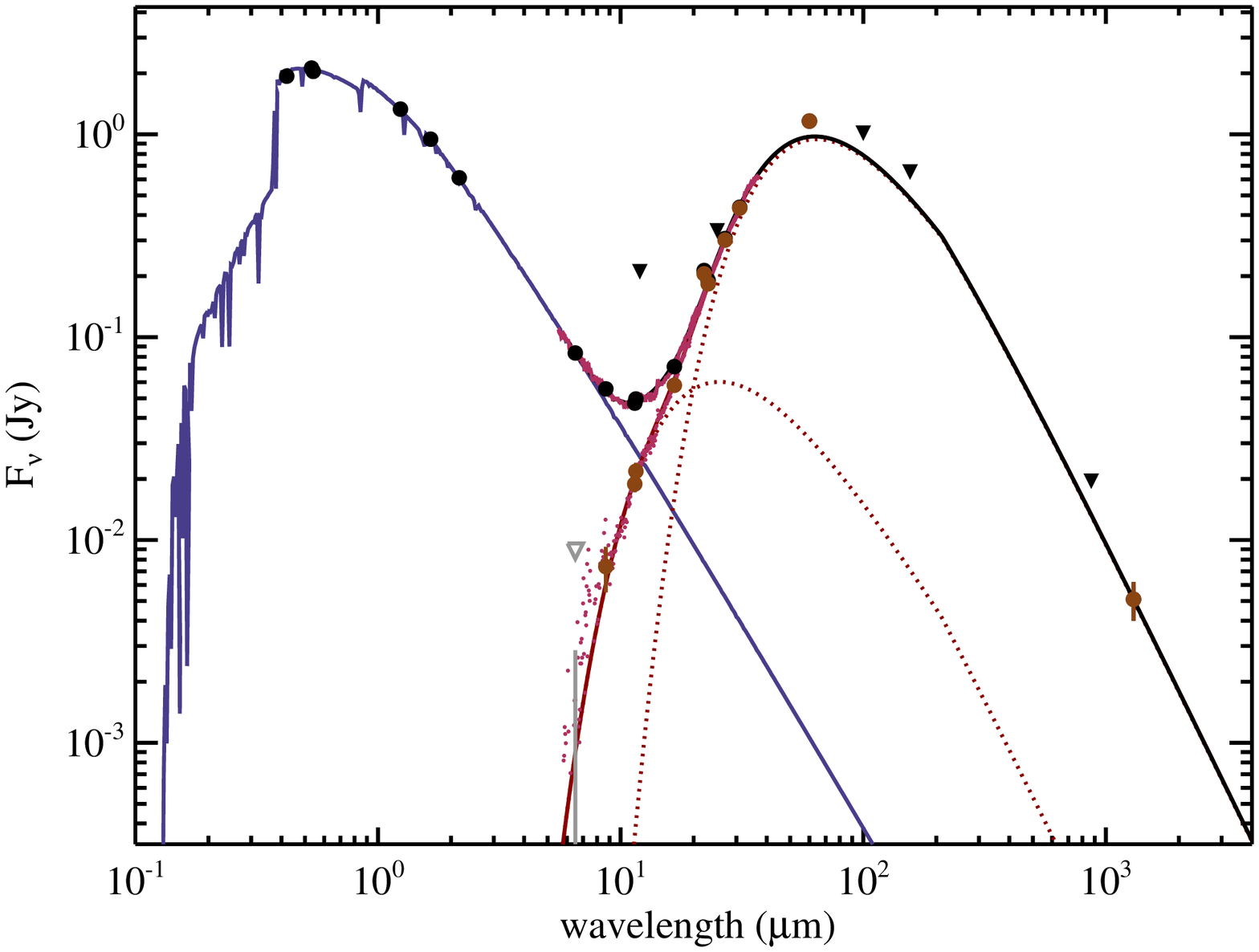}\\
  HD 38056 & HD 38206 \\
  \includegraphics[width=0.5\textwidth]{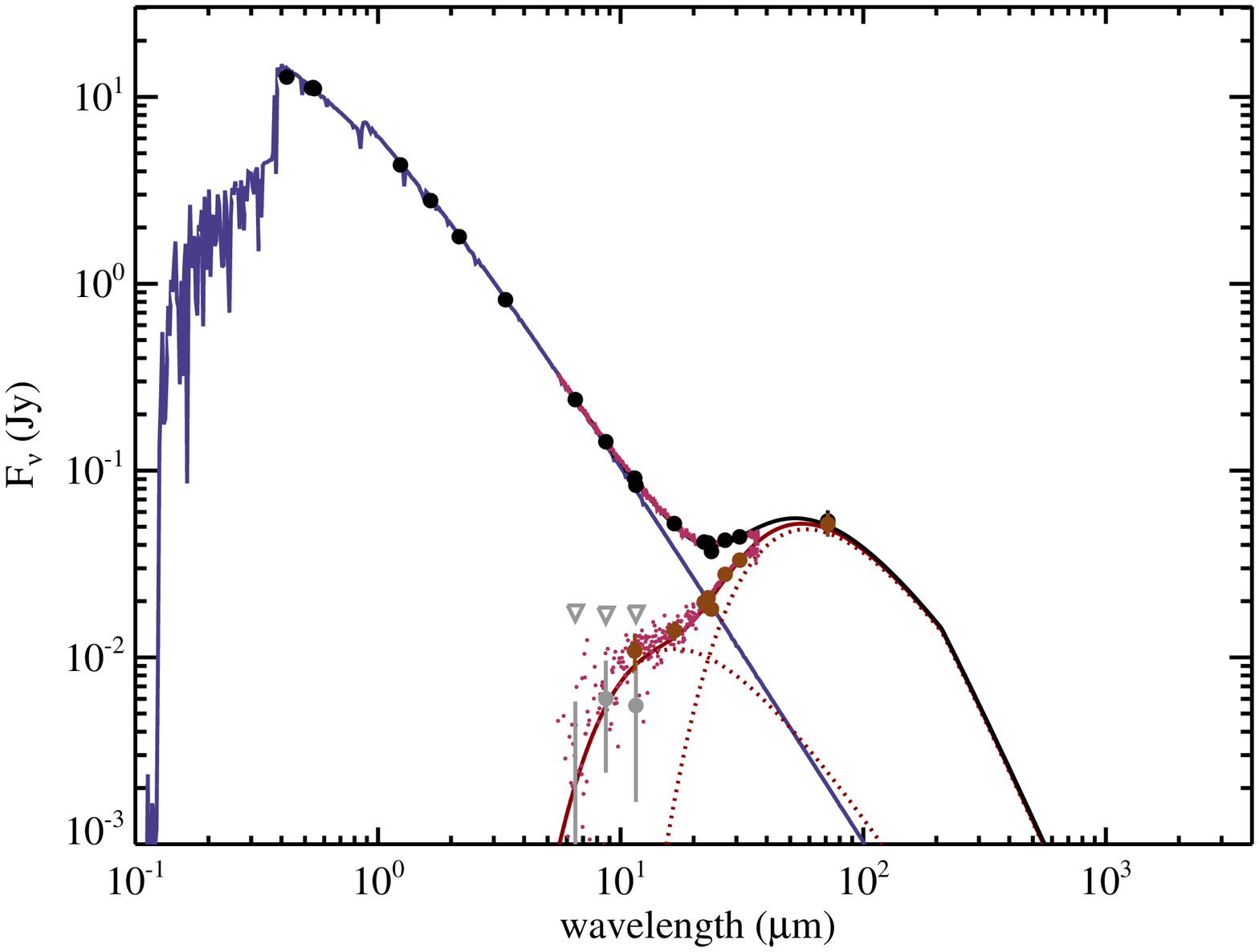}&
  \includegraphics[width=0.5\textwidth]{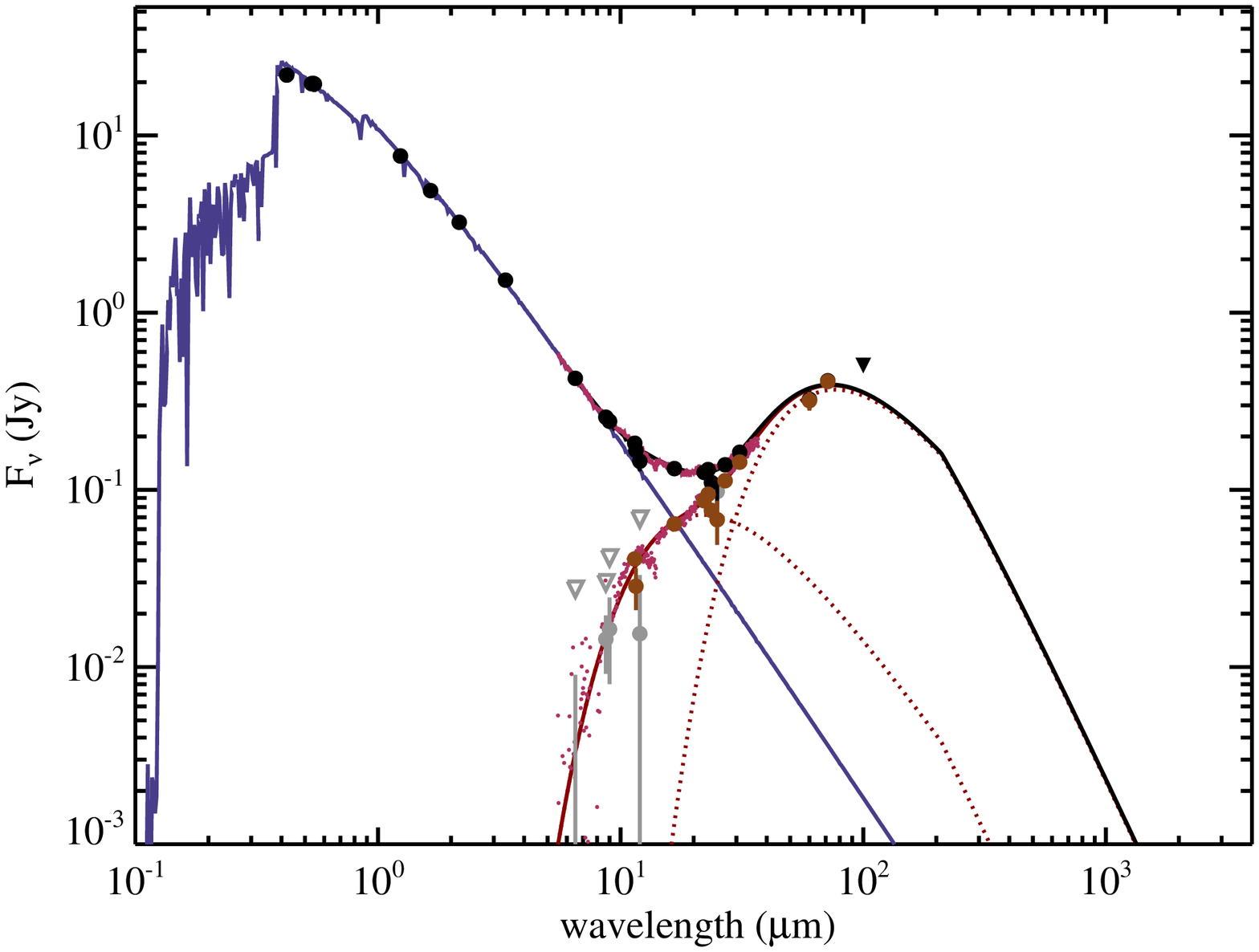}\\
\end{tabular}
\end{center}
\begin{center}
\begin{tabular}{cc}
  HD 38207 & HD 39060 \\
  \includegraphics[width=0.5\textwidth]{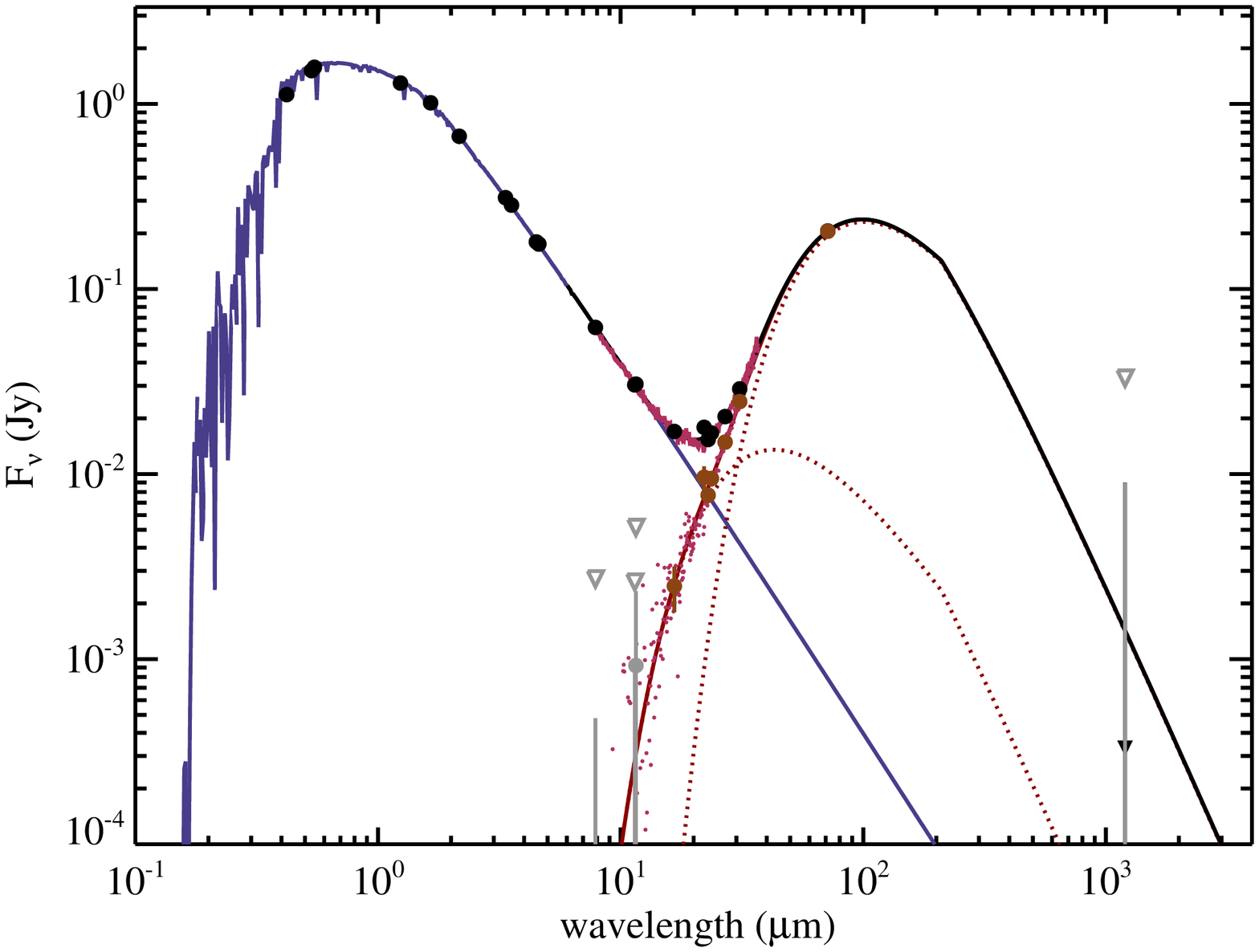}&
  \includegraphics[width=0.5\textwidth]{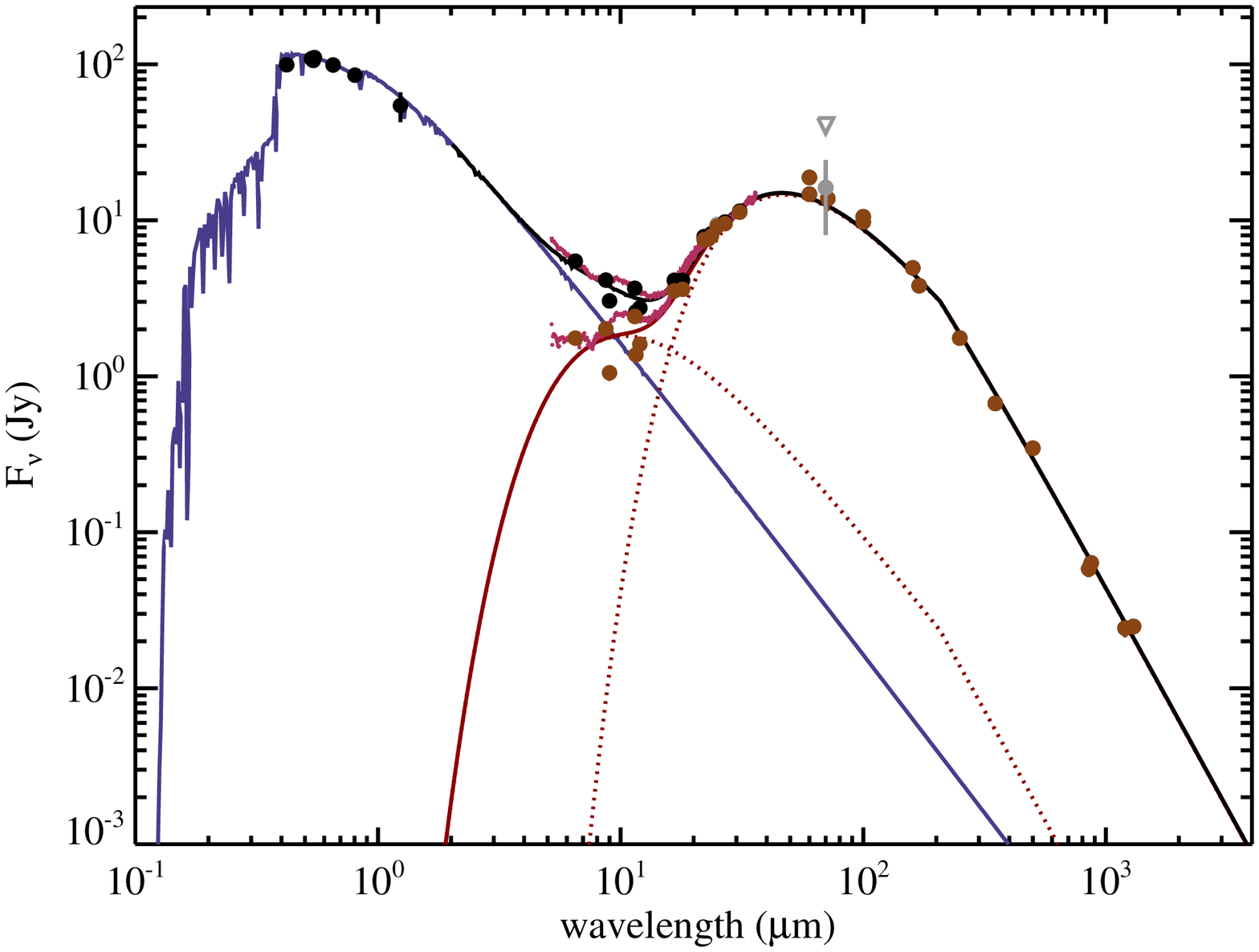}\\
  HD 61005 & HD 70313 \\
  \includegraphics[width=0.5\textwidth]{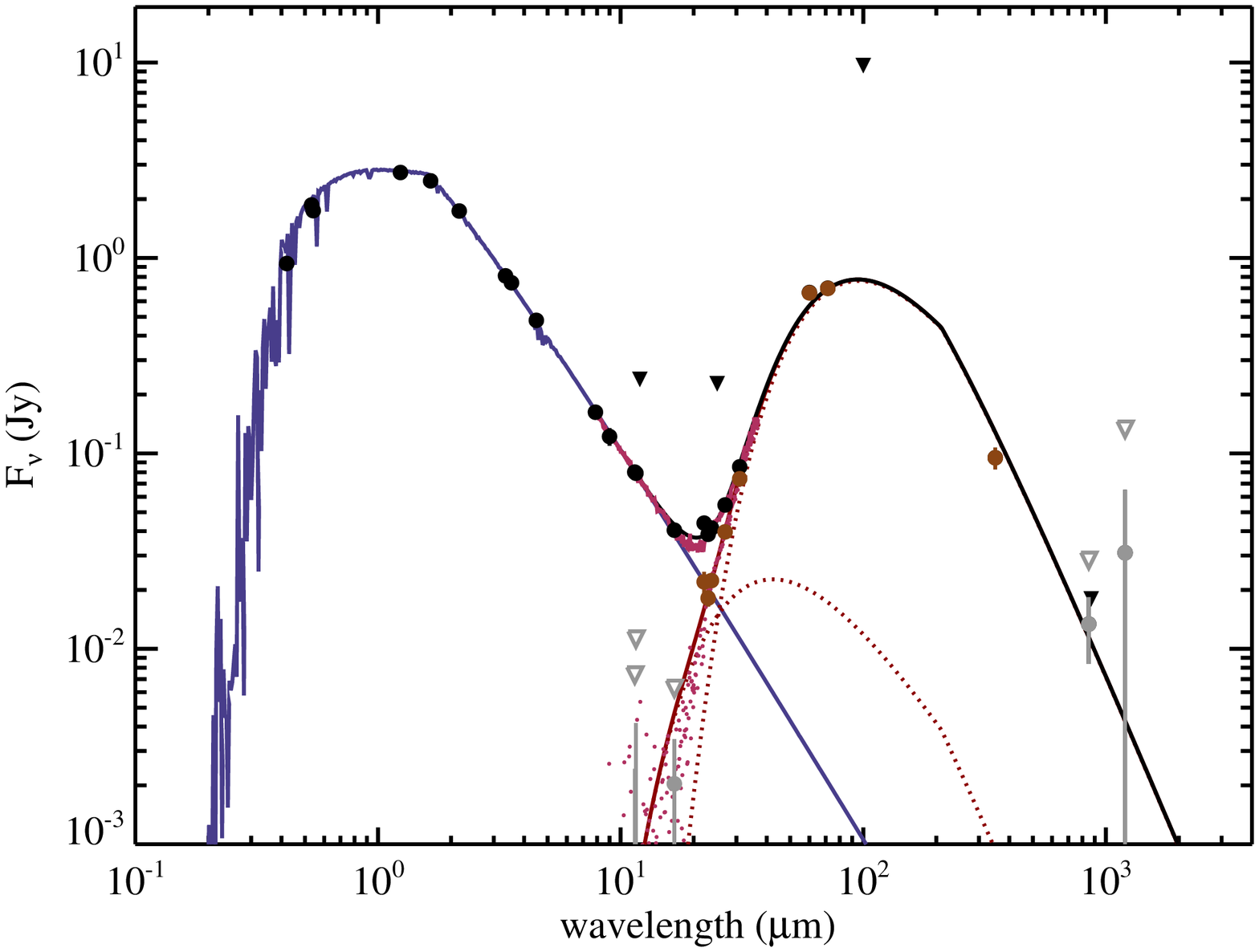}&
  \includegraphics[width=0.5\textwidth]{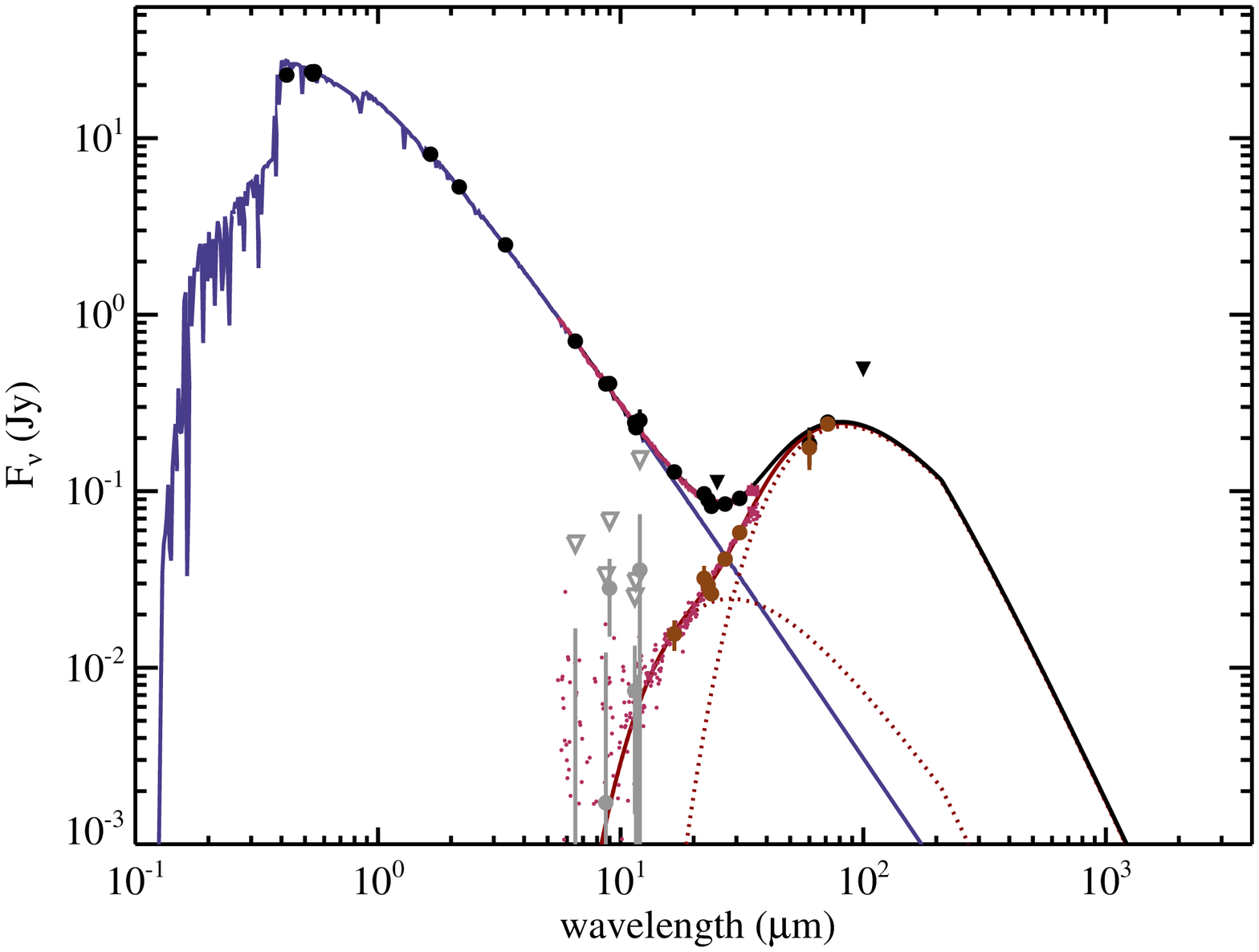}\\
  HD 71722 & HD 79108 \\
  \includegraphics[width=0.5\textwidth]{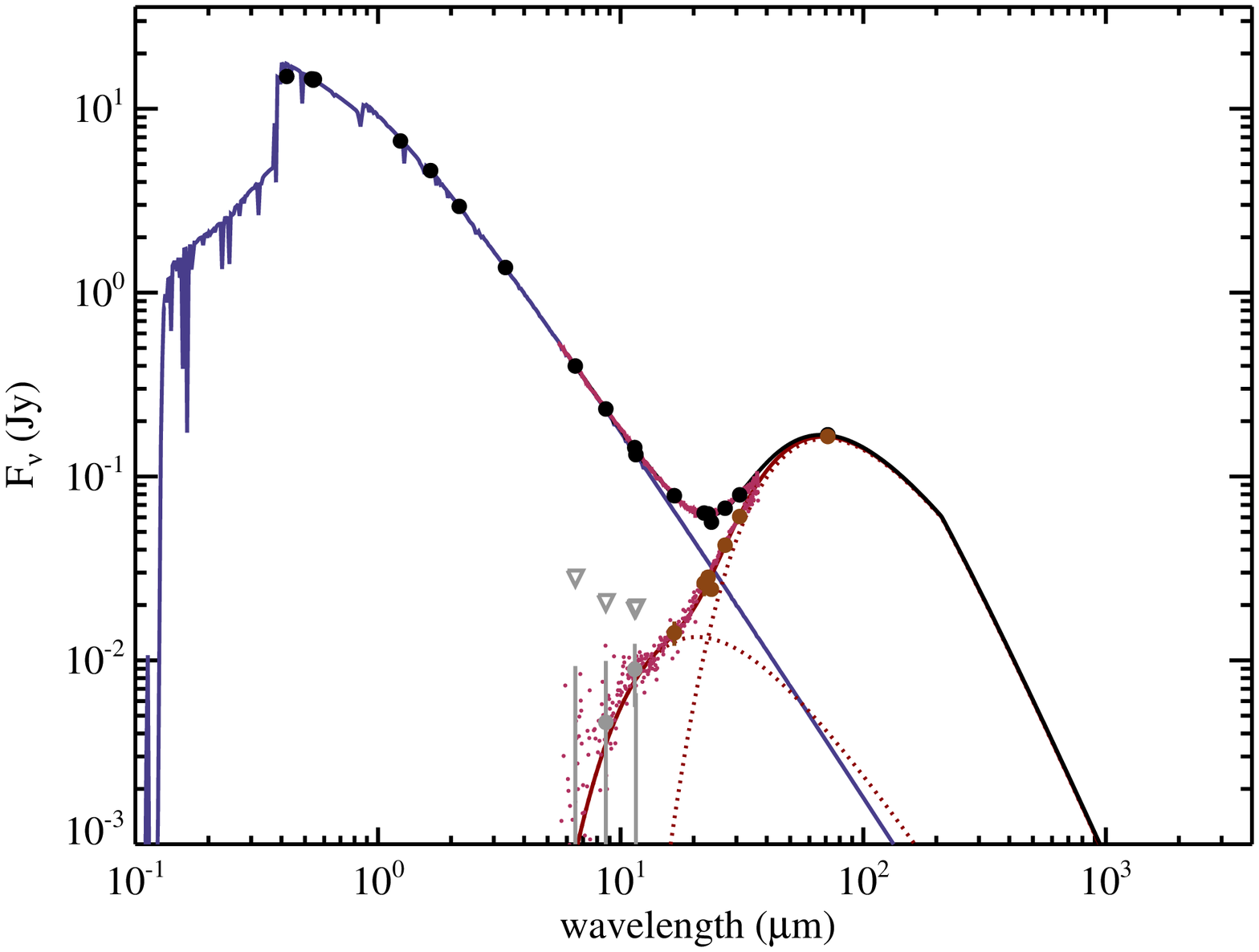}&
  \includegraphics[width=0.5\textwidth]{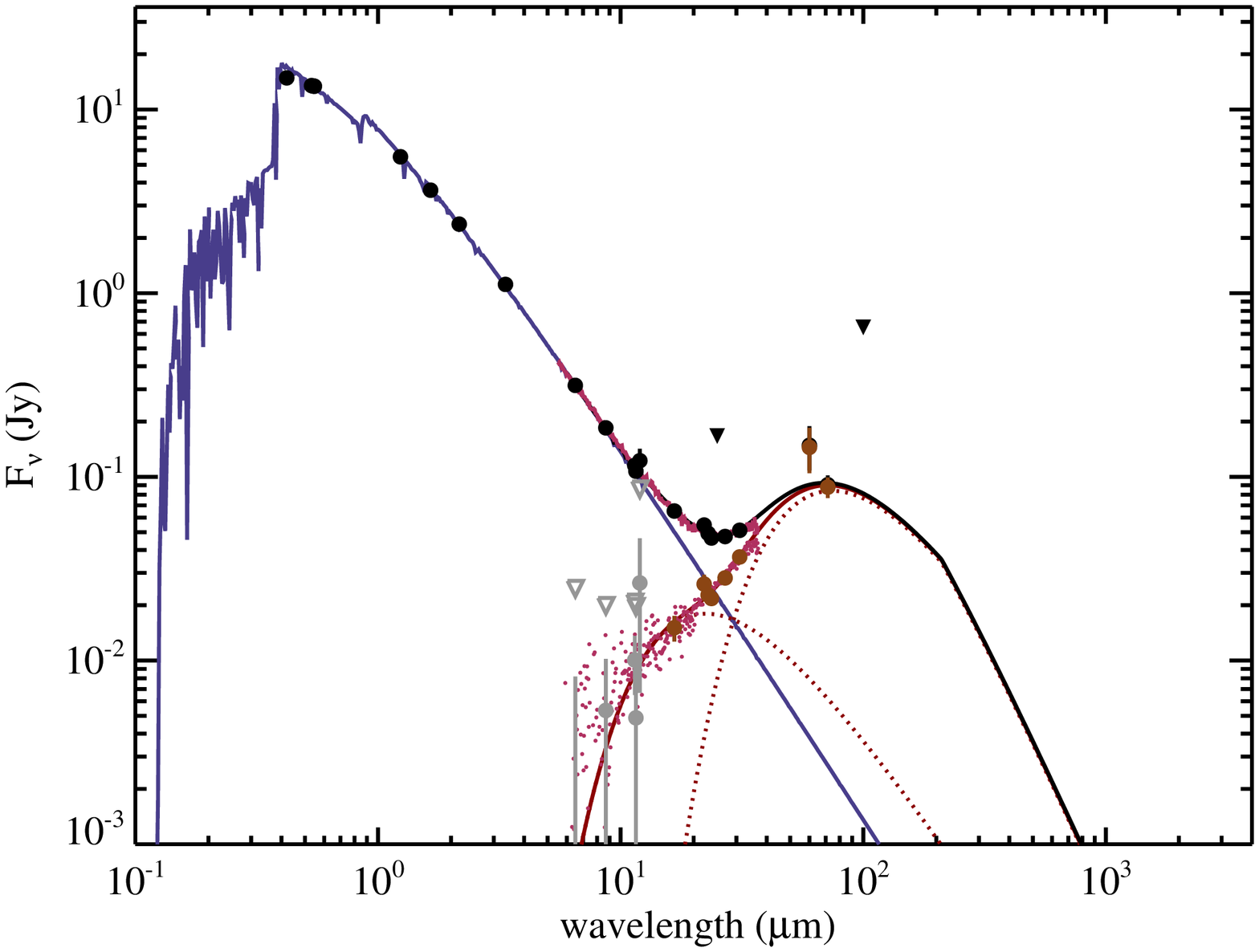}\\
\end{tabular}
\end{center}
\begin{center}
\begin{tabular}{cc}
  HD 80950 & HD 98673 \\
  \includegraphics[width=0.5\textwidth]{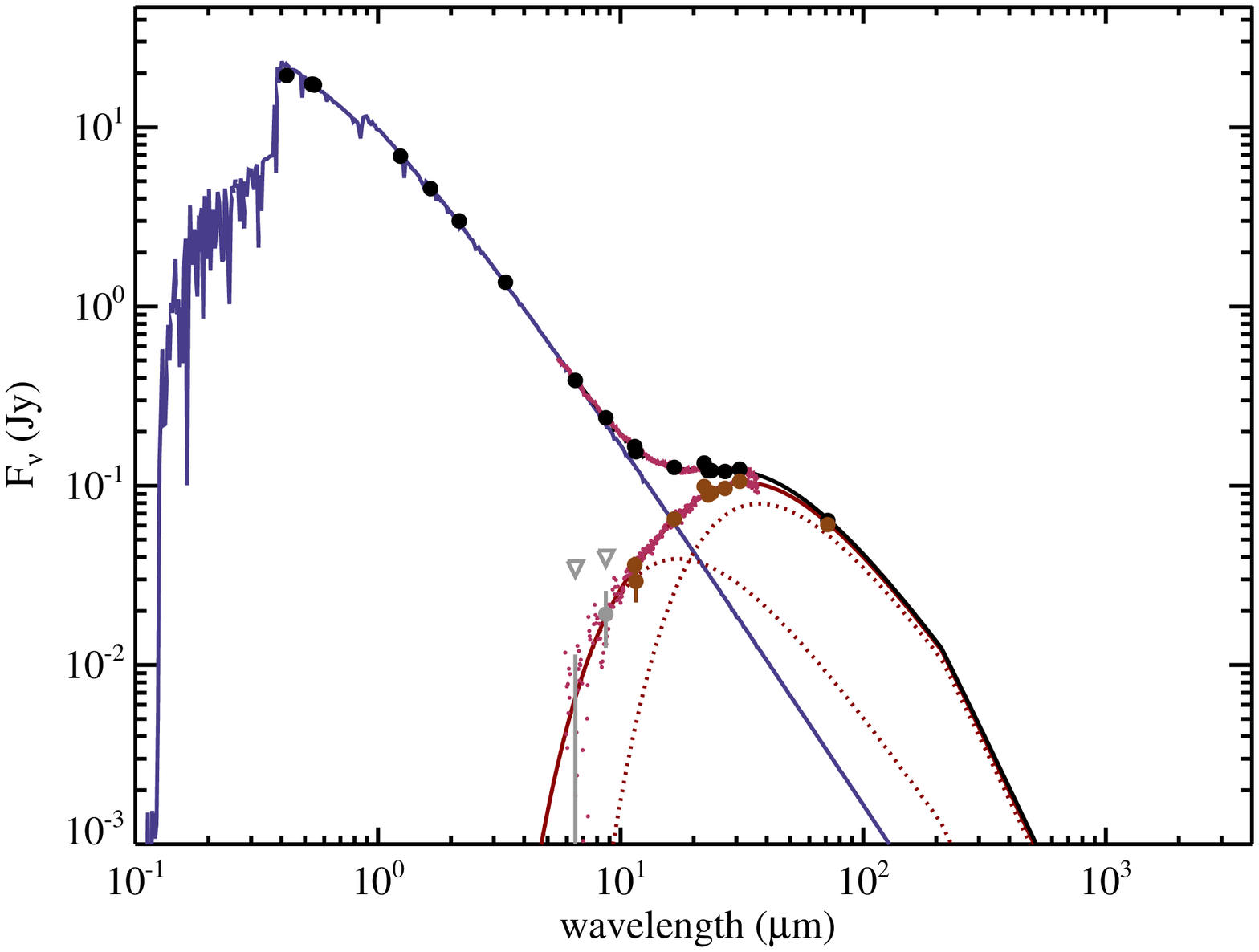}&
  \includegraphics[width=0.5\textwidth]{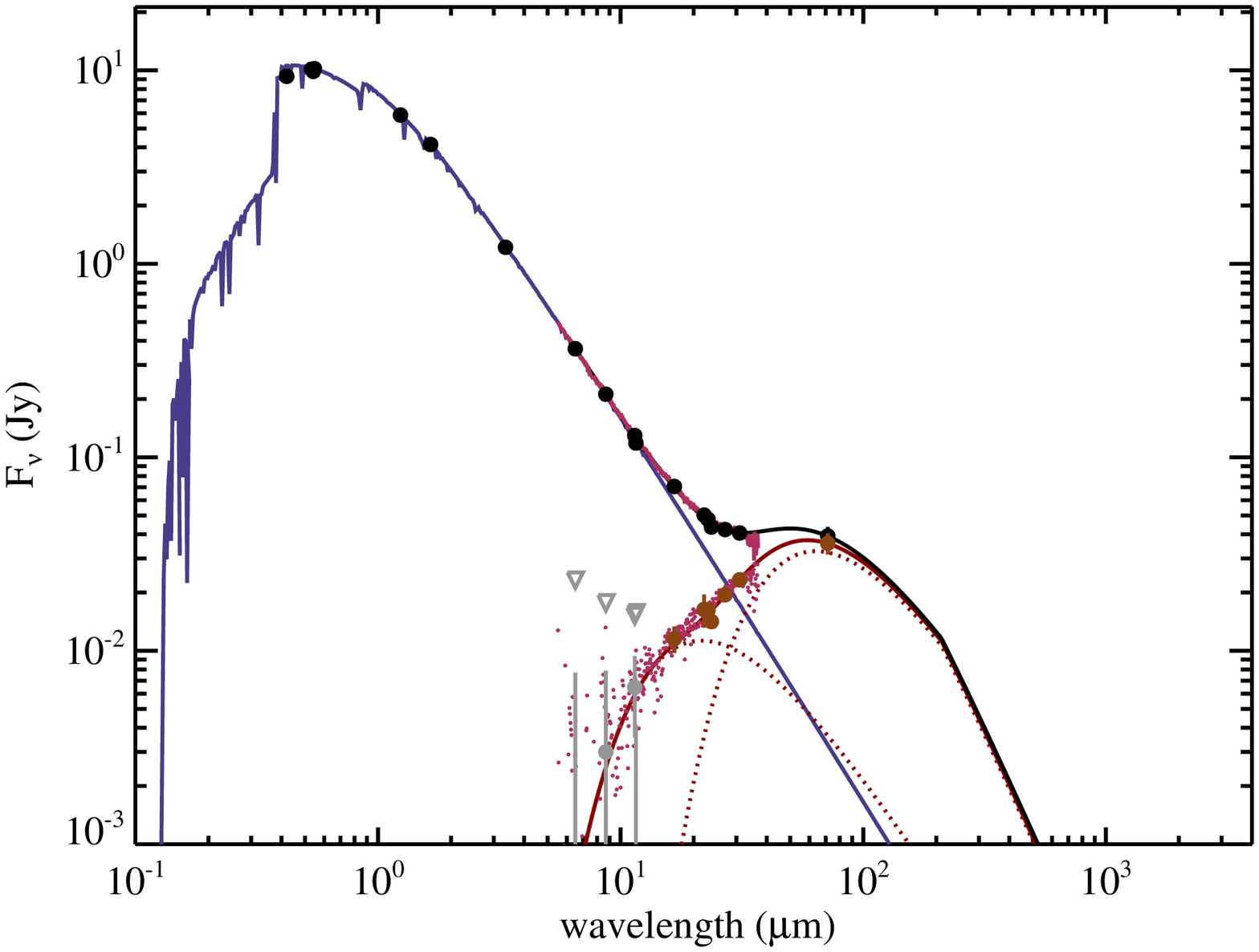}\\
  HD 107146 & HD 109085 \\
  \includegraphics[width=0.5\textwidth]{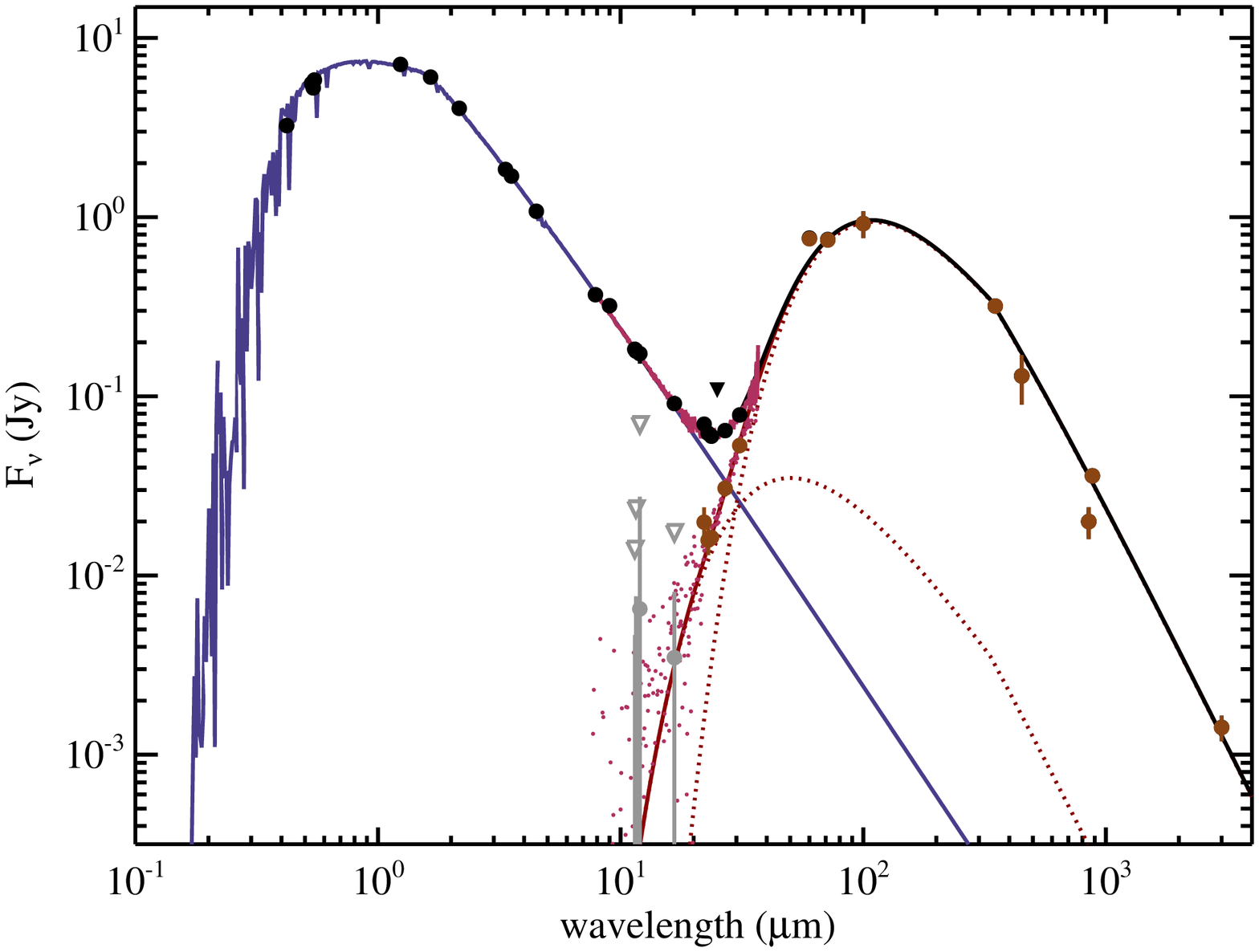}&
  \includegraphics[width=0.5\textwidth]{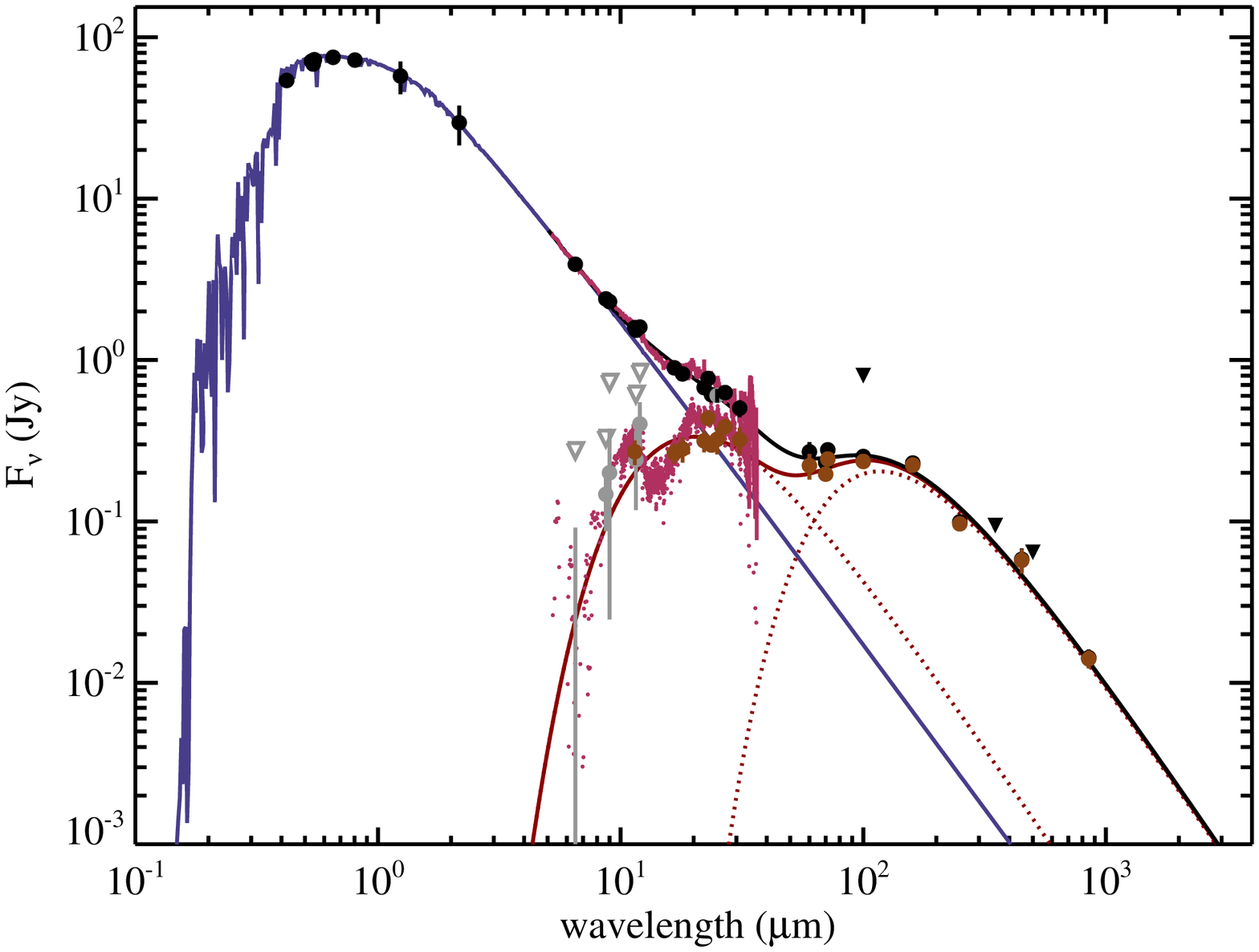}\\
  HD 110411 & HD 125162 \\
  \includegraphics[width=0.5\textwidth]{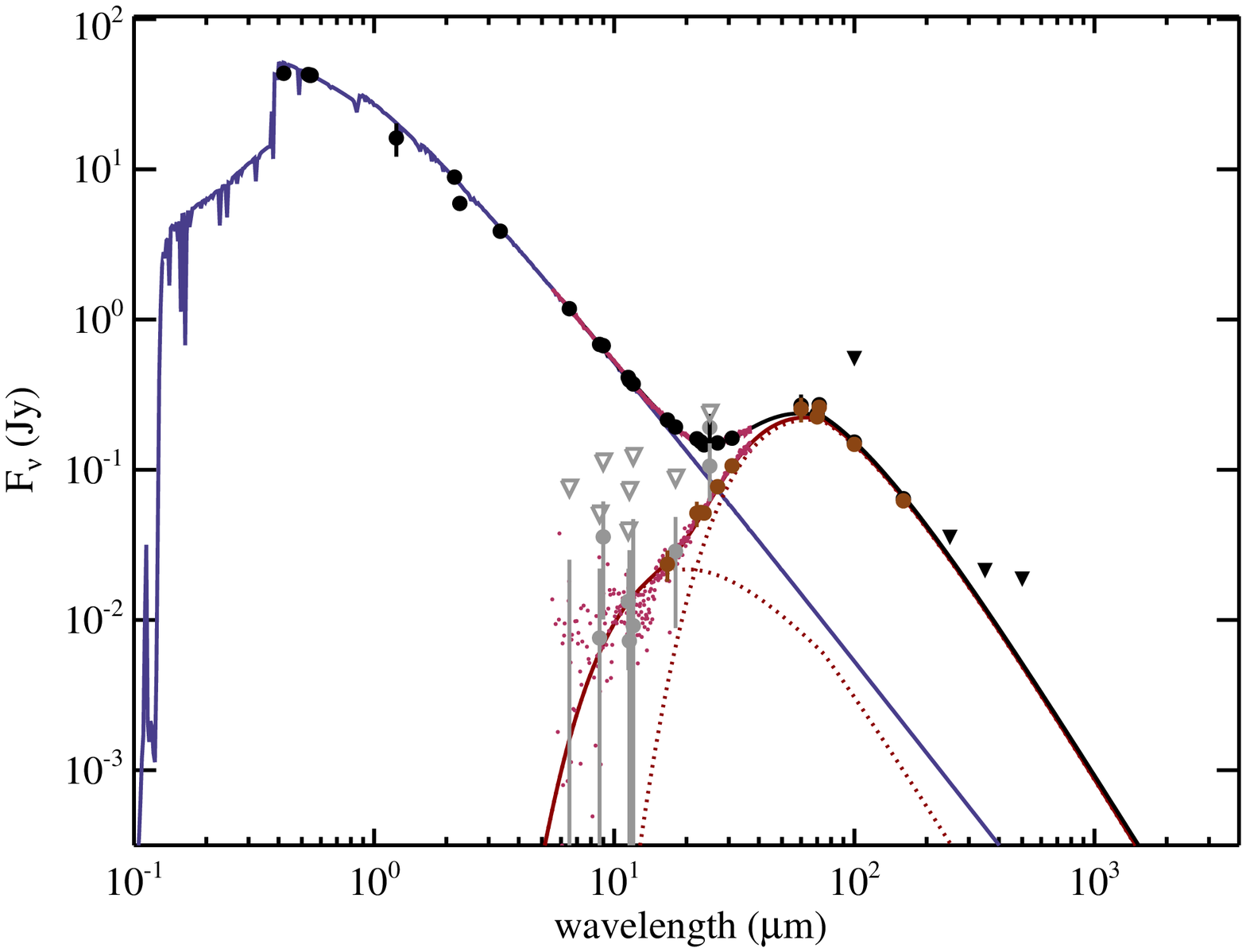}&
  \includegraphics[width=0.5\textwidth]{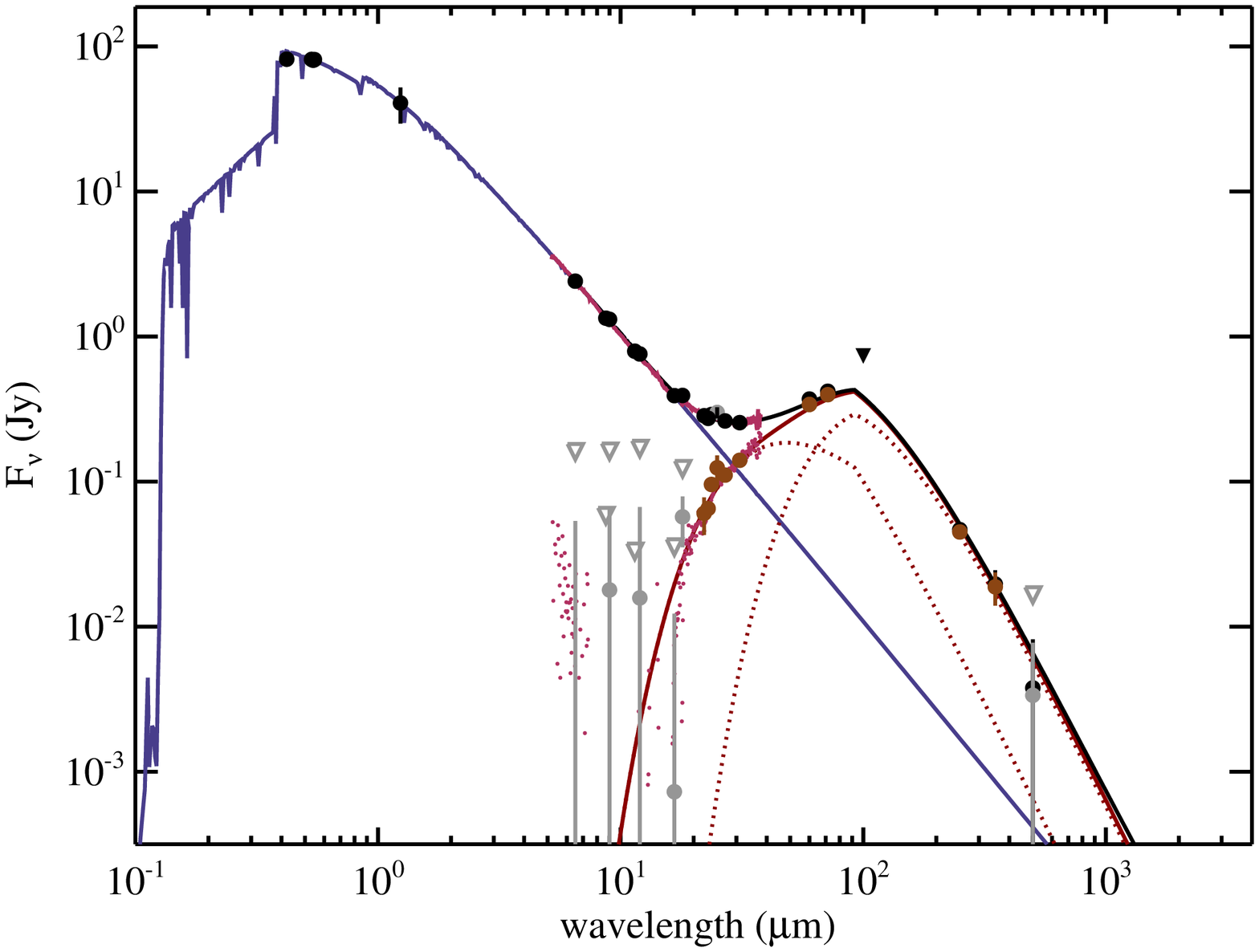}\\
\end{tabular}
\end{center}
\begin{center}
\begin{tabular}{cc}
  HD 136246 & HD 136482 \\
  \includegraphics[width=0.5\textwidth]{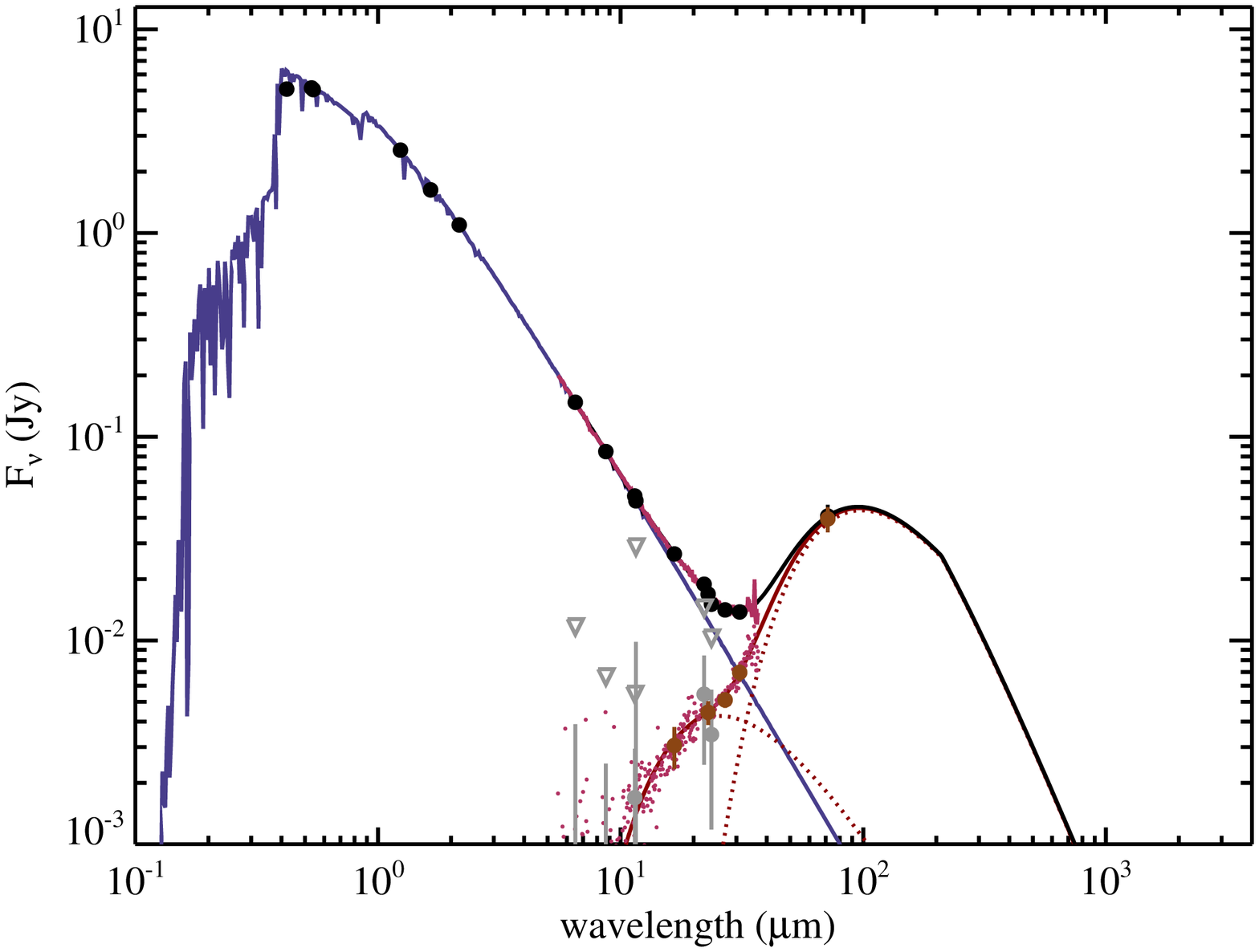}&
  \includegraphics[width=0.5\textwidth]{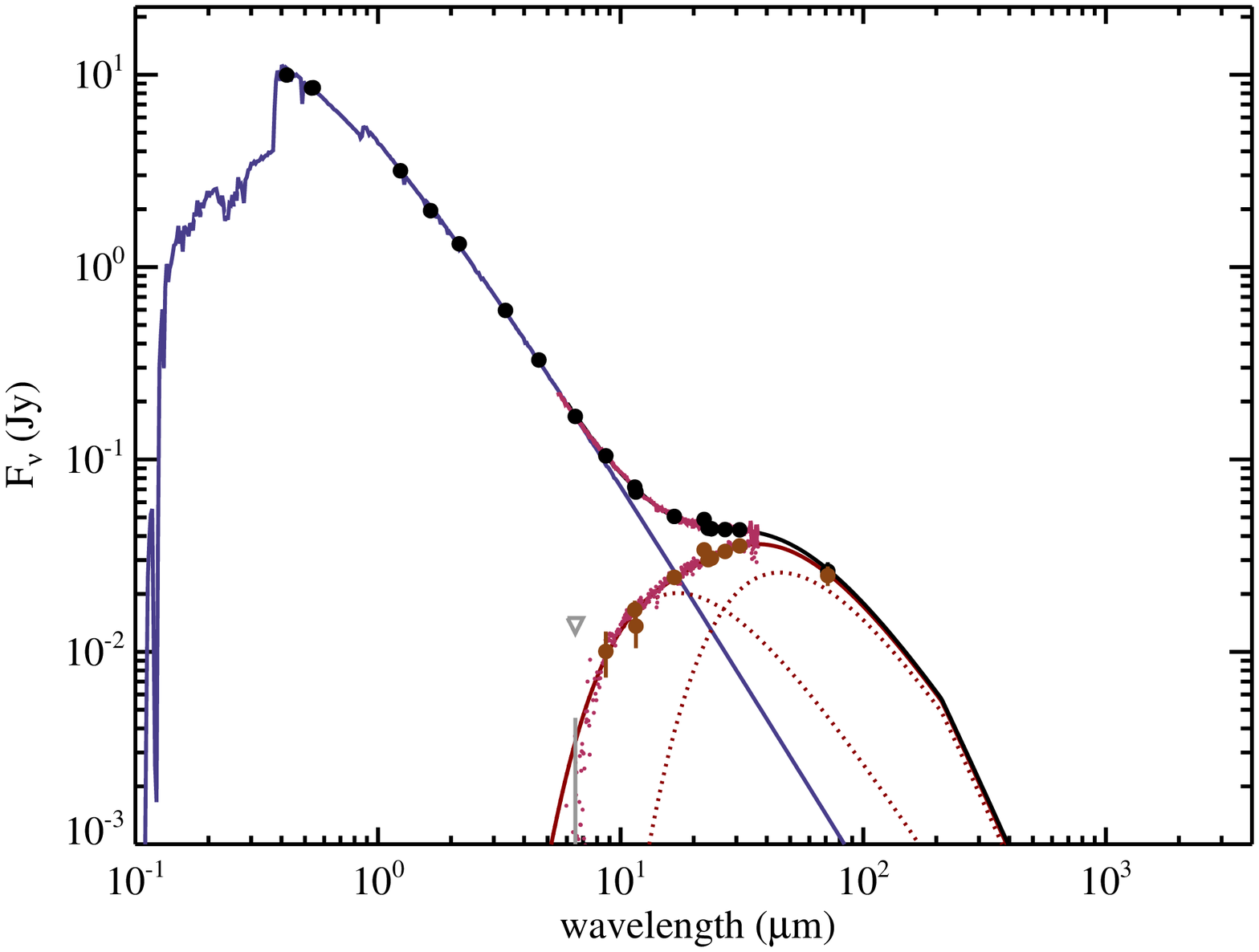}\\
  HD 138965 & HD 141378 \\
  \includegraphics[width=0.5\textwidth]{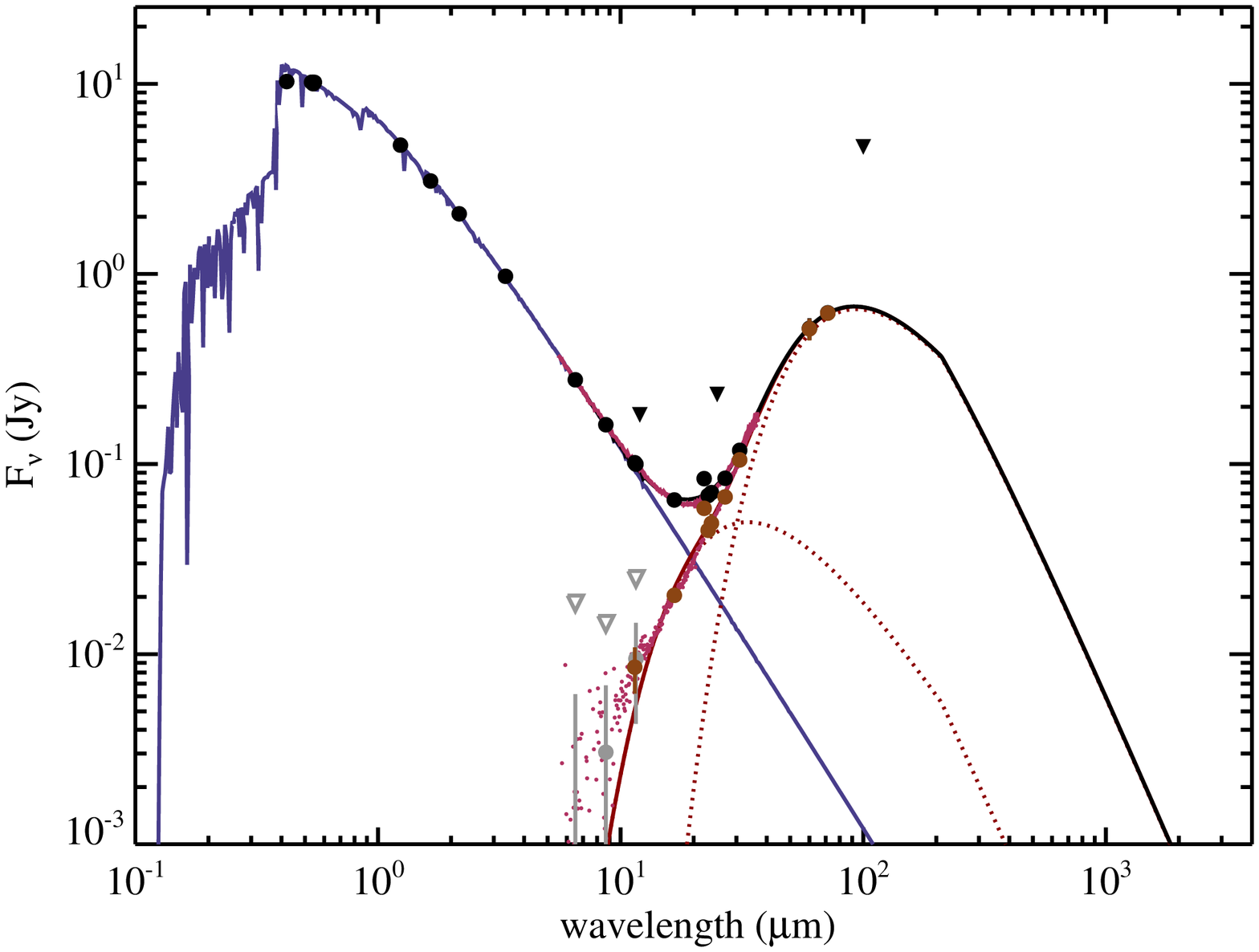}&
  \includegraphics[width=0.5\textwidth]{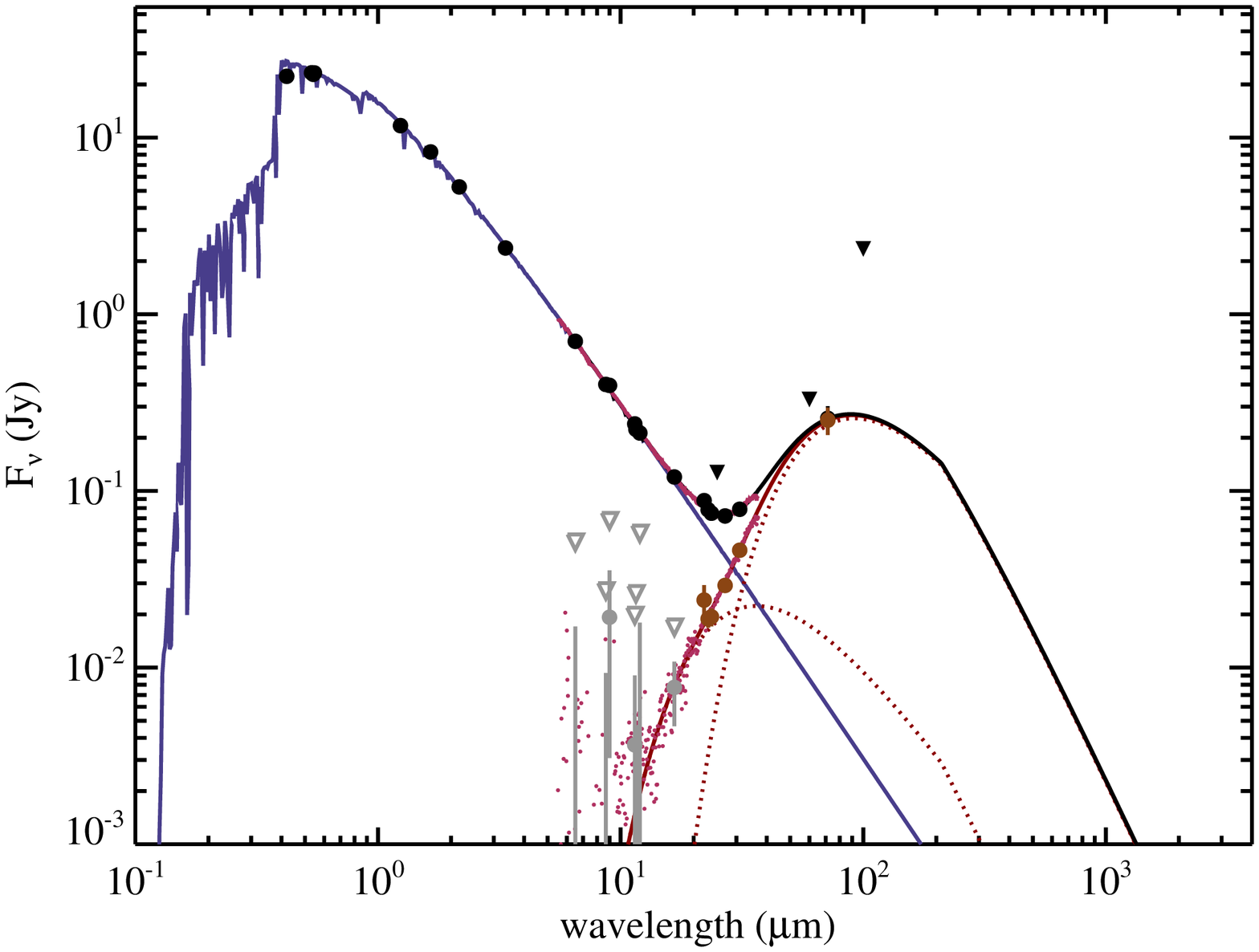}\\
  HD 153053 & HD 159492 \\
  \includegraphics[width=0.5\textwidth]{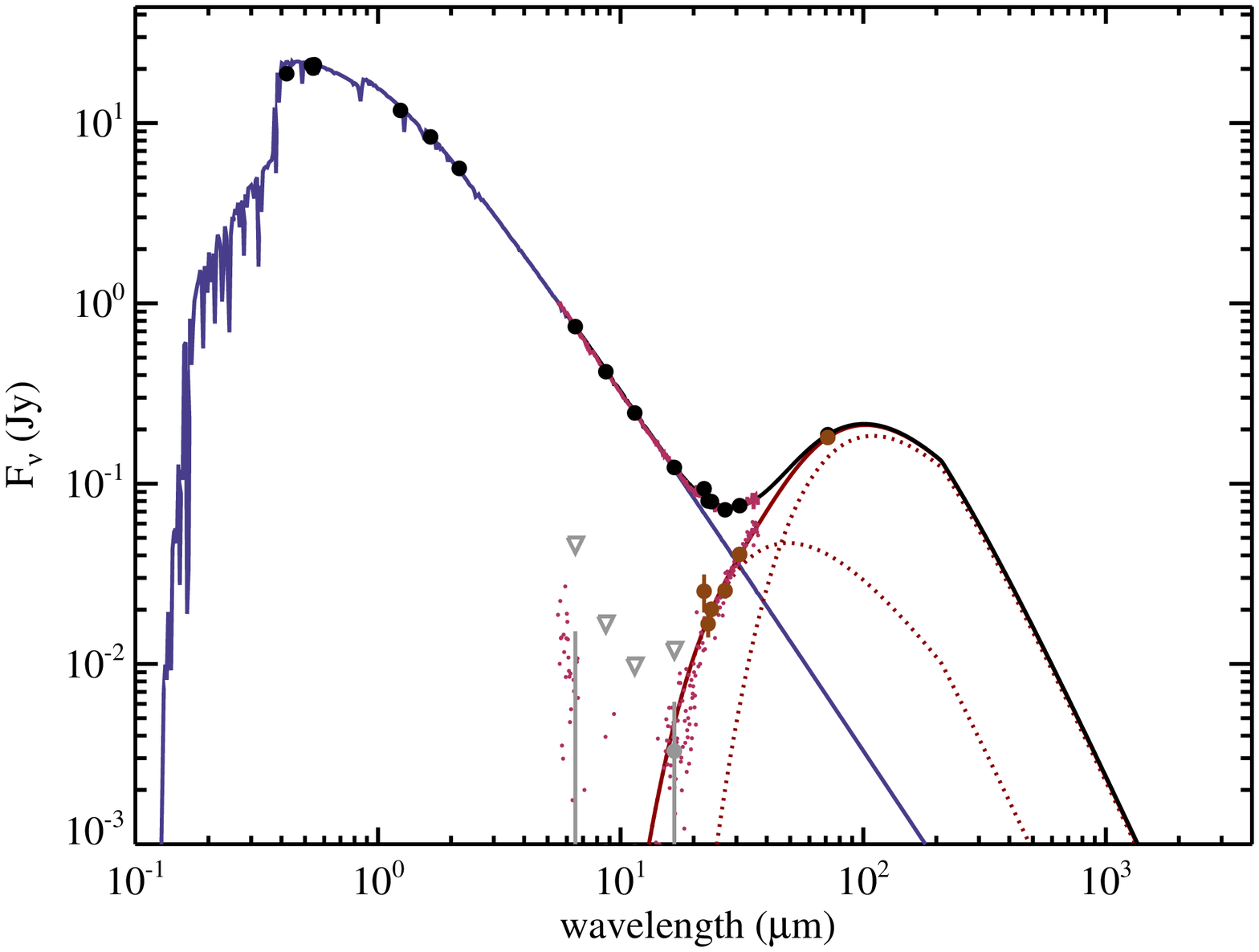}&
  \includegraphics[width=0.5\textwidth]{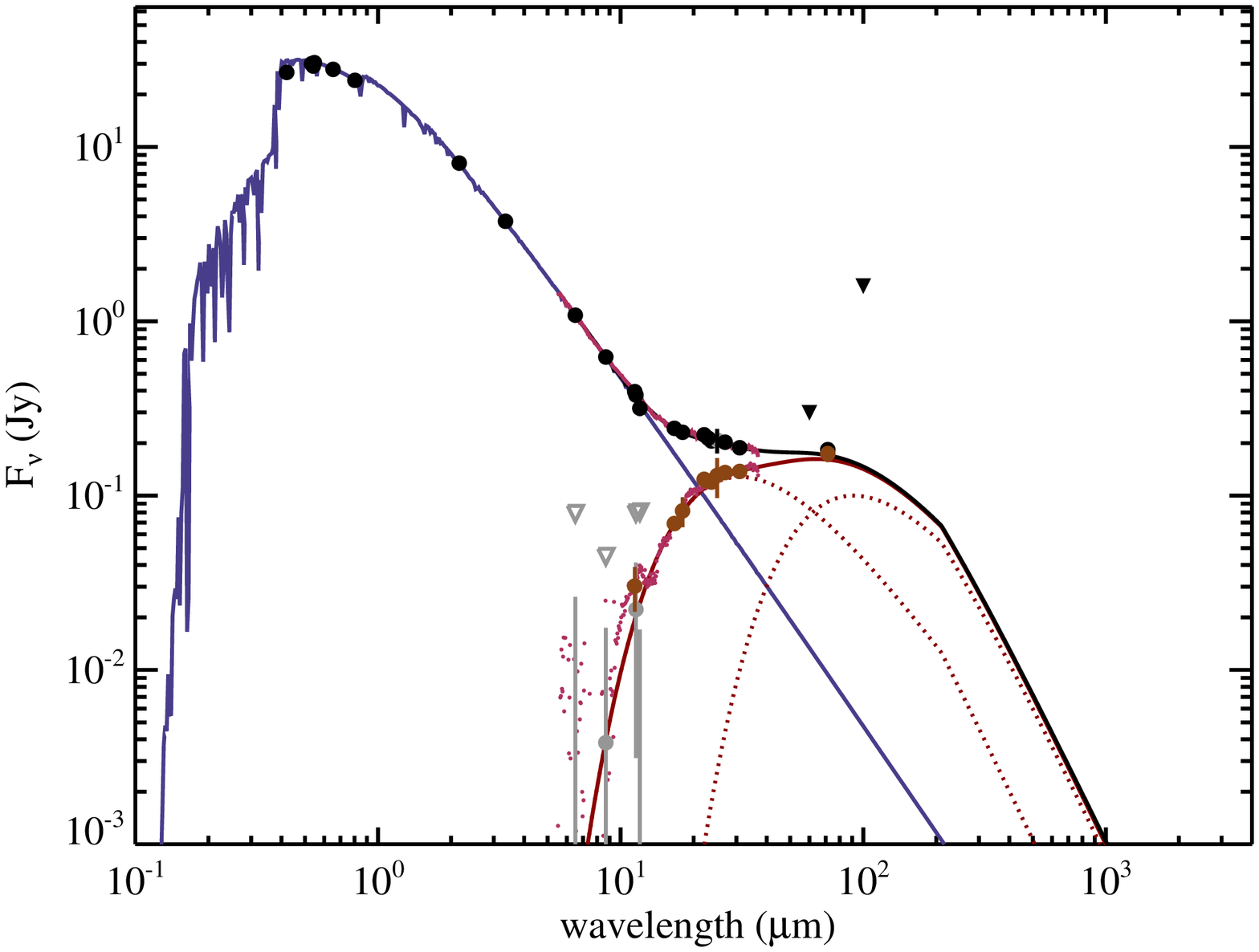}\\
\end{tabular}
\end{center}
\begin{center}
\begin{tabular}{cc}
  HD 161868 & HD 172167 \\
  \includegraphics[width=0.5\textwidth]{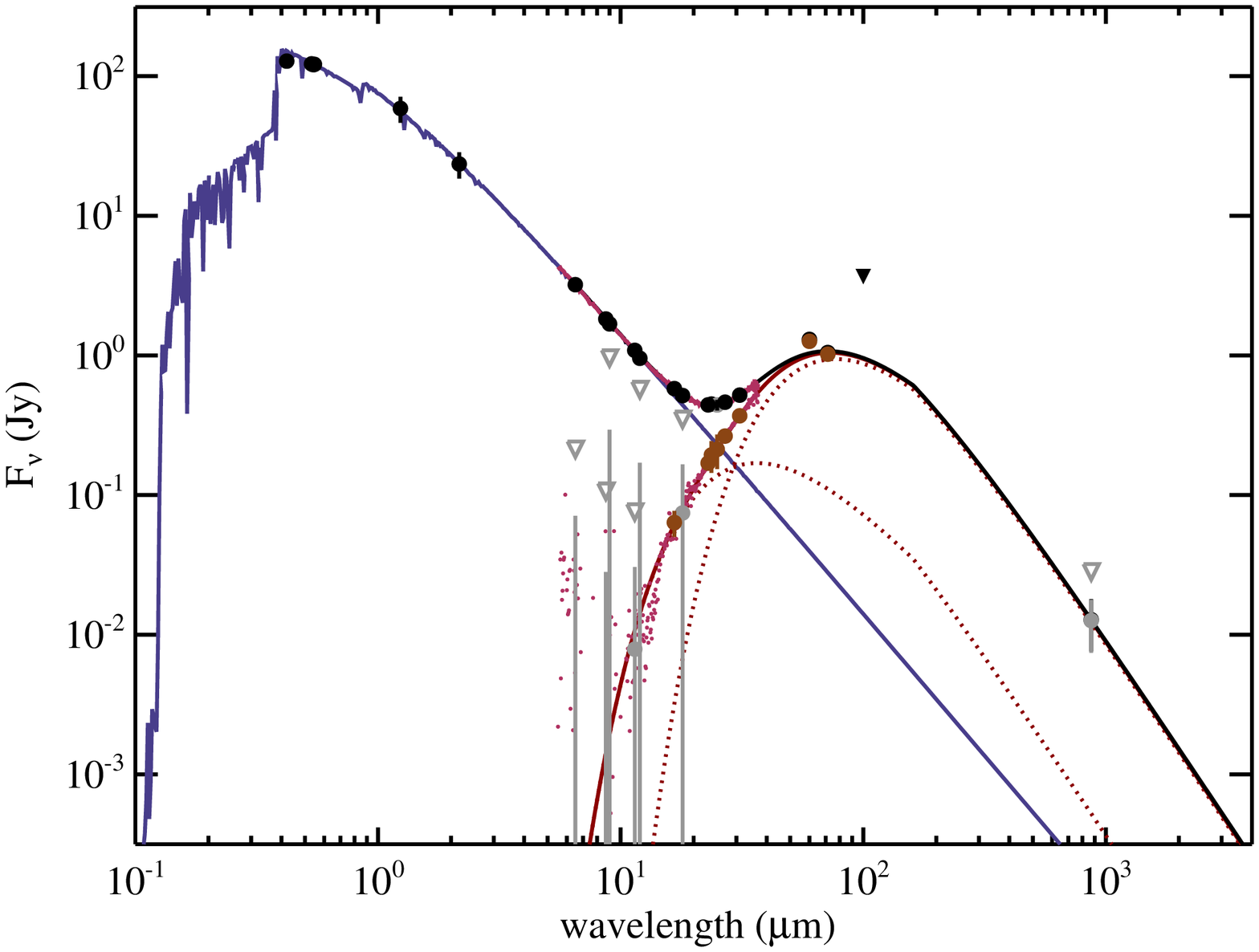}&
  \includegraphics[width=0.5\textwidth]{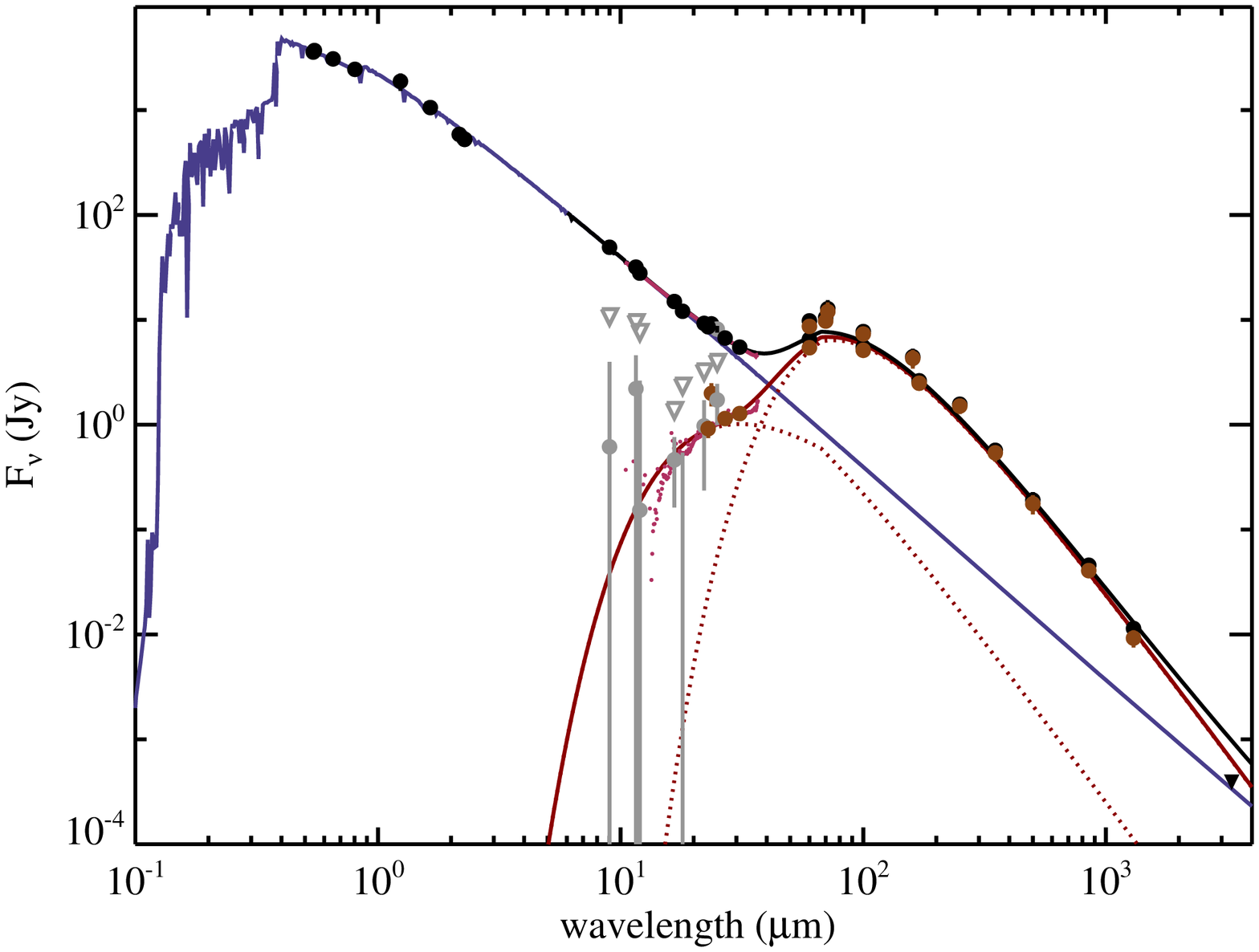}\\
  HD 181296 & HD 181327 \\
  \includegraphics[width=0.5\textwidth]{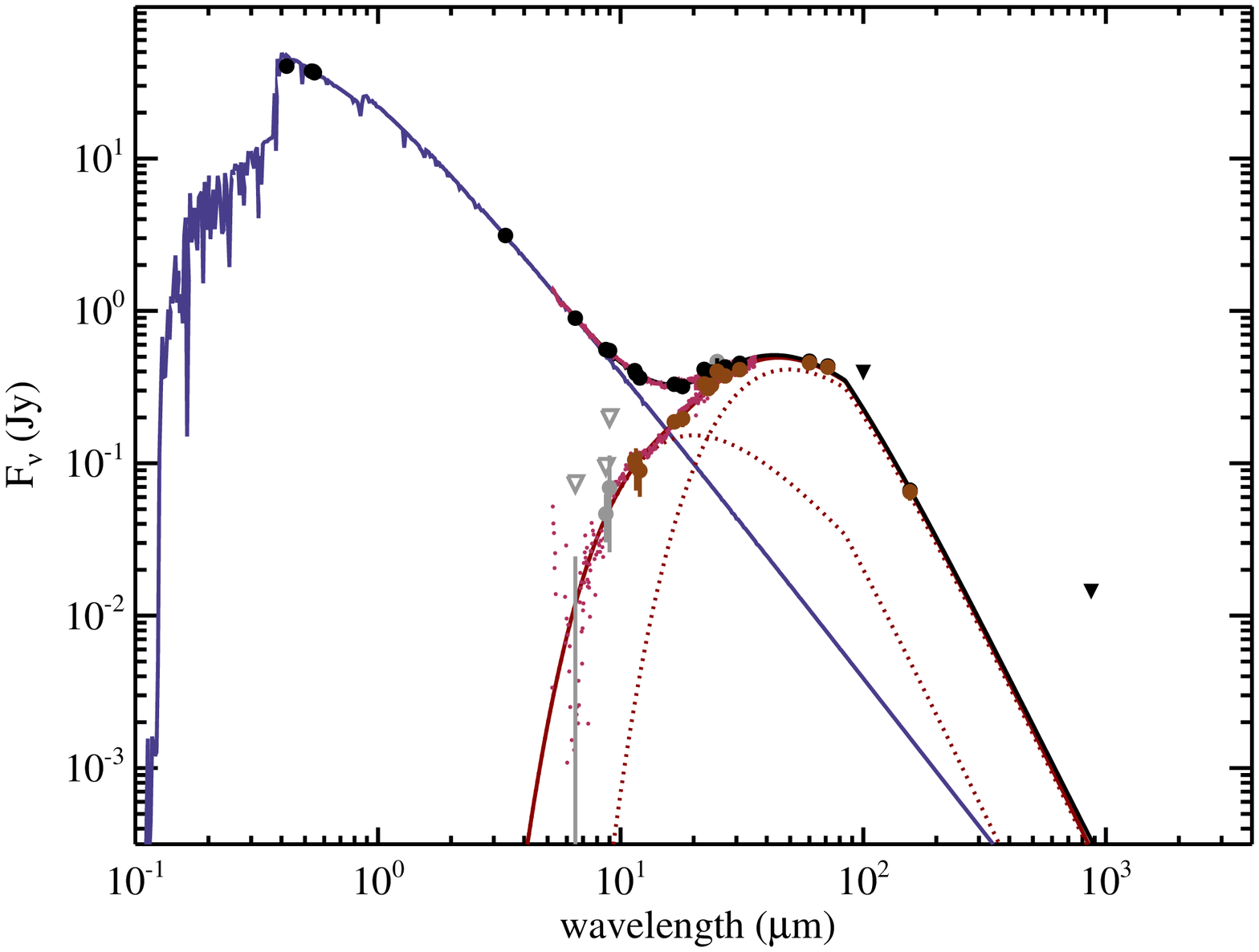}&
  \includegraphics[width=0.5\textwidth]{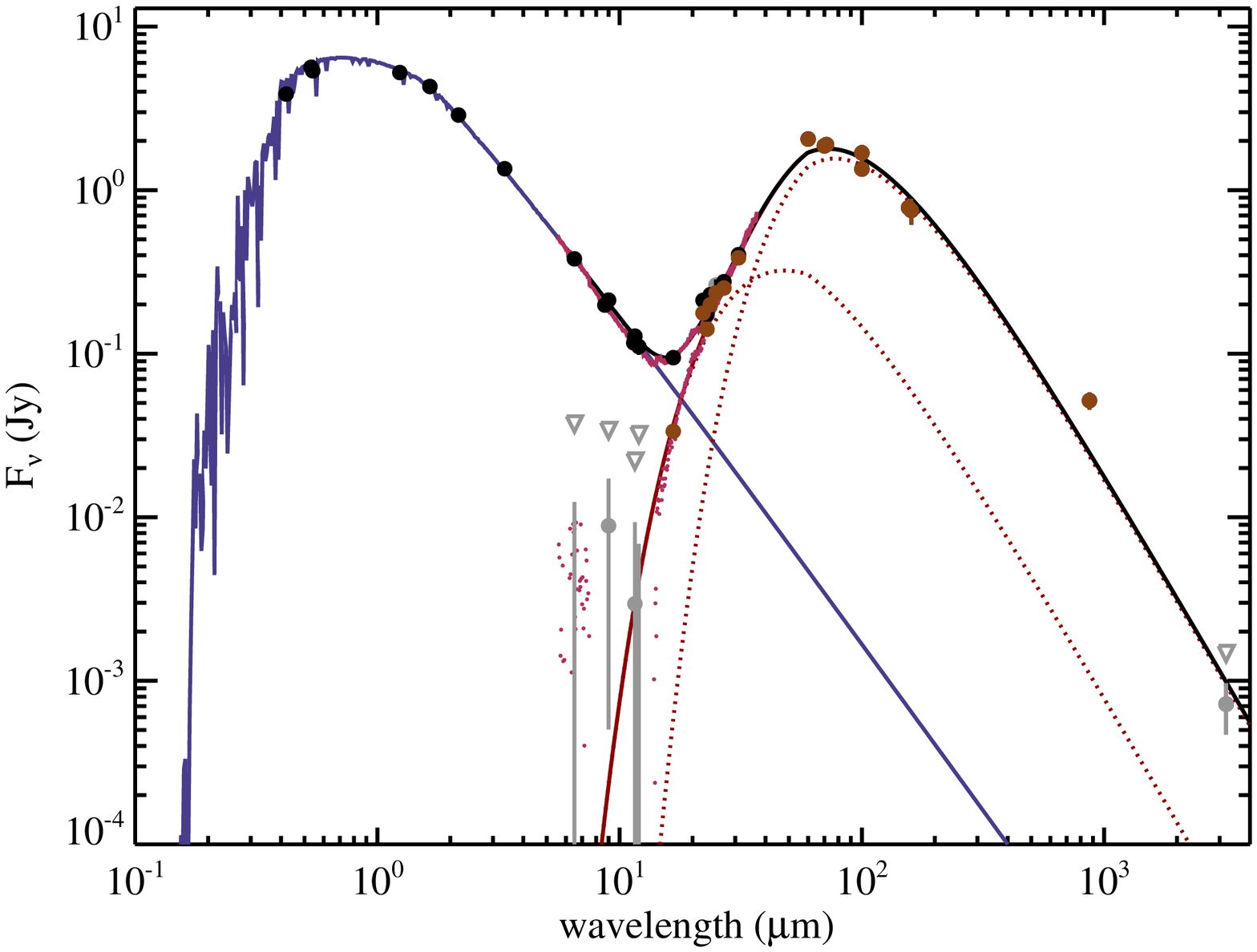}\\
  HD 182919 & HD 191174 \\
  \includegraphics[width=0.5\textwidth]{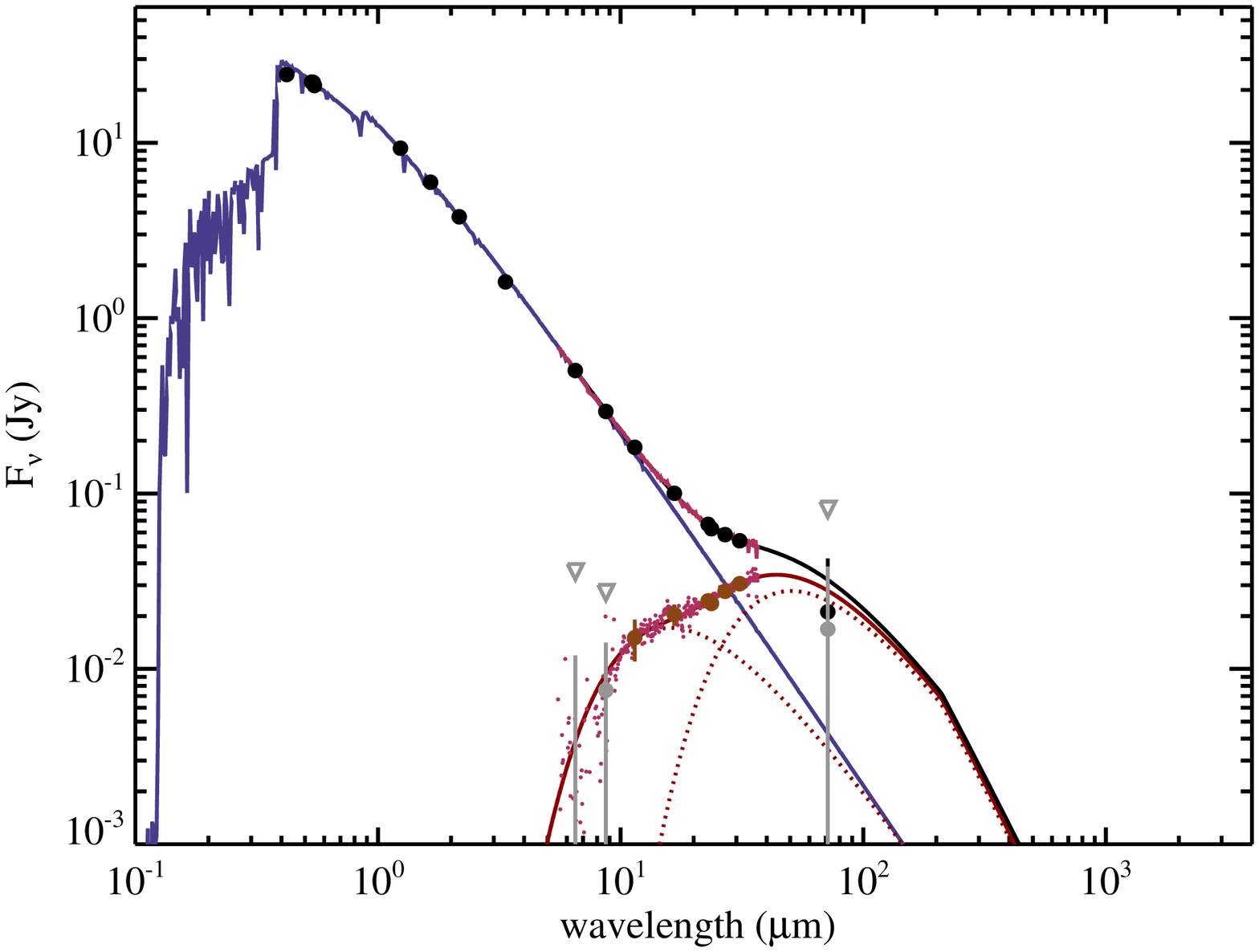}&
  \includegraphics[width=0.5\textwidth]{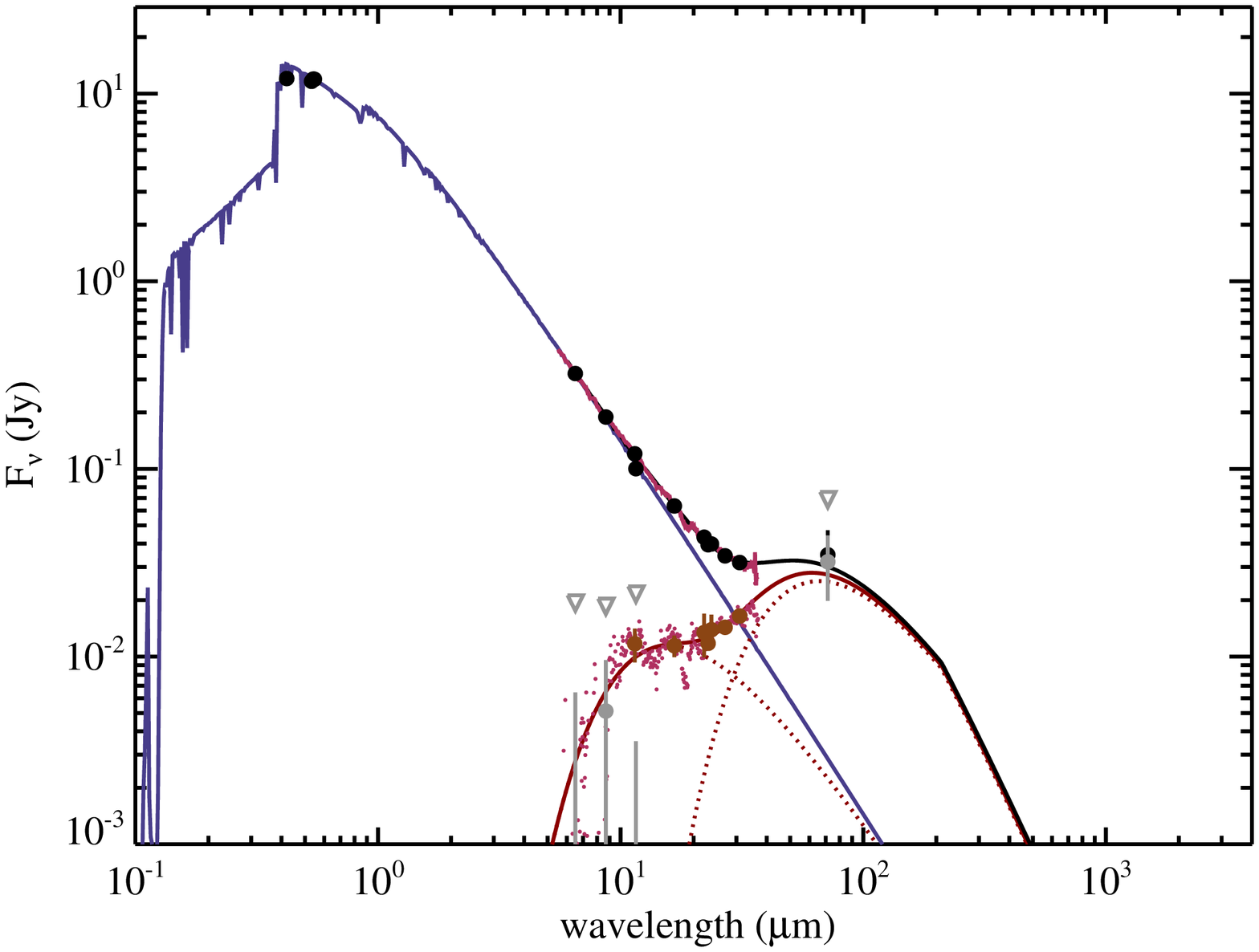}\\
\end{tabular}
\end{center}
\begin{center}
\begin{tabular}{cc}
  HD 192425 & HD 205674 \\
  \includegraphics[width=0.5\textwidth]{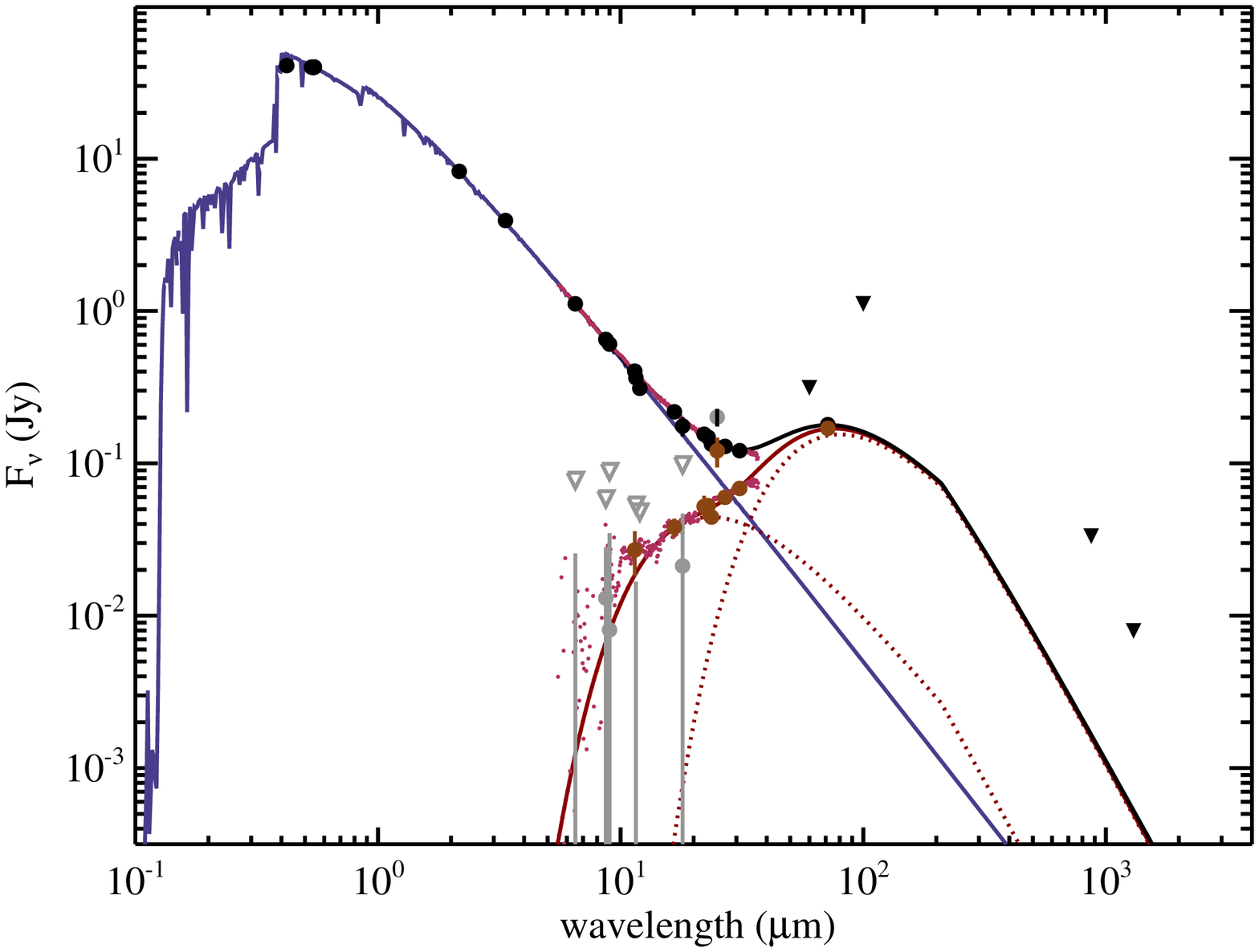}&
  \includegraphics[width=0.5\textwidth]{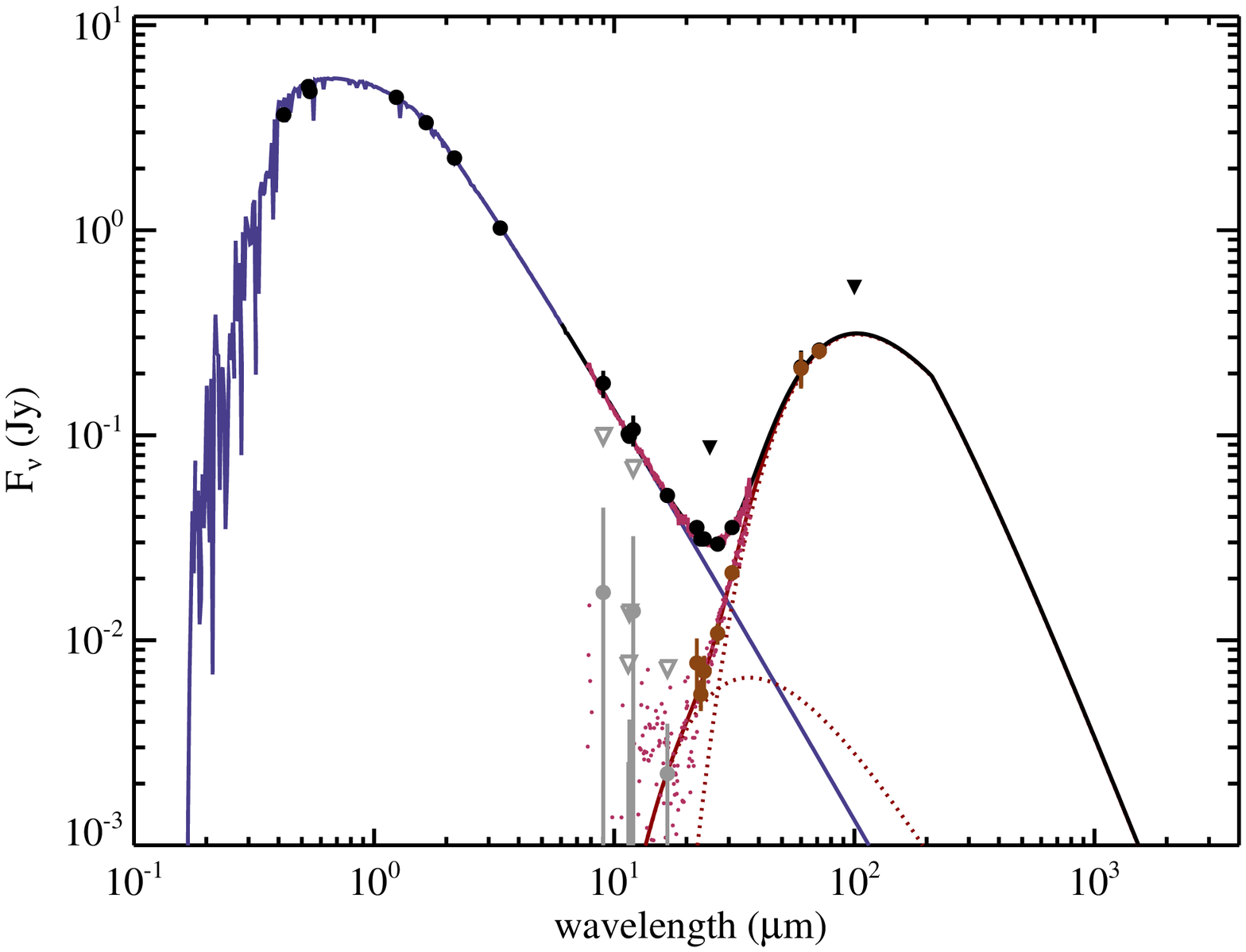}\\
  HD 216956 & HD 218396 \\
  \includegraphics[width=0.5\textwidth]{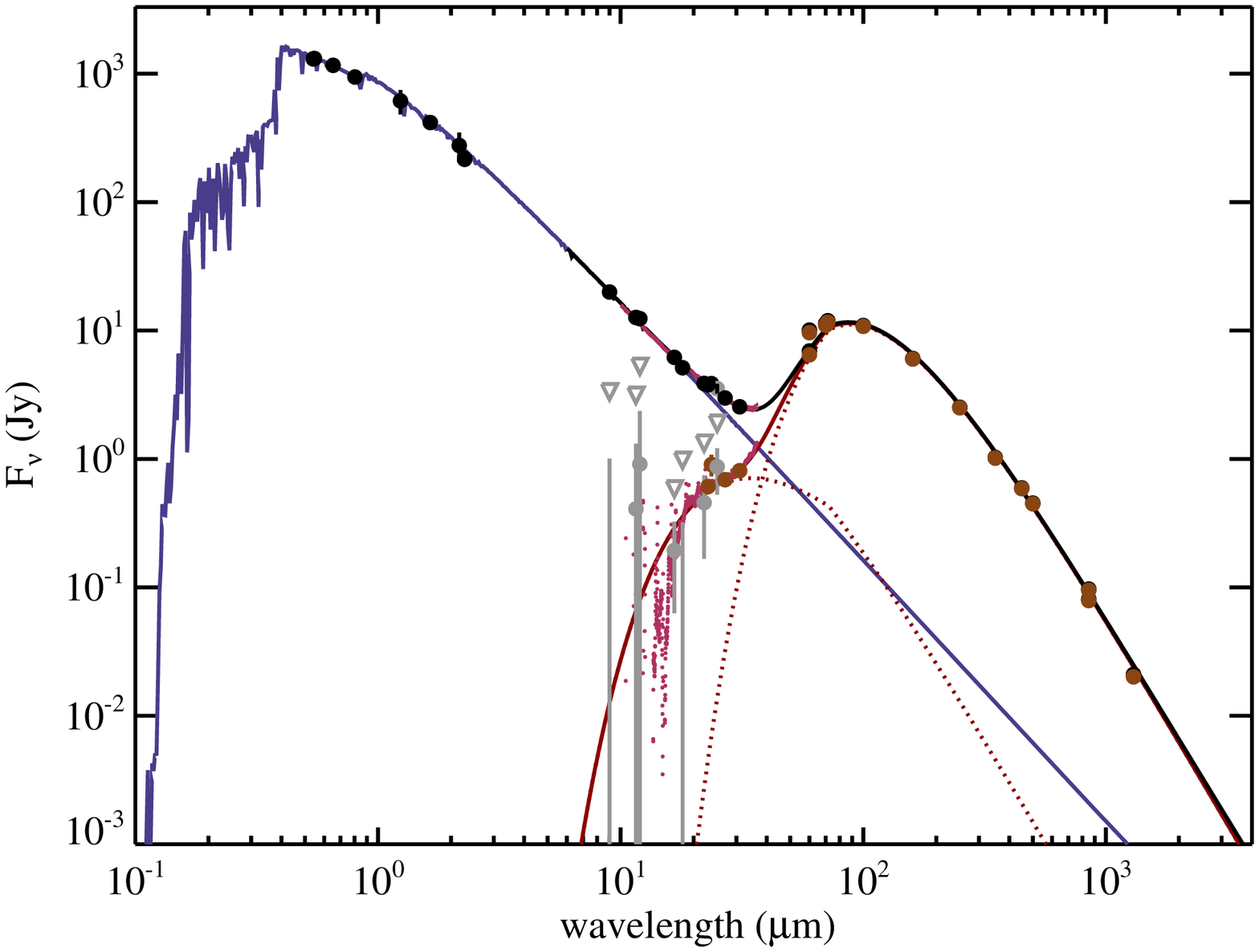}&
  \includegraphics[width=0.5\textwidth]{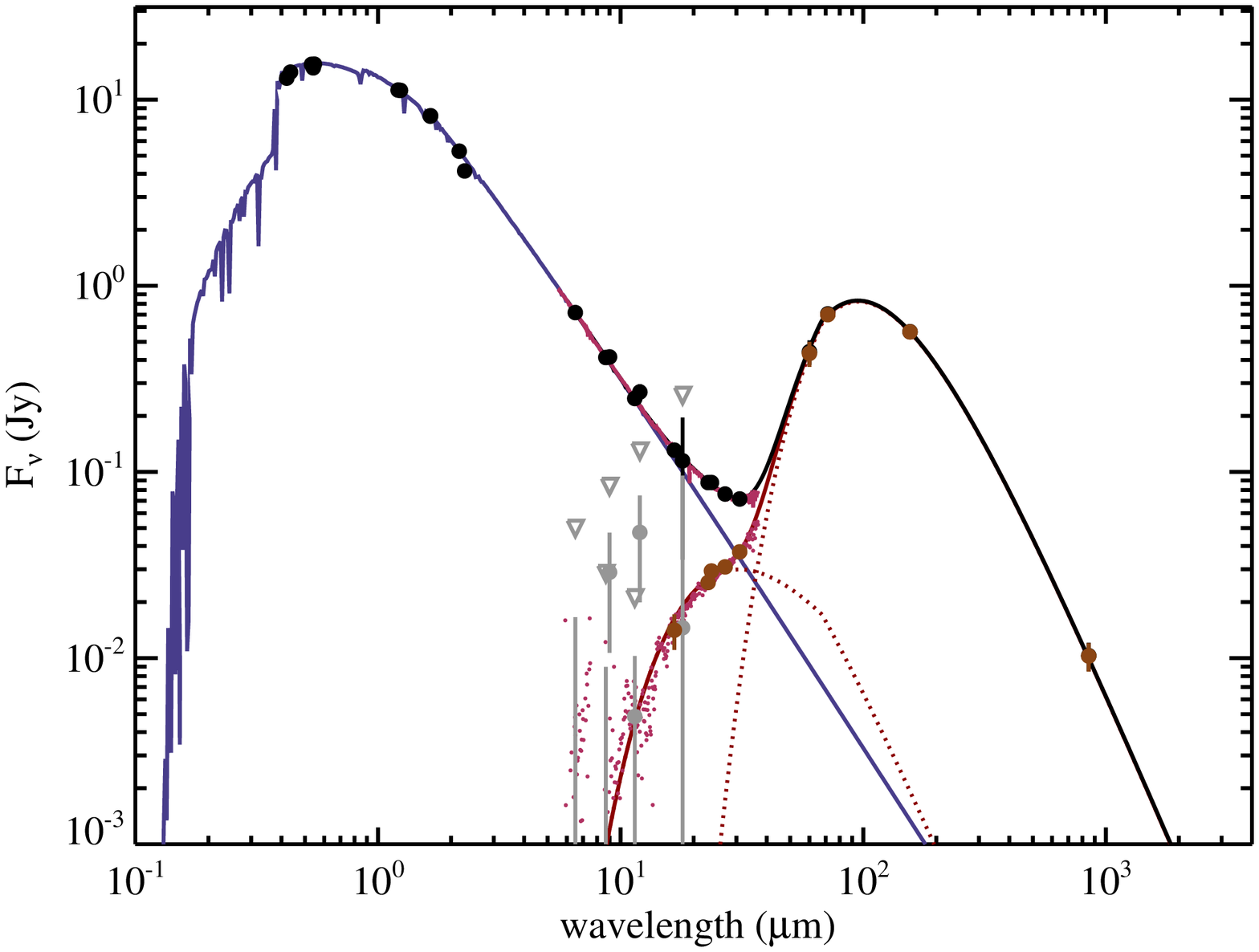}\\
  HD 221853 & HD 225200 \\
  \includegraphics[width=0.5\textwidth]{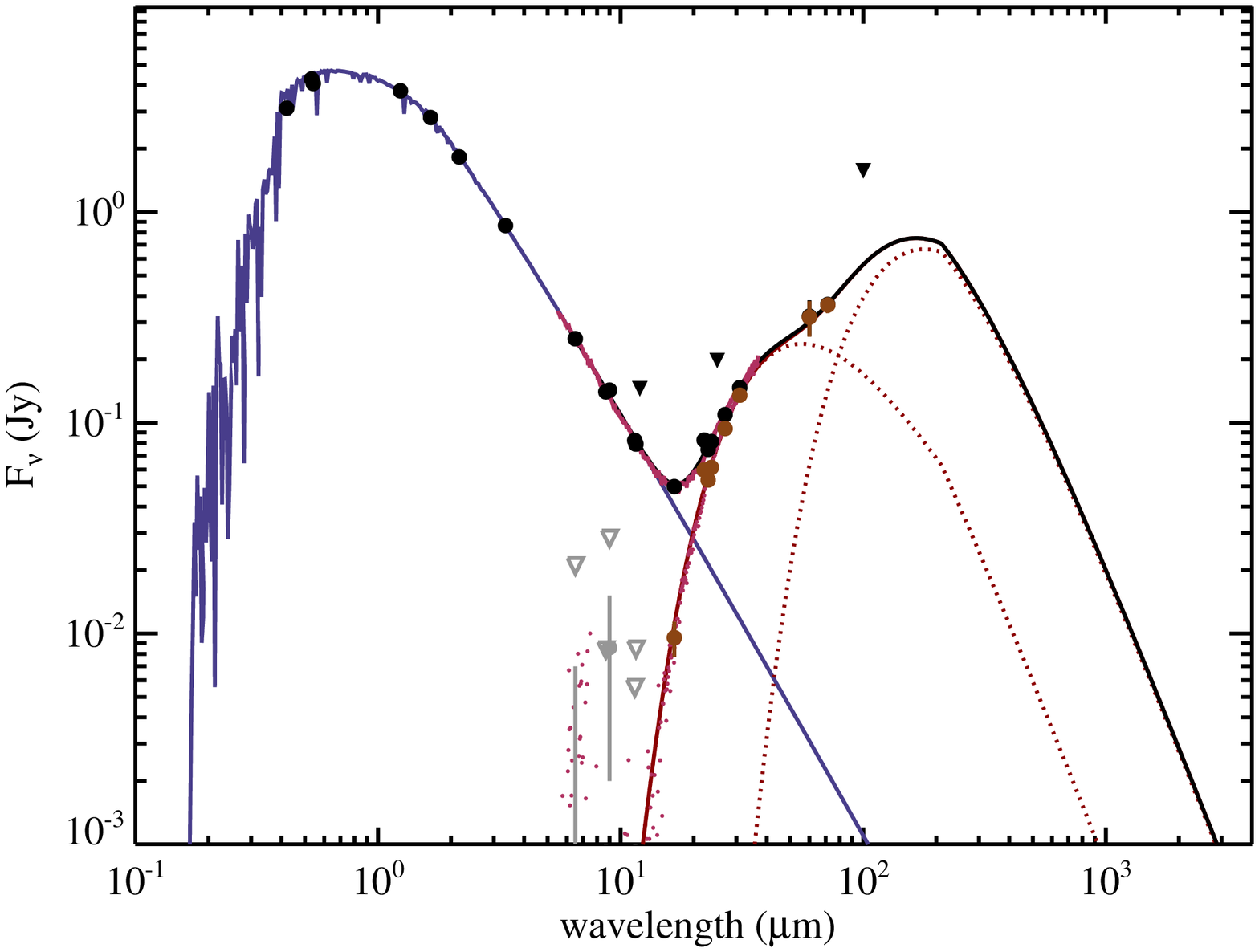}&
  \includegraphics[width=0.5\textwidth]{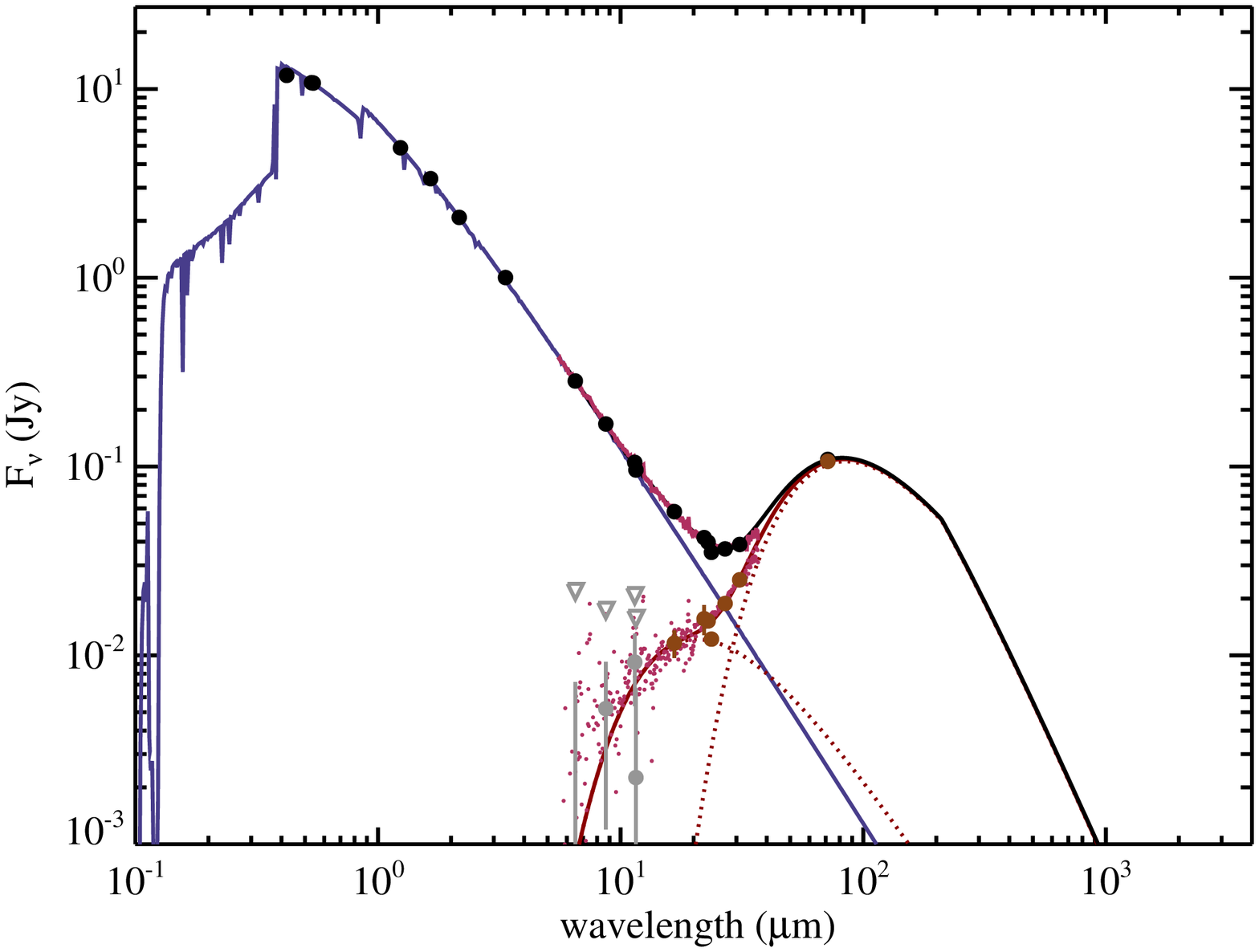}\\
\end{tabular}
\end{center}

\end{document}